\documentclass[twocolumn, times]{aastex631}
\usepackage{amsmath, amssymb, amsfonts}
\usepackage{enumitem}
\usepackage{cancel}
\usepackage{url}
\usepackage{graphicx}
\usepackage{pgfplots}
\usepackage[caption=false]{subfig}
\usepackage{tikz}
\tikzset{>=latex} 
\usepackage{ifthen}
\usepackage{xcolor}
\usepackage{physics}
\usepackage{algorithm}
\usepackage[noend]{algpseudocode}
\usepackage{soul}

\usetikzlibrary{patterns}
\pgfplotsset{compat=1.18}

\colorlet{wavecol}{orange!35!black}
\colorlet{freqcol}{green!25!black}
\colorlet{enercol}{blue!35!black}
\pgfdeclareverticalshading{rainbow}{100bp}{
  color(0bp)=(red); color(25bp)=(red); color(35bp)=(yellow);
  color(45bp)=(green); color(55bp)=(cyan); color(65bp)=(blue);
  color(75bp)=(violet); color(100bp)=(violet)
}

\newcommand{\minpts}{\text{minPts}} 
\newcommand{\figref}[1]{#1}
\newcommand{\cc}[1]{#1}

\newcommand{\Sphere}[2]{
  {sin(#1)*cos(#2)}, 
  {cos(#1)*cos(#2)}, 
  {sin(#2)}          
}

\newcommand{\Circle}[2]{
  {#1}, 
  {sqrt(1-(#1)^2)*cos(#2)}, 
  {sqrt(1-(#1)^2)*sin(#2)}          
}

\newcommand{\bfv}{
  \mathbf{v}
}
\newcommand{\RR}{\mathbb{R}}
\DeclareMathOperator{\arcsinh}{arcsinh}

\def\RotationX{-30}
\def\RotationY{-20}
\def\angle{6}

\newenvironment{conditions}
  {\par\vspace{\abovedisplayskip}\noindent\begin{tabular}{>{$}l<{$} @{${}={}$} l}}
  {\end{tabular}\par\vspace{\belowdisplayskip}}
\shorttitle{NEOWISE Flux Anomaly Extraction}
\shortauthors{M. Paz}
\graphicspath{{./}{ }}

\begin{document}

\title{A Sub-Millisecond Fourier and Wavelet Based Model to Extract Variable Candidates from the NEOWISE Single-Exposure Database}

\author[0009-0008-2229-3138]{Matthew Paz}
\affiliation{Pasadena High School \\ 
2925 E. Sierra Madre Blvd \\
Pasadena, California 91107, USA}

\affiliation{California Institute of Technology \\ 
1200 E. California Blvd \\
Pasadena, California 91125, USA}



\received{12 Apr 2024}
\revised{13 Aug 2024}
\accepted{23 Sept 2024}

\begin{abstract}
\cc{This paper presents VARnet, a capable signal processing model for rapid astronomical timeseries analysis. VARnet leverages wavelet decomposition, a novel method of Fourier feature extraction via the Finite-Embedding Fourier Transform (FEFT), and deep learning to detect faint signals in light curves, utilizing the strengths of modern GPUs to achieve sub-millisecond single-source runtime. We apply VARnet to the NEOWISE Single-Exposure Database, which holds nearly 200 billion apparitions over 10.5 years of infrared sources on the entire sky. This paper devises a pipeline in order to extract variable candidates from the NEOWISE data, serving as a proof of concept for both the efficacy of VARnet and methods for an upcoming variability survey over the entirety of the NEOWISE dataset. We implement models and simulations to synthesize unique light curves to train VARnet. In this case, the model achieves an F1 score of $0.91$ over a 4-class classification scheme on a validation set of real variable sources present in the infrared. With $\sim2000$ points per light curve on a GPU with 22GB of VRAM, VARnet produces a per-source processing time of $<53\mu s$. We confirm that our VARnet is sensitive and precise to both known and previously undiscovered variable sources. These methods prove promising for a complete future survey of variability with WISE, and effectively showcase the power of the VARnet model architecture.}

\end{abstract}





\section{Introduction} \label{sec:intro}
The Wide-field Infrared Survey Explorer (WISE) Space Telescope \citep{wisepaper} is an orbiting space telescope observatory, \cc{having recently collected its final piece of data on July 31st}. It is unique in the timescale of the data collected: WISE initially operated as a four-band infrared survey from December 2009 until it had exhausted its supply of solid hydrogen cryogen by late 2010. It then collected data briefly for 4 months as a two-band survey, since the two shortest wavelength bands do not require coolant to operate. After a hibernation period from 2011 to December 2013, the mission was reactivated as NEOWISE. After this reactivation, data was collected continuously in its two shortest wavelength bands. Due to this long timescale and infrared sensitivity, WISE presents a unique opportunity for time-domain astronomy.

Machine learning techniques have increasingly been useful to the field of astronomy \citep{pearsonpaper, kaitlynpaper, techniquesforlcc}, especially in time-domain astronomy and in large datasets where human analysis is too slow for the data ingress. Particularly, convolutional neural networks (CNN) have been the go-to choice for machine learning algorithms. \citet{pearsonpaper} explored such an application on Kepler mission data to detect exoplanet transits, leveraging its relatively short cadence and sensitivity to such events to detect likely signals throughout the time series. Other studies demonstrate the effectiveness of phase-folding to extract periodic signals in stars. For example, \citet{kaitlynpaper} implements a parallelized yet finely sampled phase-folding implementation along with a CNN also on Kepler data. Such studies have much merit where a high sampling rate is guaranteed, but methods such as phase folding become less powerful as the sampling rate decreases and varies, such as in WISE, and are also prohibitively slow at this magnitude of data. There has also been interest in the postprocessing of NEOWISE data to make analysis more feasible in light of its size. Projects such as unWISE \citep{unwise} seek to improve the data quality and sensitivity via image coadditions and post-processing of raw WISE images, but eliminate time domain information in the process. More recently came unTimely \citep{untimely}, a NEOWISE product which offers a compromise between improved data quality and sensitivity via coadditions, but only performing them on temporally close exposures. \citet{untimely} reduce the total amount of data and improve the quality since most single-frame noise is erased during image co-addition, but this also degrades or sometimes erases transient events with very short windows.

There has been a lack of progress in studying the WISE database in its "raw" form (single-exposure photometry). Its size, in the hundreds of billions of rows, is no doubt a limiting factor. This paper demonstrates a novel machine learning model and pipeline to tackle this problem within a realistic time frame. Its objective is to analyze the light curve of any source in the sky and classify it as a static, nova/bright transient, eclipsing system, or pulsating light source. Rather than relying on phase-folding methods, such as forms of Box-Fitting least squares \citep{bls} or even the fast-BLS method \citep{fbls}, which are potent but prohibitively slow for the magnitude of data present in WISE, we make primary use of convolutions and transforms to extract high-detail features. Therefore, we can detect both periodic and aperiodic variability in general. 

In this paper, we first discuss the nuances and challenges of using the WISE mission for this study [\ref{subsec:telescope}, \ref{subsec:neowise-db}] and equally the potential of such an analysis on the whole database. We then must take some steps to collect and preprocess the data. These include spatially clustering apparitions \footnote{\cc{An apparition is a single observation of a light source, with associated data such as photometry, location and time. Each row in the single-exposure source catalog is an apparition.}} from the single-exposure \cc{source} catalog with a density-based approach [\ref{subsec:dbscan}], and transformations to construct a tensor which is best fit for the model to analyze [\ref{subsec:data-preprocessing}]. Subsequently we devise VARnet, a signal-processing model which is able to rapidly identify legitimate variability in source time series data [\ref{sec:model}]. VARnet leverages a one-dimensional wavelet decomposition in order to minimize the impact of spurious data on the analysis [\ref{subsec:wavelet-decomposition}], and a novel modification to the Discrete Fourier Transform to quickly detect periodicity and extract features of the time series [\ref{subsec:feft}]. VARnet integrates these analyses into a prediction of type for the source by leveraging machine learning [\ref{subsec:ml-background}], primarily convolutional neural networks [\ref{subsec:complete-model}]. In order to train VARnet, an abundance of training data is required. Therefore we engineer an accurate synthetic light curve generator for each class of light source we aim to detect in order to create an unlimited source of training examples for a complex model [\ref{subsec:datagen}]. This methodology \cc{(Figure \figref{1})} achieves an extremely rapid run time and strong performance with high precision on the test set, producing a high-quality list of anomalies [\ref{sec:results}]. After testing the model on a limited portion of the full NEOWISE database, we cover a few interesting detections produced by VARnet [\ref{subsec:firstfoundsources}].

\begin{figure*}
    \centering
    \includegraphics[scale=1.15]{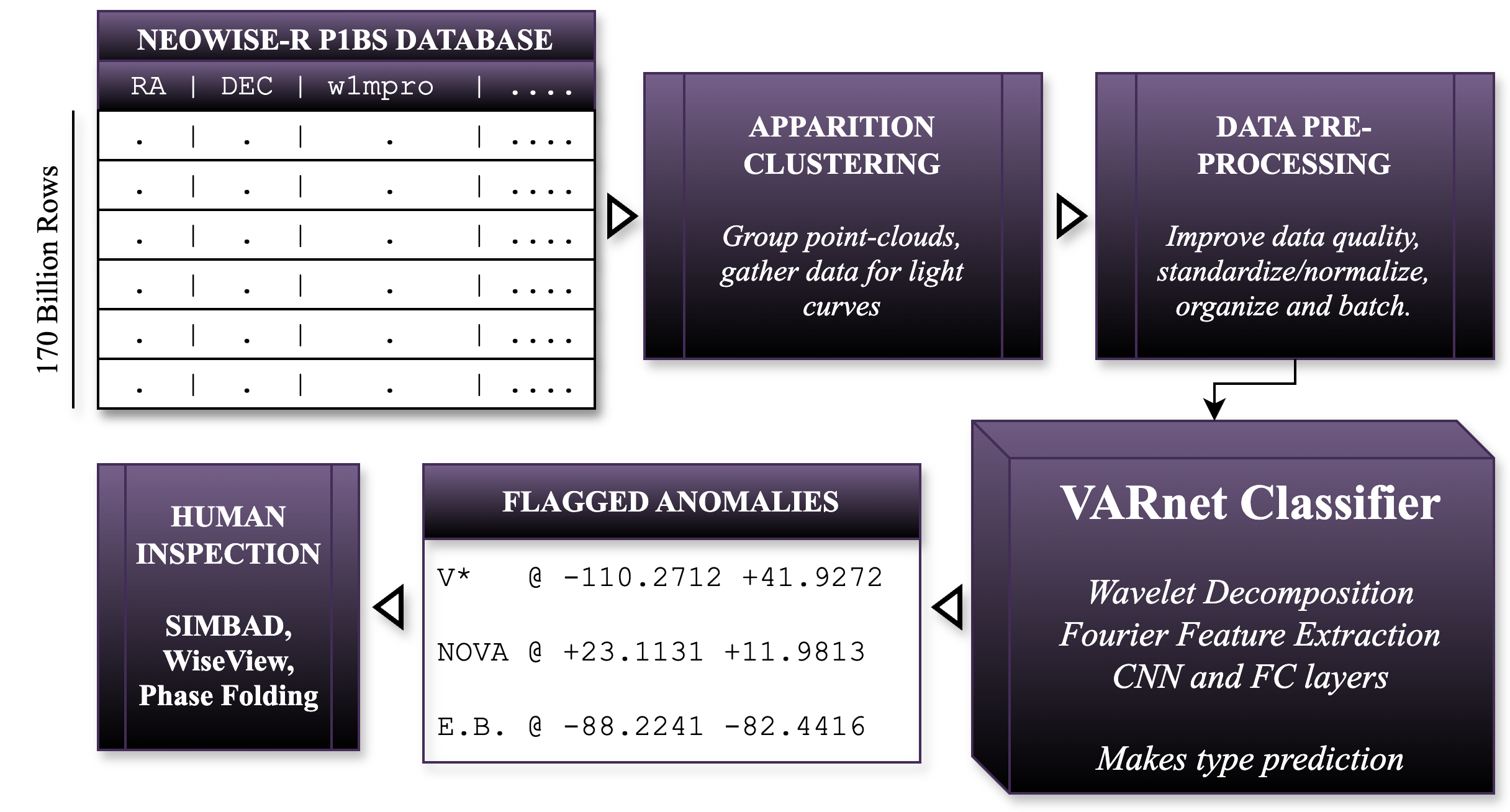}
    \vspace{1.25em}
    \caption{The anomaly extraction pipeline}
    \label{fig:flowchart}
\end{figure*}

\subsection{The WISE Space Telescope} \label{subsec:telescope}
The WISE space telescope \cc{was} a mission funded by NASA and operated by Caltech/JPL, whose objective was to provide an all-sky survey in the infrared at greater resolutions and sensitivity than previous surveys such as IRAS \citep{irastele} \citep{akaritele} and AKARI \citep{akaritele}. The Infrared Processing and Analysis Center (IPAC) at Caltech is in charge of processing and archiving the data.

For the first few months of the mission (2009-2010), 4 sensors were used to scan 4 bands in the infrared: W1, W2, W3, W4 \cc{(Figure \figref{2})}. However, in late 2010 the cryogen required for the operation of the W3 and W4 sensors had been exhausted, and the first stage of the mission was complete \footnote{The telescope did operate for a few more months as a 2-band survey prior to the hibernation. In December 2013 the NEOWISE Reactivation survey began. For simplicity, this paper refers to the actual instrument as the WISE telescope and to its data produced after the reactivation as NEOWISE data.}. After a brief hibernation period, the mission was reactivated as NEOWISE in December $2013$ \citep{neowisepaper} now repeatedly scanning the sky in the W1 and W2 bands, in search of dim objects near the earth. WISE is in low Earth orbit about the day-night terminator, oriented orthogonally to the earth-sun line. Thus WISE scans a great circle of the sky (centered at the sun), completing a full-sky scan every 6 months as the earth-sun orientation flips. However, due to the scanning being done via great circles centered at the sun, there is significant overlap in each orbital pass nearing ecliptic latitudes of $\pm 90^\circ$. As a result, sources near the ecliptic poles are much more finely sampled than those near the ecliptic equator as seen. In addition, each great circle has significant overlap. \cc{Figure \figref{3}} illustrates this process. Therefore, our sampling pattern for a given light source is as follows: 
\begin{enumerate}
    \item A few exposures within hours of each other as long as great circles overlap. For the vast majority of sources, this group is composed of 12-16 apparitions. Very close to ecliptic latitude $\pm 90^\circ$, they always overlap, and thus there is never a large gap. These sources can have up to $\sim 30000$ samples.
    
    \item Repetitions of (1) every $6$ months once the earth orbits $180^\circ$ around the sun and the great circle scans once again begin to cover the object.
\end{enumerate}

\begin{figure}
    \label{fig:spectrum}
    \centering
    \begin{tikzpicture}[scale=1.5]
    \pgfdeclarehorizontalshading{irgradient}{100bp}{
        color(0bp)=(white);
        color(100bp)=(orange)
    }
    
    \fill [shading=irgradient] (0,0) rectangle (4.8,0.5);
    \draw (0,0) rectangle (4.8, 0.5);
    
    
    \foreach \x/\itemlabelx in {0/1, 0.8/5, 1.6/10, 2.4/15, 3.2/20, 4/25, 4.8/30} {
        \draw (\x, 0) -- (\x, -0.1);
        \node[below] at (\x, -0.15) {\itemlabelx $\mu$};
    }
    \node[below] at (2.4,-0.45) {$\lambda$ [$m$]};

    \draw[blue, line width=0.2mm] (0.368, 0.65) -- (0.368, 0.75) -- (0.64, 0.75) -- (0.64, 0.65);
    \node[above] at (0.48, 0.76) {\small W1};

    \draw[green, line width=0.2mm] (0.7, 0.65) -- (0.7, 0.75) -- (1.06, 0.75) -- (1.06, 0.65);
    \node[above] at (0.89, 0.76) {\small W2};

    \draw[orange, line width=0.2mm] (1.12, 0.65) -- (1.12, 0.75) -- (2.8, 0.75) -- (2.8, 0.65);
    \node[above] at (1.9, 0.76) {\small W3};

    \draw[purple, line width=0.2mm] (3.04, 0.65) -- (3.04, 0.75) -- (4.48, 0.75) -- (4.48, 0.65);
    \node[above] at (3.75, 0.76) {\small W4};
    
    \end{tikzpicture}
    \caption{\centering \emph{Wavelength ranges of the infrared spectrum of each of the four initial WISE bands. While W3, W4 have the broadest ranges, they require onboard cryogen which depleted in mid-2013.}}
\end{figure}
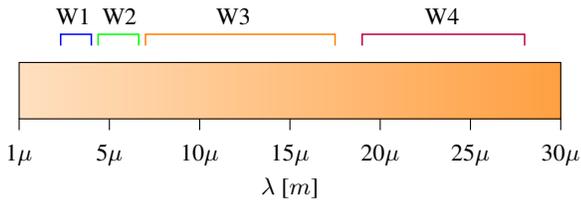

\begin{figure}%
    \label{fig:wiseorbit}
    \centering
    \subfloat[\centering Orbit and orientation of the WISE telescope, perpendicular to its fixed orbit about the terminator.]{{\includegraphics[width=4.75cm]{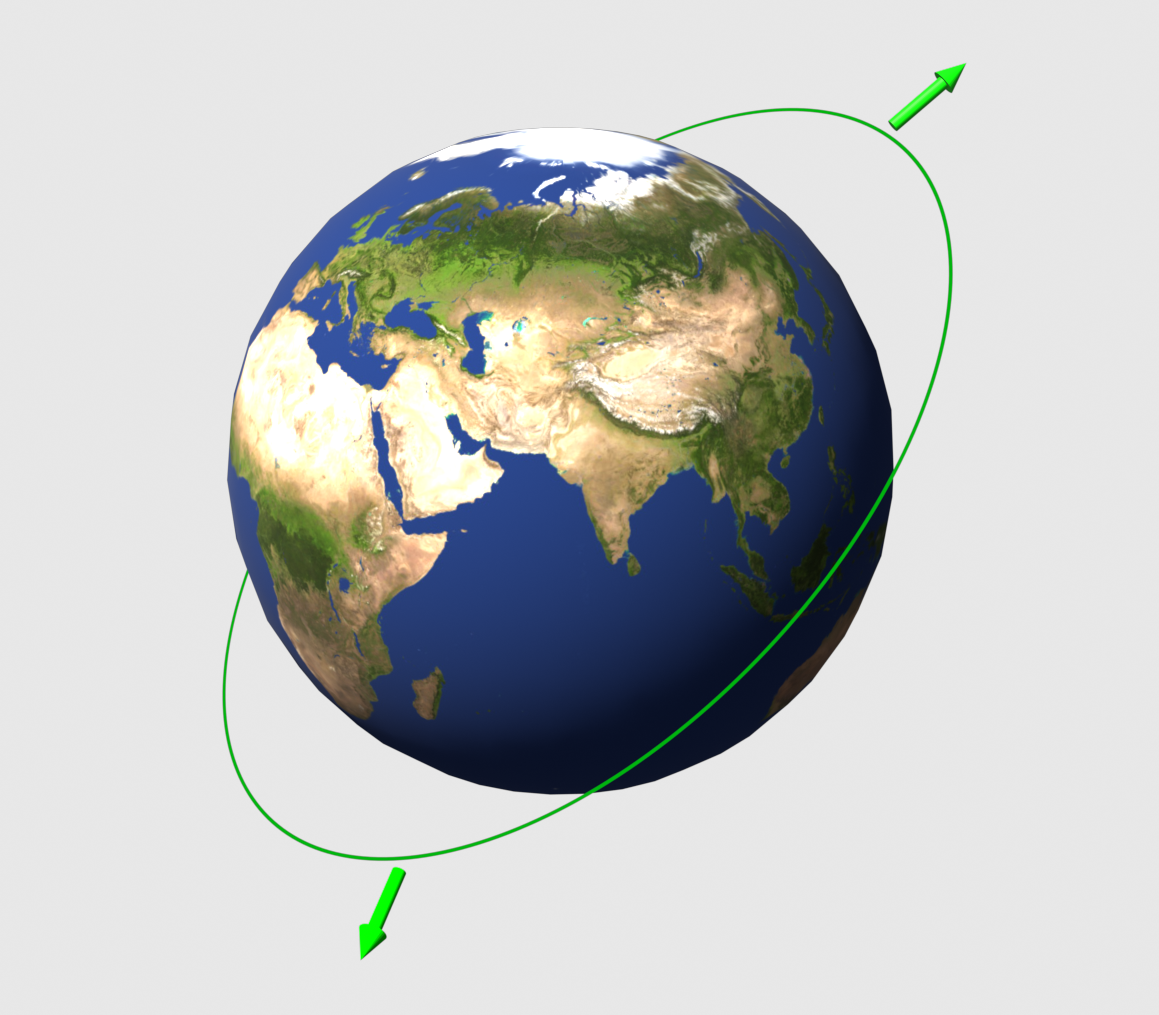} }}%
    \qquad
    \subfloat[\centering Overlap in exposures taken in close orbits. One on-sky source will yield several apparitions in this time period.]{
        \begin{tikzpicture}[scale=0.75]
            \foreach \r in {0} {
                \foreach \x in {0,...,4} {
                    \draw[fill=black, opacity=0.07] ({\r},{\x + \r}) -- ({\r + 1},{\x + \r}) -- ({\r + 1},{\x + \r + 1}) -- ({\r},{\x + \r + 1});
                    \draw[gray, line width=0pt] ({\r},{\x + \r}) -- ({\r + 1},{\x + \r}) -- ({\r + 1},{\x + \r + 1}) -- ({\r},{\x + \r + 1}) -- cycle;
                }
            };
            \foreach \r in {0, 0.25, 0.5} {
                \foreach \x in {0,...,4} {
                    \draw[fill=black, opacity=0.07] ({\r + 2},{\x + \r}) -- ({\r + 3},{\x + \r}) -- ({\r + 3},{\x + \r + 1}) -- ({\r + 2},{\x + \r + 1});
                    \draw[gray, line width=0pt] ({\r + 2},{\x + \r}) -- ({\r + 3},{\x + \r}) -- ({\r + 3},{\x + \r + 1}) -- ({\r + 2},{\x + \r + 1}) -- cycle;
                }
            };
            \draw[opacity=0] (-1,-0.5) -- (4,-0.5);
        \end{tikzpicture}
    }%
    \subfloat[\centering WISE takes exposures in great circles, centered at the sun. Notice overlap is proportional to ecliptic latitude, with high ecliptic latitude sources sampled more frequently.]{
        \tikzset{viewport/.style 2 args={
      x={({cos(-#1)*1cm},{sin(-#1)*sin(#2)*1cm})},
      y={({-sin(-#1)*1cm},{cos(-#1)*sin(#2)*1cm})},
      z={(0,{cos(#2)*1cm})}
        }}
      \begin{tikzpicture}[scale=1.6]
        \shade [ball color=white] (0,0) circle [radius=1cm];
        \draw[opacity=0] (-1.33, -1.45) -- (1.5, -1.45);
    
        \begin{scope}[viewport={\RotationX}{\RotationY}]
          \draw[variable=\t, smooth] plot[domain=90-\RotationX:-90-\RotationX, rotate around y=5] (\Sphere{\t}{0});
          \draw[densely dashed, variable=\t, smooth] plot[domain=90-\RotationX:270-\RotationX, rotate around y=5] (\Sphere{\t}{0});
    
          \foreach \p in {-70,-60,...,110}:
            \draw[variable=\t, smooth, red] plot[domain=-sin(\angle):sin(\angle), rotate around y=5] (\Circle{\t}{\p});
    
          \draw[variable=\t, smooth, red] plot[domain=90-\RotationY:-90-\RotationY, rotate around y=5] (\Circle{sin(\angle)}{\t});
          \draw[variable=\t, smooth, red] plot[domain=90-\RotationY:-90-\RotationY, rotate around y=5] (\Circle{sin(\angle)}{\t});
          \draw[densely dashed, variable=\t, smooth, red] plot[domain=90-\RotationY:270-\RotationY, rotate around y=5] (\Circle{sin(\angle)}{\t});;
    
          \draw[variable=\t, smooth, red] plot[domain=90-\RotationY:-90-\RotationY, rotate around y=5] (\Circle{-sin(\angle)}{\t});
          \draw[densely dashed, variable=\t, smooth, red] plot[domain=90-\RotationY:270-\RotationY, rotate around y=5] (\Circle{-sin(\angle)}{\t});
    
          \def\RotationY{-30}
    
          \foreach \p in {-70,-60,...,110}:
            \draw[variable=\t, smooth, green] plot[domain=-sin(\angle):sin(\angle), rotate around z=8, rotate around y=5] (\Circle{\t}{\p});
          \draw[variable=\t, smooth, green] plot[domain=90-\RotationY:-90-\RotationY, rotate around z=8, rotate around y=5] (\Circle{sin(\angle)}{\t});
          \draw[densely dashed, variable=\t, smooth, green] plot[domain=90-\RotationY:270-\RotationY, rotate around z=8, rotate around y=5] (\Circle{sin(\angle)}{\t});;
    
          \draw[variable=\t, smooth, green] plot[domain=90-\RotationY:-90-\RotationY, rotate around z=8, rotate around y=5] (\Circle{-sin(\angle)}{\t});
          \draw[densely dashed, variable=\t, smooth, green] plot[domain=90-\RotationY:270-\RotationY, rotate around z=8, rotate around y=5] (\Circle{-sin(\angle)}{\t});
        \end{scope}
      \end{tikzpicture}                         
  }
    \caption{How WISE surveys the sky, based on figures seen in \citet{wisepaper}}%
\end{figure}
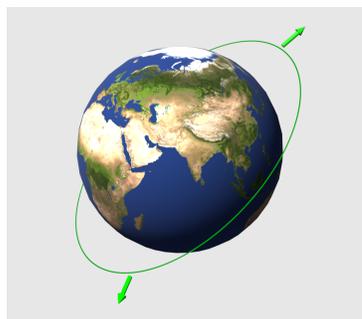
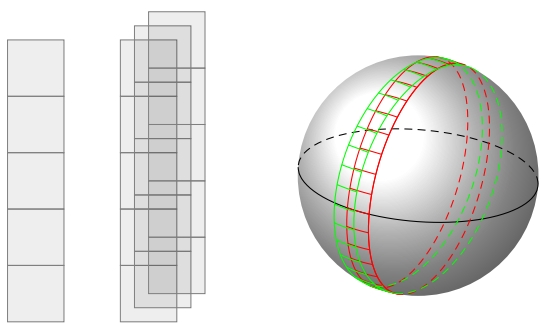

\subsection{The NEOWISE-R Single-Exposure Database} \label{subsec:neowise-db}
The database of choice for this paper is the NEOWISE Reactivation Single Exposure database, containing photometry data extracted from every single exposure taken by the telescope since its reactivation in 2013. In this way we are ignoring data from 2011-2013. This simplifies the problem, as we only have to manage data retrieval from one table and need not deal with the 2 year hibernation gap. Images received from the telescope are processed, generating positions and band magnitudes for point-sources of light. An \emph{apparition} is a single sighting of a light source in time, containing $RA, Dec, mjd, W1, W2$ data for our purposes. $RA, Dec$ are necessary for clustering and filing away the classifications, $mjd$ is normalized and used for a timestamp on each brightness value, and W1, W2 are magnitude values for their respective sensors, used for the brightness/intensity in the light curve.  Each row in the NEOWISE database is an apparition. Beyond that, the data has no structure. Therefore, all apparitions/data pertaining to a single source might be scattered in the database. When querying the database, apparitions matching the SQL conditions will be retrieved without structure, unlabeled to what source they might have originated from. In addition, the database also contains noisy apparitions which originate from cosmic rays, streaked exposures, \cc{or nebulosity}. This creates a challenge in collecting data pertaining to a single astronomical object.

\subsection{Classification of Light-Sources} \label{subsec:classes}

\cc{The goal of this paper is twofold--to present VARnet and analyze its performance, and to demonstrate a methodology for variable source extraction with WISE. Thus in this paper, we choose to categorize stars into four very broad types of light sources based on their flux variability: Null, Transient, Pulsating, Transit (Figure \figref{4}). This limited taxonomy is not exceedingly useful for comprehensive surveys and statistics on variability. However, it will allow us to analyze the performance of VARnet fairly, as the classes we choose are morphologically quite distinct from one another. For a more expansive taxonomy with classes that are more subtly distinct from one another, slower but more powerful analyses such as phase folding and model fitting should be run as a secondary classification, expanding the final number of classes\footnote{This is the planned methodology for the future complete WISE variability survey.}. This way, VARnet may detect low-amplitude variability which is potentially thrown out by more blunt traditional methods, while filtering out bogus variability so that time is not wasted further analyzing them with the secondary classification step. \emph{It is important to note that VARnet is a model architecture, and its utility is defined by the data and task that it is trained on. VARnet could feasibly be trained on a single type of phenomenon and act as a binary classifier, or potentially be trained to extract properties of timeseries it is given and act as a regression model.}}

\begin{figure*}
    \label{fig:class-examples}
    \centering
  \subfloat[\centering { Pulsating Variable - AS Mus}]{\includegraphics[scale=0.1]{ 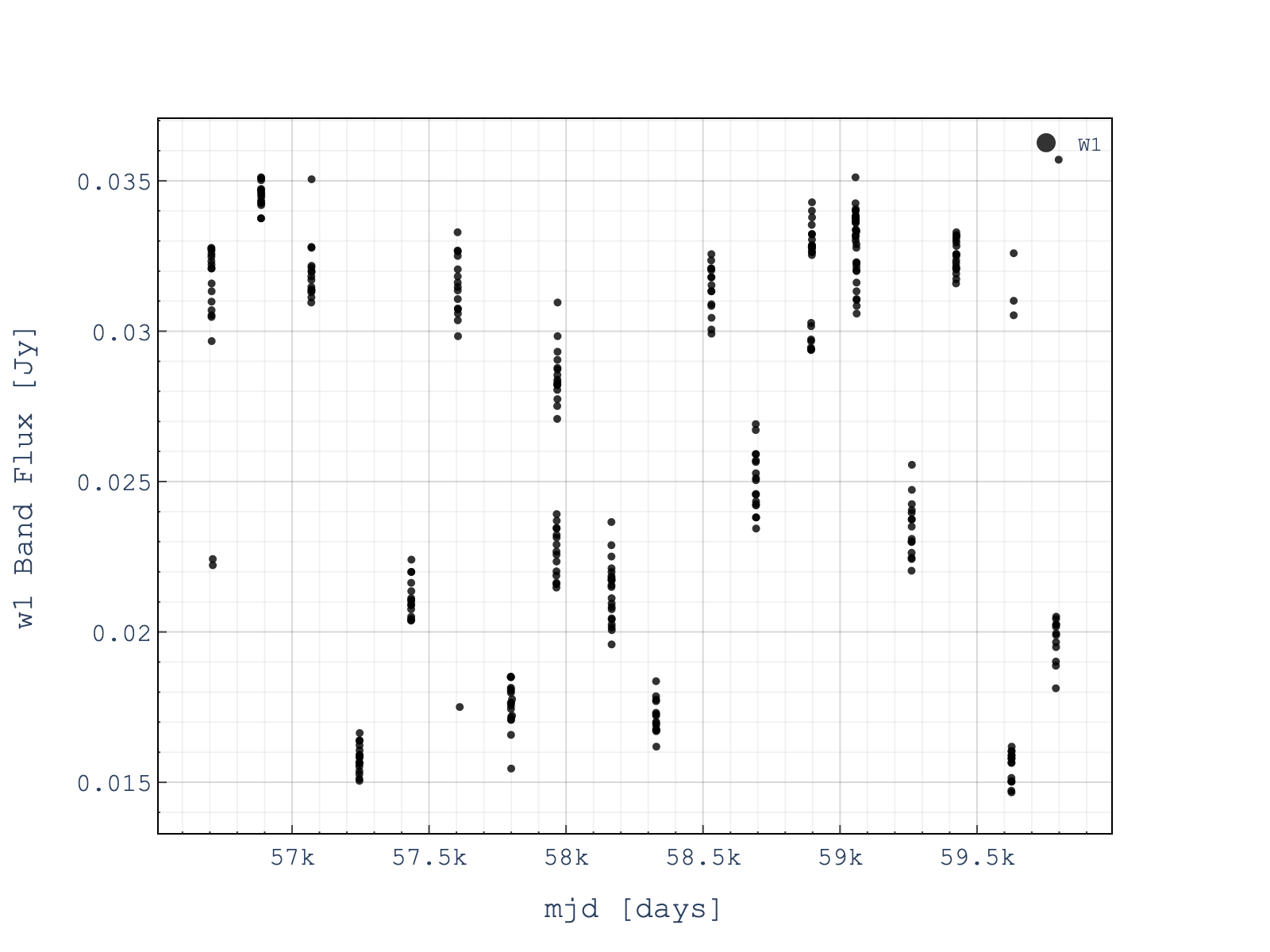}}
  \subfloat[\centering {Transit - CV Boo}]{\includegraphics[scale=0.1]{ 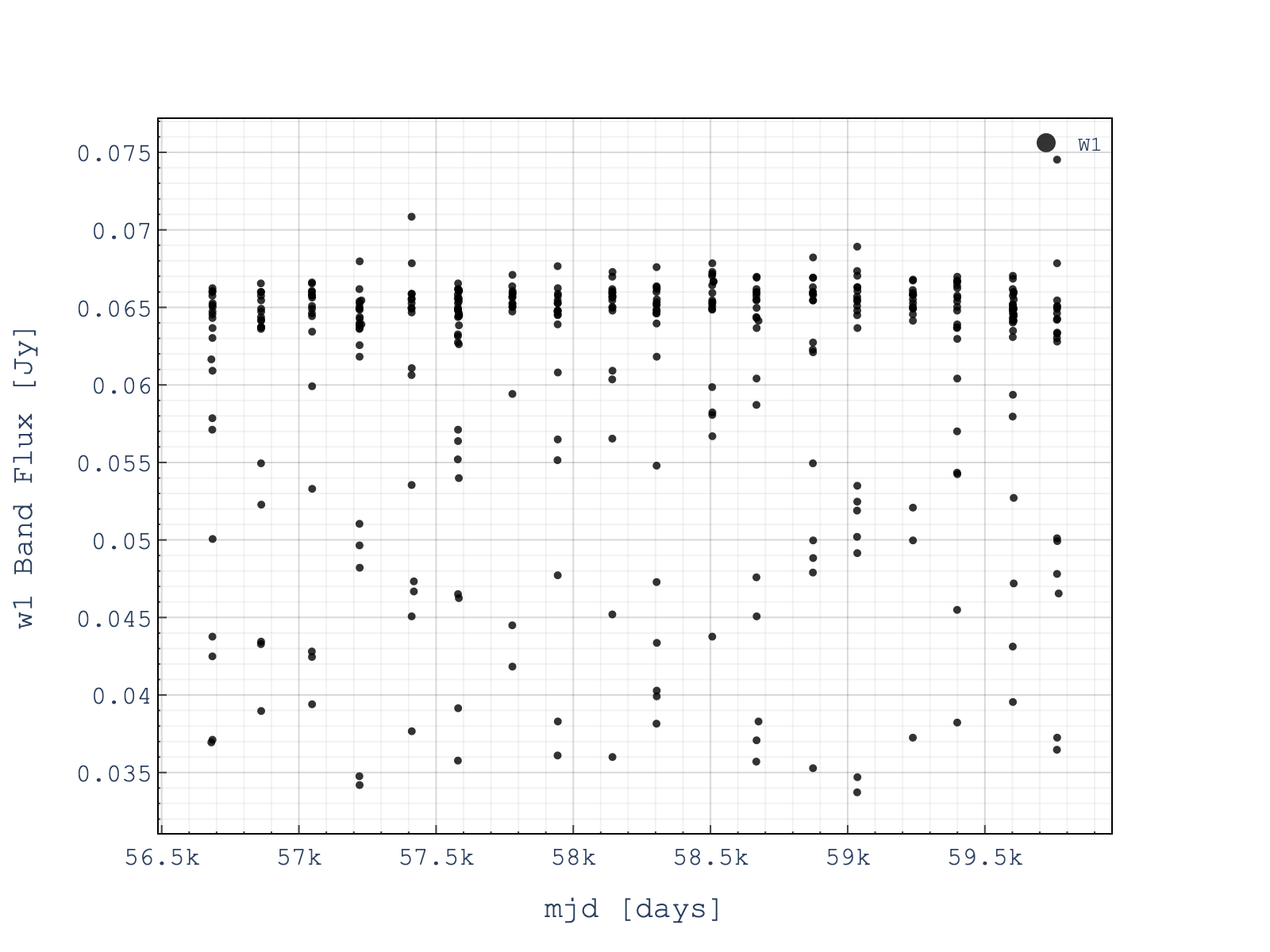}}
  \subfloat[\centering {Null - HD 165459}]{\includegraphics[scale=0.1]{ 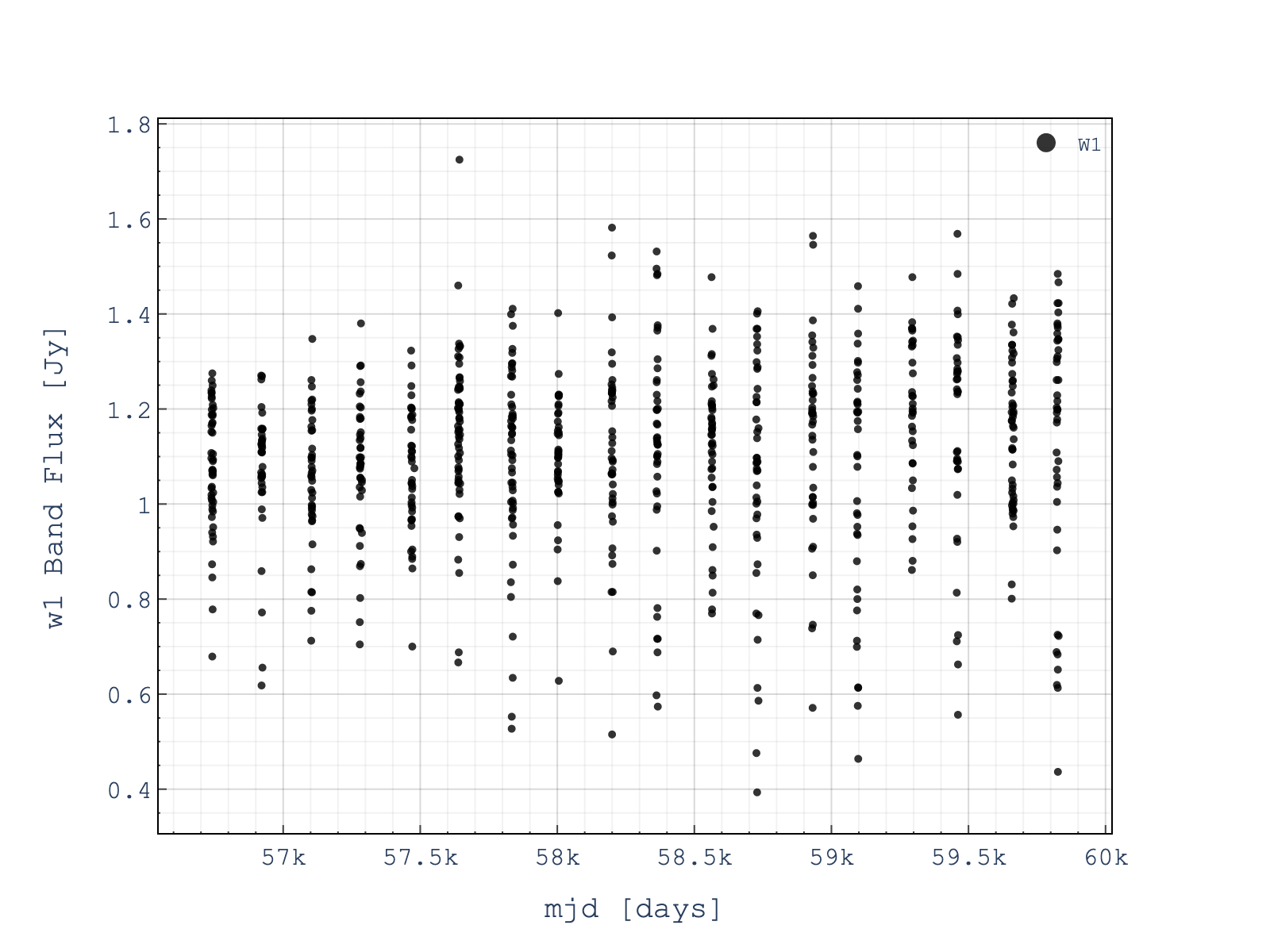}} \\
  \subfloat[\centering {Nova - V2860 Ori}]{\includegraphics[scale=0.1]{ 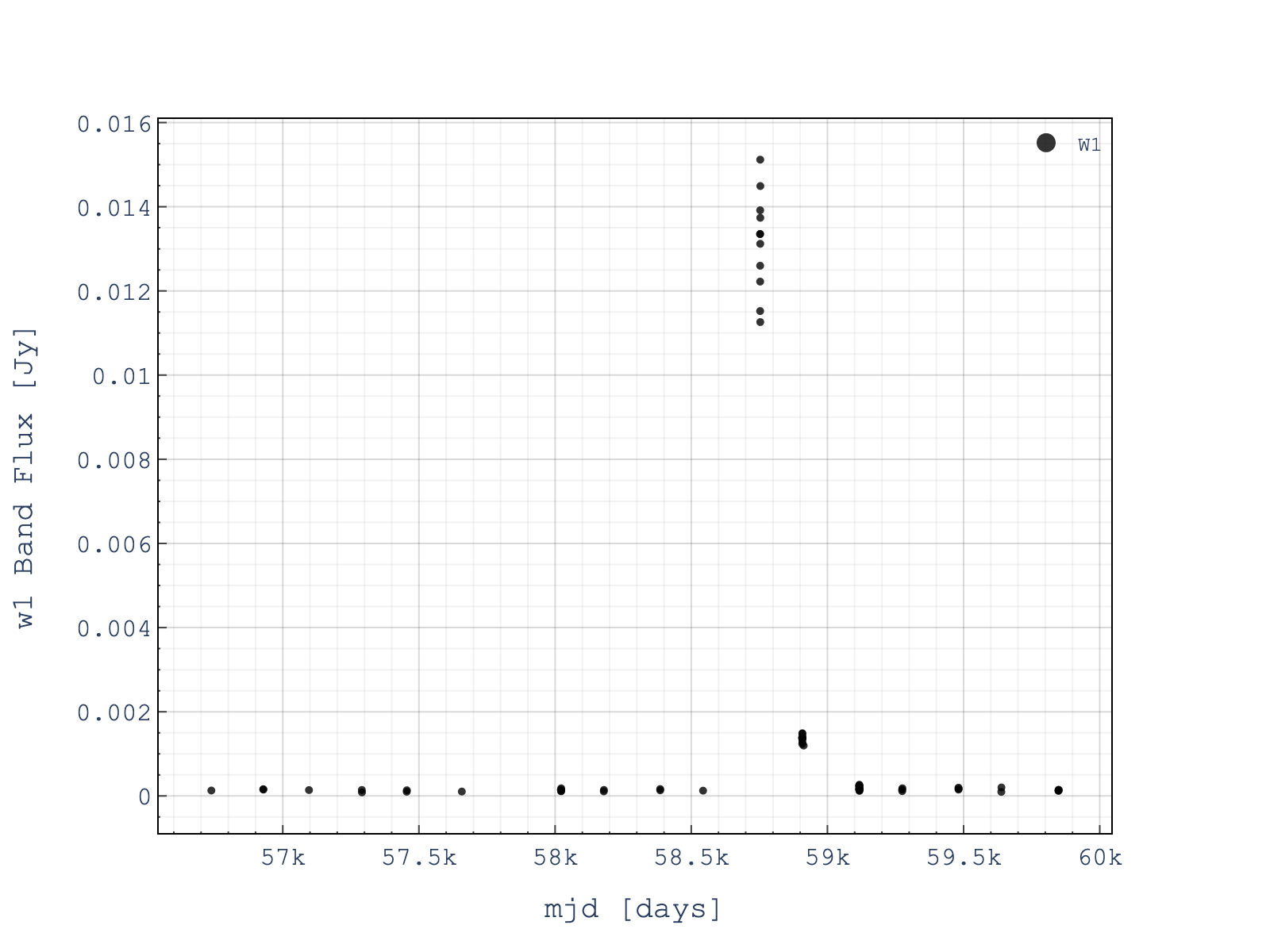}}
  \subfloat[\centering {Transit - IM Per}]{\includegraphics[scale=0.1]{ 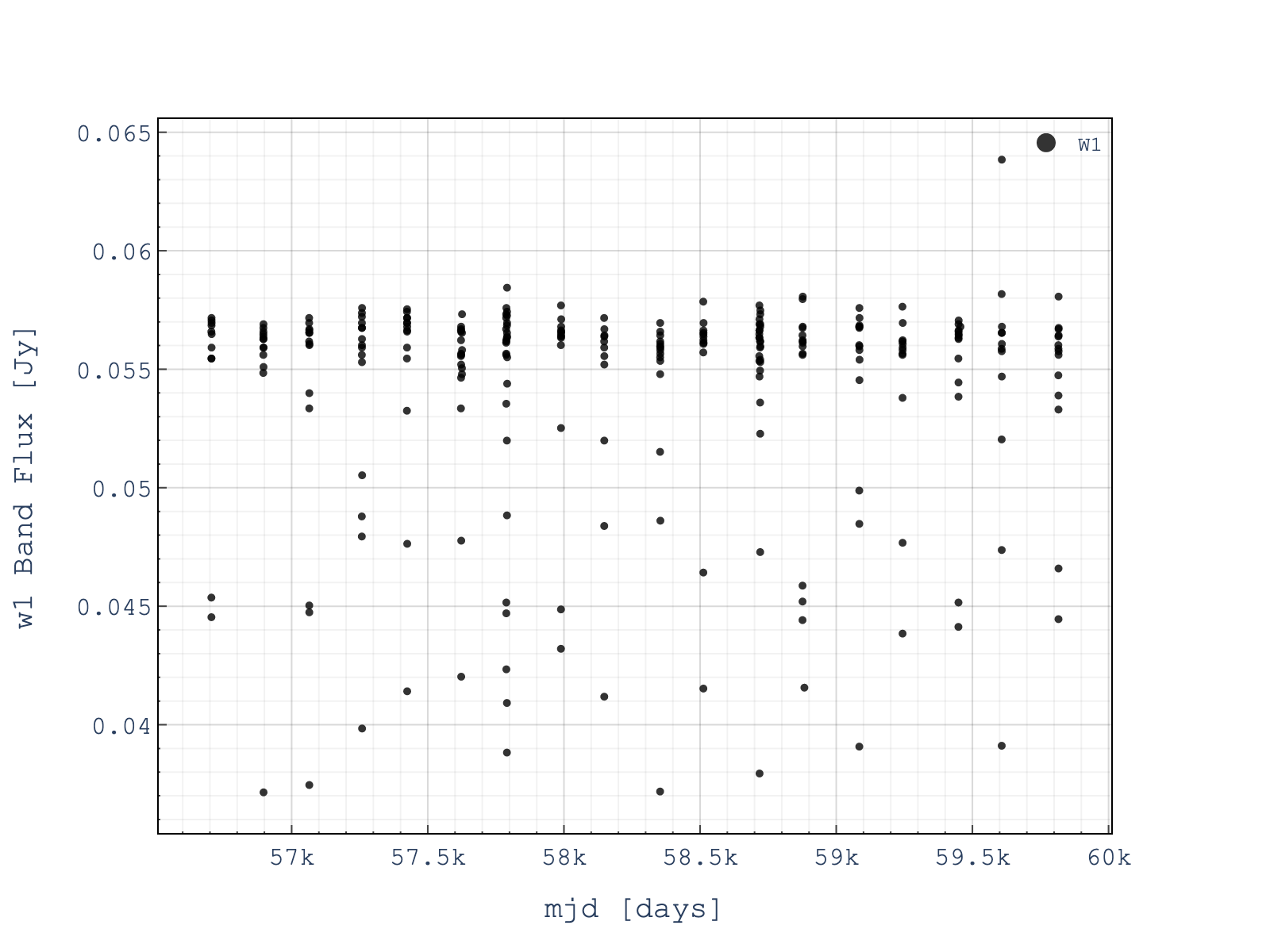}}
  \subfloat[\centering {Null - NPM1p68\_0422}]{\includegraphics[scale=0.1]{ 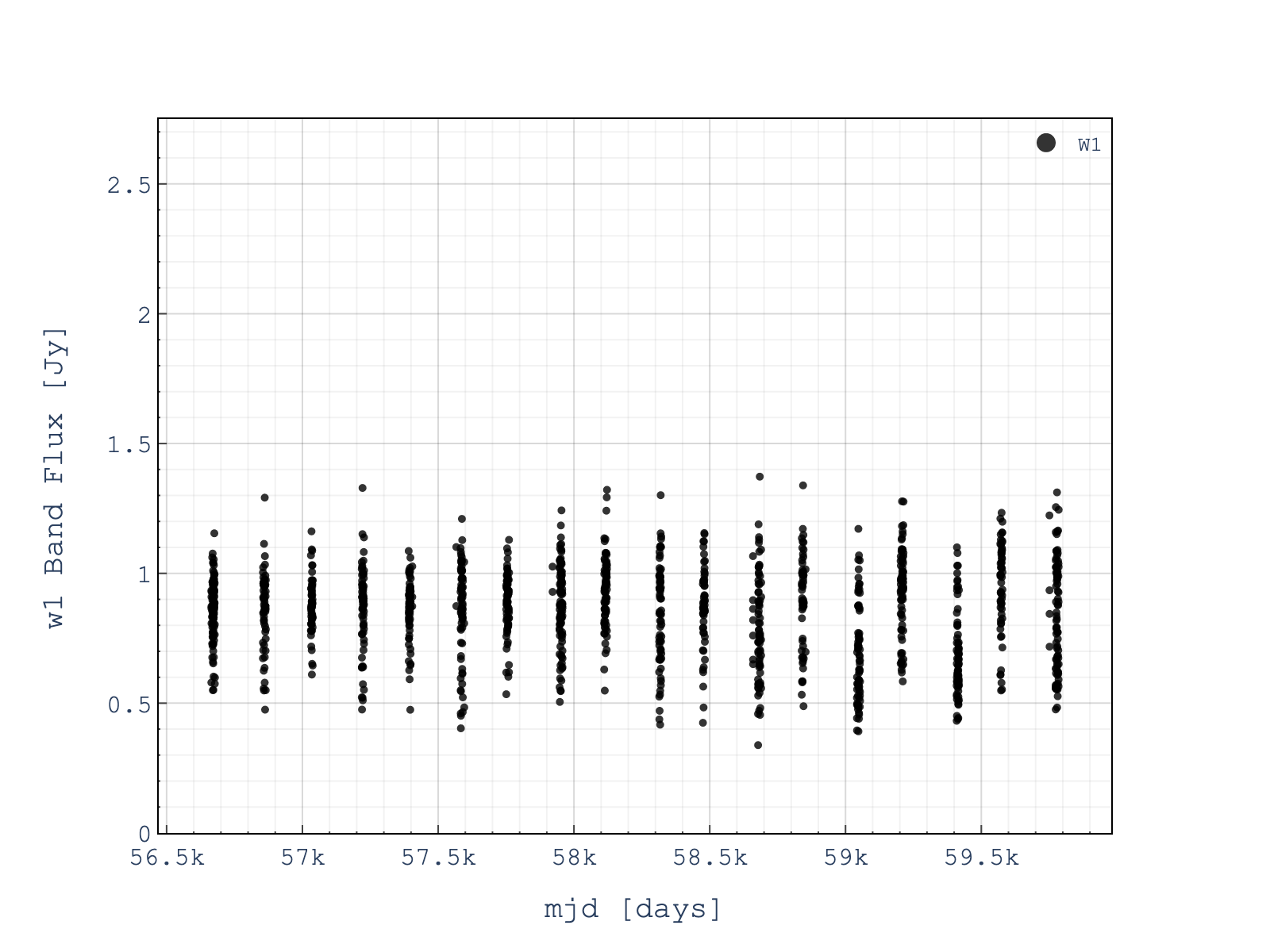}}
  \caption{Example light curves of the 4 classes \\ \emph{Note: y scales differ so that the shapes are more apparent}}
\end{figure*}

The vast majority of sources in the data are mostly unchanging; we name this the \emph{null} class. Their underlying brightness remains constant, at least to the sensitivities of WISE. \emph{Transient} events are astrophysical events in which there is a sudden and drastic change in the brightness or energy of a system. Many stars too faint to be detected by WISE become briefly visible during a transient event. This means there might only be a few apparitions of a transient light source in the database over just a few passes. \cc{Their light curves have a distinct timepoint at which the transient event occurs. It might exhibit exponential decay, or decay much more slowly}. \emph{Pulsating variables} are objects which also exhibit an intrinsic variability in brightness. However, the underlying brightness of the object is constantly varying and is often smooth, contrasting with the transient dimming of eclipsing systems. \cc{These variations are commonly periodic}. Due to the irregular sampling, their light curves often do not tend to a mean and jump around as different parts of the wave are sampled. Finally, \emph{transits} are sources which exhibit variability due to the presence of another orbital companion that eclipse them periodically, creating a sharp dip in the apparent brightness of the source. Typically, their light curves in WISE have most apparitions at a constant brightness similar to a null source, but with many outliers scattered at fainter magnitudes, originating from samplings during an eclipse

Classification over these four classes would be a more simple endeavour given a high sampling frequency. However, the cadence of WISE poses a problem. See \cc{Figure \figref{5}} as an example. The frequency of a stellar event will almost never match up with the observing frequency of WISE, leading to obfuscation of the signal. Additionally, the sampling is extremely lopsided, with dense periods of sampling, approx $0.115$ days, adjacent to very long gaps of $6$ months. Data from sources higher in the sky near ecliptic latitude $90^\circ$ are sampled with much higher frequency, thus can be classified with more precision. However, the additional data can be more difficult to handle. Thus the model must be adaptable, able to handle sequences of wildly differing lengths at differing sampling rates.

\begin{figure}
    \label{fig:varlightcurve}
    \centering
    \subfloat[\centering Light curve of V* MSX LMC 1689, which lies near the South Ecliptic Pole and is finely sampled]{\includegraphics[scale=0.14]{ 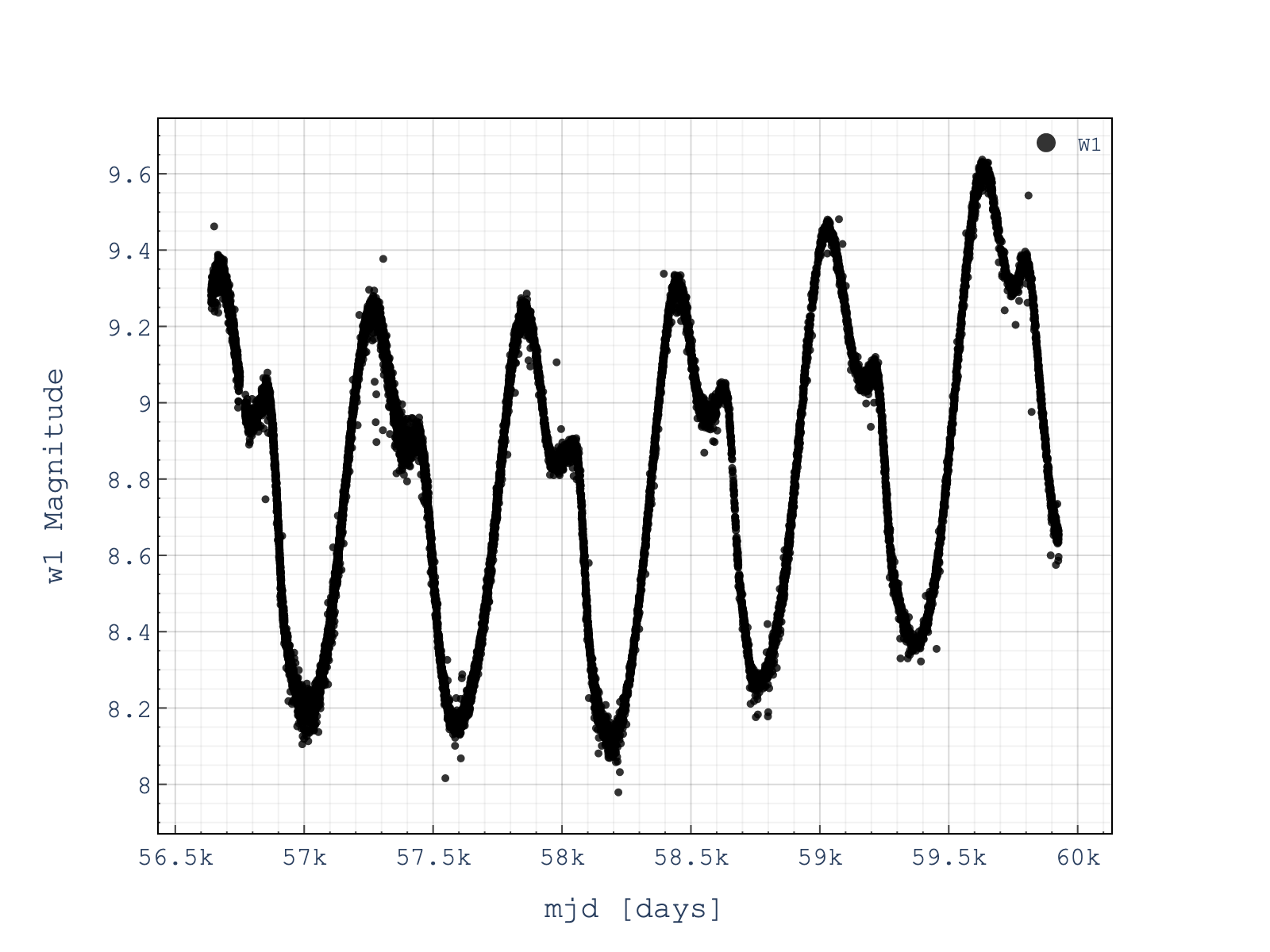}} \\
    \subfloat[\centering Simulated NEOWISE light curve if MSX LMC 1689 were located at ecliptic latitude $0^\circ$]{\includegraphics[scale=0.14]{ 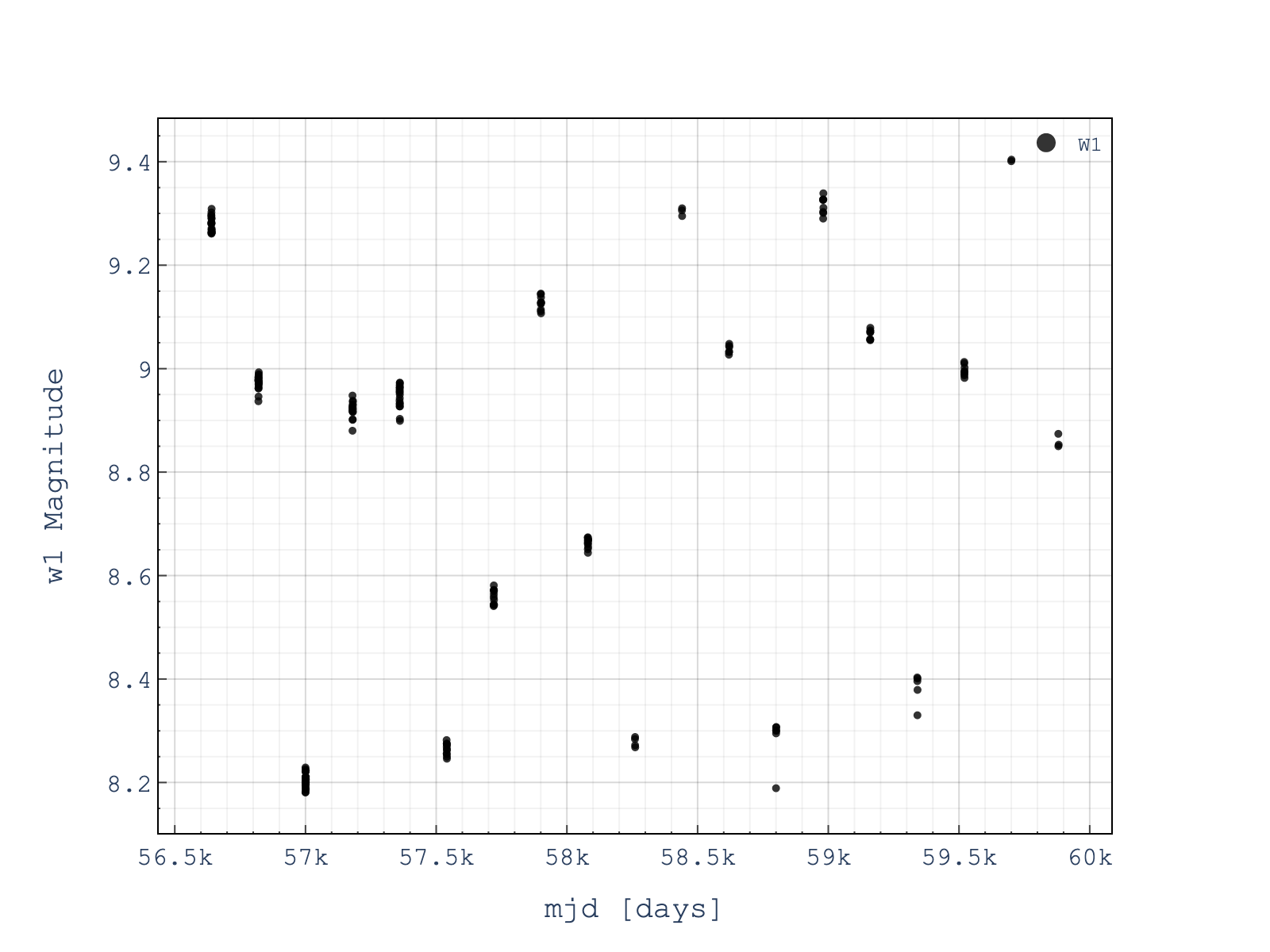}}
    \caption{Light Curve Sampling with MSX LMC 1689}
\end{figure}

\section{Clustering and Data Representation} \label{sec:clustering-datarep}

As stated before, we do not have knowledge \emph{a priori} of which rows/apparitions in the database correspond to the same source. Due to minor inconsistencies in noise and limits of the telescope precision, the center of the source might be seen at slightly different positions each pass. As a result, we obtain a point-cloud of apparitions, approximate to a 2-dimensional Gaussian about the true center, as depicted in Figure \figref{6}. Thus, to collect data about any source, a robust clustering algorithm is required that performs well on clusters that may vary in shape. Additionally, due to the presence of outliers and noise, some type of heuristic is necessary to discriminate between legitimate and extraneous detections to ensure data quality. Note that this is the most time-consuming portion of the candidate detection pipeline: most of the data manipulations and even the detection model operate in $\mathcal{O}(n)$\footnote{\cc{Big $\mathcal{O}$ notation describes the rate at which the runtime of an algorithm grows as the number of operands grows. $\mathcal{O}(n)$ means that the runtime scales linearly with the number of operands.}} time. However, as many clustering algorithms need to measure some kind of similarity between points, they operate in nonlinear time. This can be ameliorated by the use of efficient methods such as $k$-d trees \citep{kdtreespaper} to split the data in a high dimensional space and reduce the amount of distance calculations to be made, but in this case we can achieve at best $\mathcal{O}(n \log n)$ time before making compromises.

\begin{figure}
    \label{fig:point-cloud}
    \vspace{2.25em}
    \centering
    \includegraphics[trim={0 0 0 5cm},clip,scale=0.14]{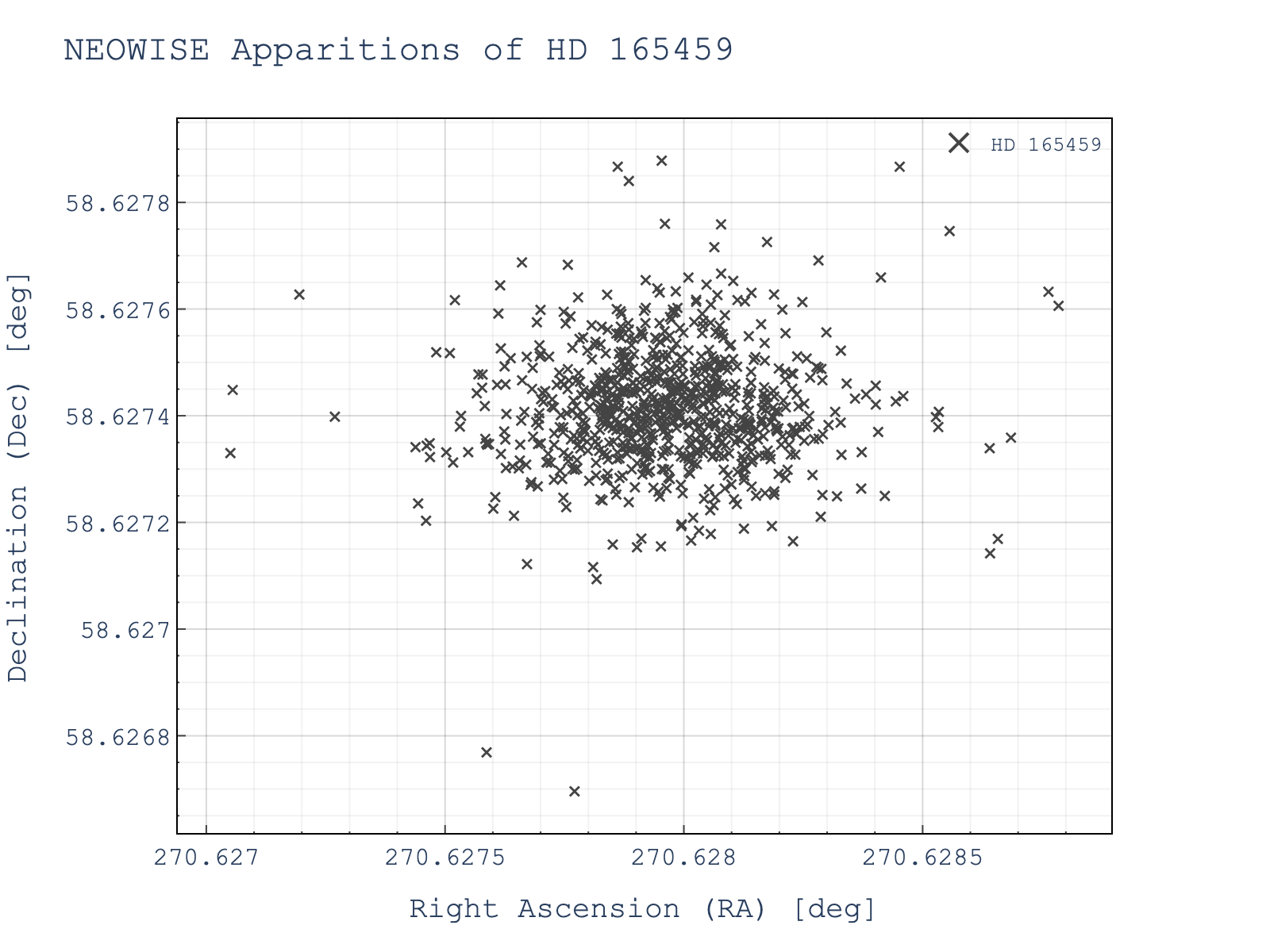}
    \caption{ \centering Example point cloud of a source: HD 165459}
\end{figure}

\subsection{DBSCAN} \label{subsec:dbscan}

In order to satisfy these requirements, we primarily make use of the DBSCAN (Density-Based Clustering of Applications with Noise) \citep{dbscan} algorithm for the reasons below:
\begin{enumerate}
    \item Works robustly on clusters with differing shapes
    \item Does not require the number of clusters to be specified
    \item Can be fine-tuned with \cc{hyperparameters}
    \item Can filter out "noise" points
\end{enumerate}

The DBSCAN algorithm leverages a notion of proximity between points to create a sort of network of reachability between points. Assuming that legitimate clusters of data are spatially denser than a possibly noisy background, it can effectively identify real phenomena. Critically it makes no assumption about the number of clusters or the shape of said clusters. Therefore, it fits well to the issue of unpredictable point clouds in the sky, pulling real groups of apparitions out of a sparse background of cosmic ray detections and instrument errors. It works as such: first choose a value for $\varepsilon$ - the distance at which points are considered "related", usually empirically determined. $\varepsilon\to\infty$ has the effect of enabling less dense clusters. Choose a value for $\minpts$, which is generally how many points you expect in the vicinity - $\varepsilon$ or less away - in the body of a cluster. Now, for each $n$-dimensional point in your dataset $\bfv \in \mathcal{V} \subset \RR^n$, identify all points in the neighborhood: 
\[B_\varepsilon(\bfv) := \{\bfv' | \bfv' \in \mathcal{V}, d(\bfv, \bfv') < \varepsilon\}\] 
If $|B_\varepsilon(\bfv)| \geq \minpts$, then all $\bfv' \in B_\varepsilon(\bfv)$ are \emph{border points}, and $\bfv$ is a \emph{core point} of the a \emph{cluster} $\mathcal{C}_i$. If the distance between two core points $d(\bfv_1, \bfv_2) < \varepsilon$, then they are both members of the same cluster:
\[B_\varepsilon(\bfv_1) \cup B_\varepsilon(\bfv_2) \subseteq \mathcal{C}_i\]
In this way we get three classes of points: core, border, and noise points. Noise points are members of the dataset that were never assigned to a neighborhood with greater than $\minpts$ members. Consequently they are assigned to no cluster:
\[\bfv^\times \text{\emph{is a noise point:}} \quad \bfv^\times \in \mathcal{V} \enspace \setminus \enspace \bigcup\limits_i \mathcal{C}_i \]
Therefore, all border points are $\varepsilon$-reachable from a core point; however, no noise points are reachable from a core point (but may be reachable from a border point). The hope is that noise points did not represent legitimate apparitions of a star, and that the clusters are not too large - e.g. two nearby light sources with reasonably separable point clouds are not erroneously merged into a single cluster. Algorithm 1 is the process by which we do this in practice.

\begin{algorithm}[H]
\caption{The DBSCAN Algorithm} \label{alg:dbscan}
\begin{algorithmic}
    \Function{DBSCAN}{$\mathcal{V}$, $\varepsilon$, $\minpts$}
        \State $c \gets 0$
        \For{unvisited $\bfv$ in dataset $\mathcal{V}$}
            \State mark $\bfv$ visited
            \State Neighbors $N \gets$ \Call{matchSet}{$\bfv$, $\varepsilon$}
            \If{$|N| \geq \minpts$}
                \State label($\bfv$) $\gets c$
                \State Branching Search Seed $S \gets N \setminus \{p\}$
                \For{$\bfv'$ in $S$}
                    \If{label($\bfv'$) is defined and not $c$}
                        \State \textbf{continue}
                    \EndIf
                    \State mark $\bfv'$ visited
                    \State label($\bfv'$) $\gets c$ 
                    \State Neighbors $M \gets$ \Call{matchSet}{$\bfv'$, $\varepsilon$}
                    \If{$|M| \geq \minpts$}
                        \State $S \gets S \cup M \setminus \{\bfv'\}$
                    \EndIf
                \EndFor
            \EndIf
        \EndFor
    \EndFunction
    \Function{matchSet}{$\bfv$, $\varepsilon$}
        \State \textbf{return} $\{\bfv' | \bfv' \in \mathcal{V}, d(\bfv, \bfv') < \varepsilon\}$
    \EndFunction
\end{algorithmic}
\end{algorithm}

\cc{In the years since DBSCAN was originally proposed, several variations on the core algorithm have also come into use, intended to address some of the issues that DBSCAN encounters in some applications. These notably include OPTICS \citep{opticspaper} and HDBSCAN \citep{hdbscanpaper}. OPTICS and HDBSCAN are generally more complex algorithms and we will not cover their mechanics here, but the key observation is that they use a hierarchical approach to clustering, which extracts clusters of varying density. In OPTICS, this gives the value of $\varepsilon$ a much looser meaning, instead mostly serving as a way to speed up the algorithm. In HDBSCAN, there is no equivalent parameter to $\varepsilon$. You may only provide the minimum cluster size, thus reducing the assumptions you place on the clustering. While this is advantageous to applications in which you might expect clusters of significantly varying densities, the process that generates clusters in our case does not vary. DBSCAN allows us to adhere strictly to what is expected of a legitimate cluster\footnote{Sources with proper motion are clustered with varying quality using our approach. We find that although their point-clouds are less compact and elongated in the direction of their motion, most are recovered by DBSCAN, save for relatively fast movers. As proper motion sources are not our objects of study, we accept the loss of these as a minority.}. As a result of error in the point-spread function fitting, we can estimate that points separated by at most $0.85$ arcseconds are related; thus we choose that value for $\varepsilon$. Knowing that each $\sim 6$ months we collect roughly between $12$-$16$ apparitions, we select $12$ for our $\minpts$ value, so that most points in an "epoch" of apparitions are considered core. These parameters in conjunction with DBSCAN produced the most consistent and desirable results in testing, with OPTICS and HDBSCAN either clustering on abnormally dense parts of nebulosity and noise, or failing to recover legitimate border points that DBSCAN was able to, even after they were tuned.}

\subsection{Data Pre-processing} \label{subsec:data-preprocessing}
The objective in this step is to construct a high quality, neural network friendly matrix representation of a light curve for analysis. From the NEOWISE database we have access to 2 infrared magnitudes for the majority of apparitions. The W1 sensor/band is the most sensitive and is mostly free of anomalies and incidents in the data, so we select it as the brightness feature. The magnitude values from the database are in log-scale, so we convert to an absolute flux reading using the equation:
\begin{align}
    \text{\emph{w1flux}} = 309.25 \times 10^{\frac{\text{\emph{W1}}}{-2.5}}
\end{align}
$309.25$ Jy is the flux of an observation with a W1 band magnitude of $0$ according to \citet{wisepaper} and $10^\frac{W1}{-2.5}$ originates from the definition of the magnitude system. This has the effect of transforming the data into a linear scale. The absolute brightness of the light curve has no bearing on the analysis necessary to classify it (other than the expected variance), so we subtract the \cc{median}.
\begin{align}
    \text{\emph{w1flux}} \gets \text{\emph{w1flux}} - \operatorname{med}(\text{\emph{w1flux}})
\end{align}
The data is then standardized with the \cc{interquartile range.}
\begin{align}
    \text{\emph{w1flux}} \gets \frac{\text{\emph{w1flux}}}{\operatorname{iqr}(\text{\emph{w1flux}})}
\end{align}
We find that the neural network performs best when values are compressed near the $[-1,1]$ range using the $\arcsinh$ function, which returns it to a modified log-scale once again.
\begin{align}
    \text{\emph{w1flux}} \gets \arcsinh \text{\emph{w1flux}}
\end{align}

\cc{The timestamp $t_i$} corresponding to each brightness level should be considered carefully. As previously stated, the sampling rate generated by WISE is highly irregular: groups of apparitions hours apart, separated by approximately 6 months with no data. We receive this time data corresponding to the apparition in Modified Julian Date. \cc{Assuming the apparitions are sorted by ascending time, we transform this value to $[0,1]$ given that there are roughly $4000$ days in our observation window.}
\begin{align*}
    t_i &\gets \text{mjd}_i - \text{mjd}_0 \\
    t_i &\gets \frac{t_i}{4000}
\end{align*}


\cc{Accompanying the measured brightness is the expected error of that measurement, expressed by the $w1sigmpro$ column. The expected error is impacted by several factors, such as instrumental error, background brightness and frame smearing from spacecraft jitter. At different error levels, the threshold for variability in measured brightness changes; thus we must include this value in some form as a feature to the model. To keep it at the same scale as the flux value, we again rescale with division by interquartile range, and take the hyperbolic inverse sine of that value.}

\begin{align}
    \text{\emph{w1sigmpro}} &\gets \arcsinh\left(\frac{\text{\emph{w1sigmpro}}}{\operatorname{iqr}(\text{\emph{w1flux}})}\right)
\end{align}

Finally, we collect these values into a time series. For a star with $T$ samplings, we have a series/matrix $L_i^k$, indexed in chronological order by $0 \leq i \leq T-1$ and by features $0 \leq k \leq 2$. Each time point $L_i$ is a vector in $\RR^3$, with first entry brightness, second entry uncertainty, third entry timestamp. In order to take full advantage of modern graphics cards, we employ batching of many series so that matrix multiplication is maximally parallelized. A batch $B^n_{ik}$ is a 3D array, first dimension $n$ indexing each time series, and $i,k$ still indexing order and feature respectively. Time series $L_i^k$ will vary in length, so we pad the shorter series in the batch with zero vectors. We feed all data to the network in batches.

\subsection{Data Generation} \label{subsec:datagen}
In this step we need to secure the data in order to perform supervised learning on the model. \cc{This is the step which distinguishes our use of VARnet to other potential applications. Our selection of training data and classes is intended as a first step to a more complete survey of variability with WISE down the line.} While searching for example sources in each of our classes, we were limited most heavily by the amount of known \cc{transients} with sufficient and acceptable data present in the NEOWISE database, around $40$ examples after some effort. There is a sizeable amount of eclipsing binaries and intrinsic pulsating variables in the dataset that have acceptable light curves, yet still an insufficient quantity to train even a medium-sized model. As a result, we opted to formulate mathematical models to generate simulated WISE light curves for each of the four classes, and delegate the real light curves for \cc{model} validation. Additionally, after performing optimizations, it could be made so that these examples were generated in batches quickly and on-demand during model training. In a pseudo-online learning setting \citep{onlinelearning}, this effectively expands the size of the training dataset to $\infty$, given enough variability is added. The simple fact that the training data does not repeat gives a strong boost to model training performance and generalization, but the process of accurately emulating real WISE light curves is not fully trivial. \cc{The steps to acquiring a complete generated example are as follows:}

\begin{enumerate}
    \item Construct an underlying brightness function $f(t)$
    \item Sample this function in the same cadence as WISE 
    \item Add Gaussian noise $y_t \sim \mathcal{N}(f(t), \sigma_{f(t)})$
\end{enumerate}
Pseudocode for these processes can be seen in Algorithm 2.

\subsubsection{Basic Source Parameters} \label{subsubsec:basicparams}
 \cc{To select the mean or base brightness of the source before we add variability}, we randomly choose a magnitude value $m \in [6,16]$.\footnote{We weight the random calculation to yield dimmer magnitudes much more often to more accurately simulate the sky distribution} We convert to flux with equation (1):
\[m \in [6,16] \enspace \tilde{\to} \enspace \omega \in [0.0001, 0.2]\]
 and assume a normal distribution of noise. To acquire the uncertainty and thus standard deviation of noise about the underlying signal, \cc{we choose a random value between $10^{-4}$ and $10^{-1}$}
\begin{align}
    \sigma_\omega \in [0.0001, 0.1]
\end{align}
\cc{Globally we expect the measurement error to correlate with the absolute brightness of that source. However there are cases in which the background flux is high, particularly in the galactic plane, which increases error for bright stars. We also cannot use SNR alone to determine this value, as for stars brighter than magnitude $\sim 11$, a slightly different version of the photometry pipeline is used, sometimes resulting in abnormally high measurement errors for sources near magnitude $11$. We choose $\sigma_\omega$ independently of absolute brightness to cover all cases and prevent overfitting.
Finally, the \emph{w1sigflux} measurement error value in the NEOWISE database does not necessarily equal one sigma of actual spread about the real brightness value, so we must rescale the uncertainties used in our synthesizer to match the values from the database. We find that multiplying $\sigma_\omega$ by some real number between $0.4$ and $0.6$, sampled at random for each synthetic light curve, provides the best performance.}

\begin{algorithm}[H]
\caption{Data Generation Helpers} \label{alg:datagen-helpers}
\begin{algorithmic}
    \Function{baseFlux}{$ $}
    \State $m \gets random([6,16])$
    \State $\omega \gets 309.54 \times 10^{-m / 2.5}$
    \State \Return $\omega$
    \EndFunction
    
    \Function{uncertainty}{$ $}
    \State $\sigma_\omega \gets random([0.0001, 0.1])$
    \State \Return $\sigma_\omega$
    \EndFunction 

    \Function{timeSampling}{$ $}
    \State $t \gets [0]$
    \State apparitions\_per\_viewing $\gets$ random(10, 50) \Comment{If $\infty$, samples a polar star}
    \While{$t[-1] < total\_days$}
        \If{length($t$) \% apparitions\_per\_viewing $== 0$}
            \State $t$.append($t[-1] + 0.115)$ \Comment{0.115 is the orbital period in days}
        \Else
            \State $t$.append($t$[-1] + 170) \Comment{About 6 months until great circle scans cover the star again}
        \EndIf
    \EndWhile
    \State \Return t
    \EndFunction 
\end{algorithmic}
\end{algorithm}

\subsubsection{Nulls} \label{subsubsec:nulls}
We generate nulls very simply. Choose a base flux value and sample the normal distribution with $\mu = \omega, \sigma = \sigma_\omega$. (See Algorithm 3.)

\begin{algorithm}[H]
\caption{Static Light Curve Generation} \label{alg:nullgen}
\begin{algorithmic}

\State $\omega \gets \Call{baseFlux}{  }$

\Function{null}{$x$}
\State \Return $\mathcal{N}(\omega, \Call{uncertainty}{ })$
\EndFunction

\State $x \gets $ \Call{timeSampling}{$ $}
\State $y \gets $ \Call{null}{$x$}
\State $err \gets \Call{uncertainty}{ }$

\end{algorithmic}
\end{algorithm}

\subsubsection{Transients} \label{subsubsection:novae}
\cc{We devise a model for the luminosity of a transient event as a function of time:} 
\begin{equation*}
\ell(t) = \frac{1}{ts\sqrt{2\pi}}\exp\left(-\frac{\left(\operatorname{arsech}\left(\frac{t}{d}\right)-u\right)^{2}}{2s^{2}}\right)
\end{equation*}
where:
\begin{conditions}
u & "decay speed" with $1 < u < 3$ \\
s & "peakiness" with $1 < u < 2$ \\
d & duration in days with $d > 0$
\end{conditions}
\cc{This equation is based off of morphology rather than astrophysical phenomena, but it is visually consistent and gives performant results. It is able to model sharper events like some novae, events with slower decay such as supernovae, and events with very slow decay such as some types of transient YSO activity \citep{fuori, exlup, class0}.} There are generally two types of transients: those which are visible before their transient event, and those which only become visible due to their transient event and thus have a shorter observation window. Some of the time, we eliminate the dimmest points of the light curve so that we have both types of transients. (See Algorithm 4.)

\begin{algorithm}[H]
\caption{Transient Light Curve Generation} \label{alg:novagen}
\begin{algorithmic}
\State $\omega \gets \Call{baseFlux}{  }$
\State $d \gets $ random($600, 3000$)
\State $t' \gets $ random($0, $total\_days$-d$)
\State $s \gets $ random($1, 2$)
\State $u \gets $ random($1, 3$)
\Function{transient}{$x$}
    \If{$t^* < x < t^* + d$}
        \State $\omega' \gets \omega\cdot \frac{1}{ts\sqrt{2\pi}}\exp\left(-\frac{\left(\operatorname{arsech}\left(\frac{(t-t')}{d}\right)-u\right)^{2}}{2s^{2}}\right) + \omega$
        \State \Return $\mathcal{N}(\omega', \Call{uncertainty}{ })$
    \Else
        \State \Return $\mathcal{N}(\omega, \Call{uncertainty}{ })$
    \EndIf
\EndFunction
\State $x \gets $ \Call{timeSampling}{$ $}
\State $y \gets $ \Call{transient}{$x$}
\If{random($0,1$) $< 0.5$}
    \State filter $y > 1.25\omega$
\EndIf

\State $err \gets \Call{uncertainty}{ }$
\end{algorithmic}
\end{algorithm}

\subsubsection{Transits} \label{subsubsec:transits}
The underlying light curve of a transit is well understood, and can be well approximated by a trapezoidal model \citep{kaitlynpaper}. We implement such a model, limiting transit depth to at least $3$ S/N and a maximum of a $60\%$ reduction in flux. We add an opposite secondary transit between $0.2$ and $0.9$ times the depth, since stellar transits are the primary object of study here. The periods vary between $0.1$ and $25$ days. Finally we vary the trough length of the transit, in effect modulating the size of the star or the orbital period of the system. (See Algorithm 5.)

\begin{algorithm}[H]
\caption{Transit Light Curve Generation} \label{alg:transitgen}
\begin{algorithmic}

\State $\omega \gets \Call{baseFlux}{  }$
\State period $\gets $ random($[0.1, 25]$)
\State depth $\gets $ random(3 S/N in decimal, 0.75)
\State secondary\_depth $\gets $ random($0.2,0.9$)

\Function{transit}{$x$}
\State $x \gets x \mod $ period
\If{$x$ in primary transit}
\State $\omega' \gets \omega(1 - \text{depth})$
\State \Return $\mathcal{N}(\omega', \Call{uncertainty}{\omega'})$
\ElsIf {$x$ in secondary transit}
\State $\omega' \gets \omega(1 - \text{depth} \cdot \text{secondary\_depth})$
\State \Return $\mathcal{N}(\omega', \Call{uncertainty}{ })$
\Else
\State \Return $\mathcal{N}(\omega, \Call{uncertainty}{ })$
\EndIf

\EndFunction
 
\State $x \gets $ \Call{timeSampling}{$ $}
\State $y \gets $ \Call{transit}{$x$}
\State $err \gets$ \Call{uncertainty}{ } 

\end{algorithmic}
\end{algorithm}

\subsubsection{Pulsating Variables} \label{subsubsec:pulsators}

Finally, the generation of intrinsic variables is the most involved. In nature they vary the most, with a period ranging from less than a day to more than a year, such as in Mira variables \citep{miravar}. Their waveform can vary greatly as well, from a sine or small Fourier series, to a sharper and more extreme waveform, often a sawtooth. Rather than concerning ourselves with any specific set of waveforms to look for, we leverage the fact that neural networks naturally gain advantage from learning on a wide variety of examples. We can generate a variety of waveforms by:
\begin{enumerate}
    \item Creating an approximating discrete grid, representing one period
    \item Adding dirac $\delta$ impulses of varying heights at some points. (See Figure \figref{7}.)
    \item Performing a discrete convolution on the grid with a gaussian filter $\sim e^{-x^2}$
\end{enumerate}
This generates a series of peaks, smoothly interpolated thanks to the filter. By padding with zeros we ensure the signal returns to the mean at the endpoints of the period. The period is then repeated for the entire observation window. (See Algorithm 6.) A sample of 64 synthetic light curves is available in appendix \ref{sec:appendixB}.

\begin{figure}
    \label{fig:pulsatorgen}
    \centering
    \subfloat[]{\includegraphics[width=\linewidth]{ 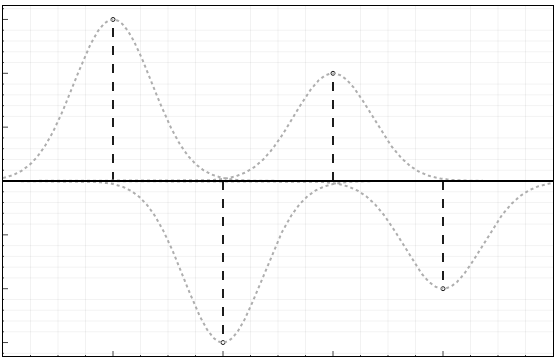}} \\
    \subfloat[]{\includegraphics[width=\linewidth]{ 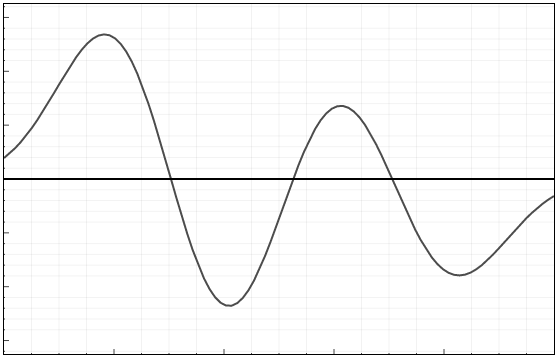}}
    \caption{Random dirac impulses, convolved with a gaussian filter produce a smooth but potentially highly unique waveform.}
\end{figure}

\begin{algorithm}[H]
\caption{Pulsator Light Curve Generation} \label{alg:pulsatorgen}
\begin{algorithmic}

\State $\omega \gets $\Call{baseFlux}{}

\State Grid $\gets [0,1, \cdots, 100]$

\State period $\gets$ random($[0.1,500]$)
\State nPeaks $\gets$ random($1,5$)
\State heights $\gets$ [random($[3 S/N, 0.75]$) for nPeaks]
\State centers $\gets$ [random($0, 99$) for nPeaks]

\State Grid[centers] $\gets$ heights \Comment{Dirac $\delta$s}

\State \textbf{envelope}(x) $=$ $\frac{1}{\sqrt{a\pi}}e^{-\frac{x^2}{a}}$
\State filter $\gets$ \textbf{envelope}(-50,50)

\State Grid $\gets$ \textbf{convolve}(filter, Grid) \Comment{Now smooth}

\Function{Pulsator}{$x$}
\State $x \gets x \mod $period
\State index\_before, index\_after $\gets \lfloor x \rfloor, \lceil x \rceil$
\State $\omega' \gets$ interpolate(Grid[index\_before], Grid[index\_after])
\State \Return $\mathcal{N}(\omega',$ \Call{uncertainty}{ }$)$
\EndFunction

\State $x \gets $ \Call{timeSampling}{}
\State $y \gets $ \Call{Pulsator}{$x$}
\State $err \gets$ \Call{uncertainty}{ } 

\end{algorithmic}
\end{algorithm}

\subsection{\cc{Validation Catalog}}
\cc{In order to measure the accuracy of the model, we construct a validation catalog of variable and nonvariable objects with labels. Objects were chosen over a wide brightness range and confirmed by eye to show variability in NEOWISE in the W1 band. We chose known eclipsing binaries and pulsating stars from the General Catalog of Variable Stars \citep{gcvs}, as updated at the HEASARC website (\url{https://heasarc.gsfc.nasa.gov/W3Browse/all/gcvs.html}). In total, 48 eclipsing binaries were chosen, along with 49 pulsators, the latter of which contained several examples in each of the following classes: Cepheids, W Vir-type, Miras, RR Lyr-type, RV Tau-type, and semi-regular variables. True positives for transient objects were drawn from Koji Mukai's list of Galactic novae (\url {https://asd.gsfc.nasa.gov/Koji.Mukai/novae/novae.html}) whose transient events occurred during the NEOWISE Reactivation mission (i.e., after 2013 Dec 13). This list includes novae, symbiotic stars, and cataclysmic variables. As with the pulsators, these transients were confirmed to show variability in the NEOWISE W1 band, resulting in a list of 34 objects. Our nonvariable examples were chosen at random from across the sky and confirmed to be nonvariable by manual inspection of the light curve data. We have exactly 100 in the final catalog. The validation catalog thus contains 231 actual infrared sources in total.}

\section{The VARnet Model} \label{sec:model}

At the core of the pipeline is the VARnet classifier. If the NEOWISE catalog were not so large, it would potentially be feasible to run more accurate methods with elements of brute-force and resampling, such as phase-folding, but in this case we have to operate primarily on the original sequence. A neural network was selected for its ability to fit well to difficult tasks while maintaining the same runtime. In particular, the issue of detecting low S/N phenomena without the ability to phase-fold makes the problem much more difficult, suitable for thorough analysis via machine learning. In addition, neural networks can be run by batch, which provides a speedup by orders of magnitude using modern GPUs. This makes sub-millisecond in-depth light curve classification possible, which is necessary in order for the processing of the entire sky to be feasible. 

The power of the model originates from the use of the Discrete Wavelet Transform \cc{(DWT)} \citep{waveletdaub} to de-noise and extract high-frequency detail with a frequency and time localized filter, and subsequently a modified Discrete Fourier Transform \cc{(DFT)} process, the Finite-Embedding Fourier Transform (FEFT). The latter intelligently extracts longer-lasting periodicities and trends in the processed data. The convolutional and fully connected layers of the predictor serve to interpret and optimize the data before and after these operations to give a meaningful prediction.

\subsection{Background} \label{subsec:ml-background}

As \cc{VARnet} is a supervised learning model, its objective is to recognize and interpret patterns and produce a maximally accurate prediction, minimizing the loss function. We measure the quality of the prediction via categorical cross-entropy loss. Each entry of the output vector $\nu$ is called a \emph{logit}, and represents the predicted likelihood that the star is of each class (we use $\nu_0=$null, $\nu_1$=transient, $\nu_2$=pulsator, $\nu_3$=transit). We compare the output vectors to one-hot target vectors $t$ with this formula:
\begin{align}
    f(\nu)_i &= \frac{e^{\nu_i}}{\sum_j e^{\nu_j}} \\
    \text{Cross-Entropy Loss} &= -\sum_i t_i \log(f(\nu)_i) 
\end{align}
Through backpropagation, the partial derivative of this value with respect to every weight in the network can be acquired. Then, the weight values move counter to their calculated gradient in order to minimize this value, thus maximizing the confidence and quality of our detections. Cross-Entropy Loss and the F1 score\footnote{The harmonic mean of precision and recall} are the main metrics we track during the training process.

The actual procedure of the network is completely composed of matrix operations. A fully connected layer for example is a single matrix $A_{ij} \in \RR^{n \times m}$, in conjunction with a nonlinearity, in our case the ReLU \citep{ReLU}, which allows the fitting of complex relationships in the data. The convolutional layers of the network are slightly more complex. Rather than operating on every entry of the vector $\nu_i$ at once, in effect it only considers a few neighboring entries at a time. These types of layers have awareness of the temporal coherence across the sequence, and are more resistant to overfitting due to their smaller count of parameters.

\subsection{Finite-Embedding Fourier Transform} \label{subsec:feft}

Often the variability present in a star is long-lasting and smooth. A DWT is less directly suited to extracting these signals due to its time-localization. It may refine the true signal and smooth noise, but it does not "identify" the periodicity in the same way a Fourier transform would. However, the DFT is perfectly frequency-localized and time-distributed, which gives it the ability to pick out sine-like frequencies which remain relatively consistent in period.

The definition of the Discrete Fourier Transform on an $N$ long signal $x(n)$ is
\[X(k) = \frac{1}{\sqrt{N}}\sum\limits_{n=0}^{N-1} x(n)e^{\frac{-2i\pi n }{N}}\]
This can be formulated as a matrix multiplication where the signal is a column vector in $\RR^N$. Set $\omega = e^{\frac{-2i\pi}{N}}$ to be the primitive $N$-th root of unity:

\begin{align}
{\displaystyle {\frac {1}{\sqrt {N}}}{\begin{bmatrix}1&1&1&1&\cdots &1\\1&\omega &\omega ^{2}&\omega ^{3}&\cdots &\omega ^{N-1}\\1&\omega ^{2}&\omega ^{4}&\omega ^{6}&\cdots &\omega ^{2(N-1)}\\1&\omega ^{3}&\omega ^{6}&\omega ^{9}&\cdots &\omega ^{3(N-1)}\\\vdots &\vdots &\vdots &\vdots &\ddots &\vdots \\1&\omega ^{N-1}&\omega ^{2(N-1)}&\omega ^{3(N-1)}&\cdots &\omega ^{(N-1)(N-1)}\end{bmatrix}}}
\end{align}

Instead of performing this matrix multiplication, which requires $\mathcal{O}(N^2)$ operations, in practice the Fast Fourier Transform \citep{fft} is performed which leverages matrix factorizations to reduce the DFT of an $N$ long signal to $\mathcal{O}(N\log N)$ time. Notice that the matrix has rank $N$ and is square. Usually this is a desired property, as it then represents a change of basis operation within the same vector space $\RR^N$.

However, the potential for variable-length feature vectors within a network is highly inconvenient, as we need to ensure the network always outputs a set-length prediction vector. Combined with the potential to have input sequences thousands of entries long as well as only tens or hundreds of entries long, the efficacy of a recursive model such as a \cc{Recurrent Neural Network (RNN) or a Long-Short Term Memory model (LSTM)} in this case was shown to be inferior to other methods of embedding the sequence into a definite-length vector. Attention-based methods resulted in a significant speed impact and were quite difficult to train.

Therefore, we modify the original Discrete Fourier Transform in order to extract features from the variable length sequences. By creating a rectangular matrix $\RR^{m \times N}$, we create a DFT-like mapping from the Euclidean basis in $N$ dimensions to a coarser basis in $m$ dimensions. It is defined in this way: Notice that after taking the natural logarithm of the original Vandermonde matrix of equation (8), the resulting is equal to the outer product of two vectors. Set $z = \frac{-2i \pi }{N} = \ln(\omega)$

\[
\ln\left(\begin{bmatrix}1&1&1&1&\cdots &1\\1&\omega &\omega ^{2}&\omega ^{3}&\cdots &\omega ^{N-1}\\1&\omega ^{2}&\omega ^{4}&\omega ^{6}&\cdots &\omega ^{2(N-1)}\\1&\omega ^{3}&\omega ^{6}&\omega ^{9}&\cdots &\omega ^{3(N-1)}\\\vdots &\vdots &\vdots &\vdots &\ddots &\vdots \\1&\omega ^{N-1}&\omega ^{2(N-1)}&\omega ^{3(N-1)}&\cdots &\omega ^{(N-1)(N-1)}\end{bmatrix}\right) 
\]
\[
= \begin{bmatrix}0&0&0&\cdots &0\\
0& z & 2z&\cdots &(N-1)z\\
0&2z&4z&\cdots &2(N-1)z\\
\vdots &\vdots &\vdots &\vdots &\vdots \\
0&(N-1)z& 2(N-1)z & \cdots &(N-1)^2 z \\
\end{bmatrix} 
\]
\[
= u \otimes v
\]
\begin{align}
    u_j &= j \quad 0 \leq j < N \\
    v_j &= zj \quad 0 \leq j < N
\end{align}
Thus we can define a family of DFT-like transformations given two vectors $u,v$ by the form:
\[\mathfrak{F}(k) = \frac{1}{\sqrt{N}} \left(\exp{uv^T}\Vec{a} \right)_k\]
With the restriction that $v \in \RR^N$. Thus the dimension of $u$ defines the codomain vector space. When $u_k$ ranges from $0$ to $N-1$, the DFT samples those $N$ frequencies in the data. However if we want to fix $u$ to a certain size so that we achieve a finite embedding of that information, we inevitably have to prioritize sampling some frequencies over others. Rather than explicitly finding frequencies which best fit the problem, which might vary depending on how earlier network layers transform the data, we can directly introduce $u$ as a parameter for the model, with its dimension as a hyperparameter. We name this hyperparameter \emph{samples}, and initialize the FEFT by 
\[\alpha_k \in \RR^\text{samples} \quad \alpha_k = \frac{k}{\text{samples}}\]
\[u_k = \alpha_k(N-1)\]
In the final model, we choose samples=$700$. We do not modify the vector $v$; rather, we simply construct it by integer multiples of $z$ up to the length of the input vector. The outer product and elementwise exponentiation of the matrix are relatively fast operations, and since we do not vary $u$ depending on the size of the input vector, it has time complexity $\mathcal{O}(samples \times n)$, potentially faster than a full Fast Fourier Transform since we are projecting into a definite vector space.

\subsection{Wavelet Decomposition} \label{subsec:wavelet-decomposition}

A \emph{wavelet} is a wave-like oscillation that is time-localized, in that it is near zero for most of the real line. There are many varieties, often characterized by desirable mathematical properties. The study of wavelets is a relatively new field, pioneered by Ingrid Daubechies in the 1990s after the discovery of an orthonormal basis that is both frequency and space localized \citep{waveletdaub}. Thereafter it has merited applications in a variety of fields, such as image processing \citep{imagewavelets}, signal processing \citep{waveletsigprocess}, audio processing \citep{audiowavelets} and even found use in solving partial differential equations \citep{pdewavelets1} \citep{pdewavelets2}. Daubechies first proposed her family of orthonormal wavelets $db_n$, named by having $n$ vanishing moments, AKA points where the wave crosses the $x$ axis. From there, many other families have been discovered or engineered for particular use cases. See Table 1 for a survey of the utility of different wavelets for this use case.

\begin{center}
\label{tbl:wavelets}
\begin{tabular}{||c c c||} 
 \hline
 Wavelet Family & Min. Cross Entropy & @ Epoch \\ [0.5ex] 
 \hline\hline
 haar & 0.307 & 28 \\
 db4 & 0.32 & 51 \\
 db8 & 0.354 & 203 \\
 symlet4 & 0.281 & 38 \\
 discrete\_meyer & 1.05 & 2 \\
 reverse\_biorthogonal5.5 & 0.308 & 206 \\
 biorthogonal2.2 & 0.252 & 36 \\
 biorthogonal1.3 & 0.394 & 58 \\
 coiflet1 & 0.351 & 30 \\
 coiflet3 & 0.316 & 112 \\ [1ex] 
 \hline
\end{tabular}
\vspace{0.75em}

\textbf{Table 1} \emph{Survey of different wavelet families, minimum cross entropy loss achieved over 500 epochs. Lower is better}
\end{center}
\vspace{0.5em}

There exists a myriad of other families with other features and characteristics best suited to some problems more than others \citep{waveletfamilies}. However, the process of the Discrete Wavelet Transform (DWT) is the same for all. The DWT consists of two operations: a high-pass filtering and a low-pass filtering. Essentially, it is a series of convolutions with daughter wavelets of the mother wavelet function $\Psi$. These daughter wavelets are simply rescaled and shifted copies of the mother wavelet:
\[\psi(\alpha, \beta) = \frac{1}{\sqrt{\alpha}}\Psi\left(\frac{x - \beta}{\alpha}\right)\]
The result of this operation is two vectors, one named the \emph{approximation} and one the \emph{detail}. For our purposes, this can be thought of as a down-scaled and smoothed version of the light curve in the approximation, with the detail vector containing high-frequency, momentary impulses that are similar to the waveform chosen. Importantly, there is no loss in information after passing data through a DWT. After surveying the performance of a variety of wavelets, the biorthogonal wavelet 2.2 achieved the lowest loss, thus we implement the DWT making use of it. Appendix \ref{sec:appendixA} depicts the biorthogonal/reverse biorthogonal family of wavelets.

\begin{figure*}
    \label{fig:model}
    \centering
    \subfloat[Model Phase 1]{\includegraphics[width=\textwidth]{ 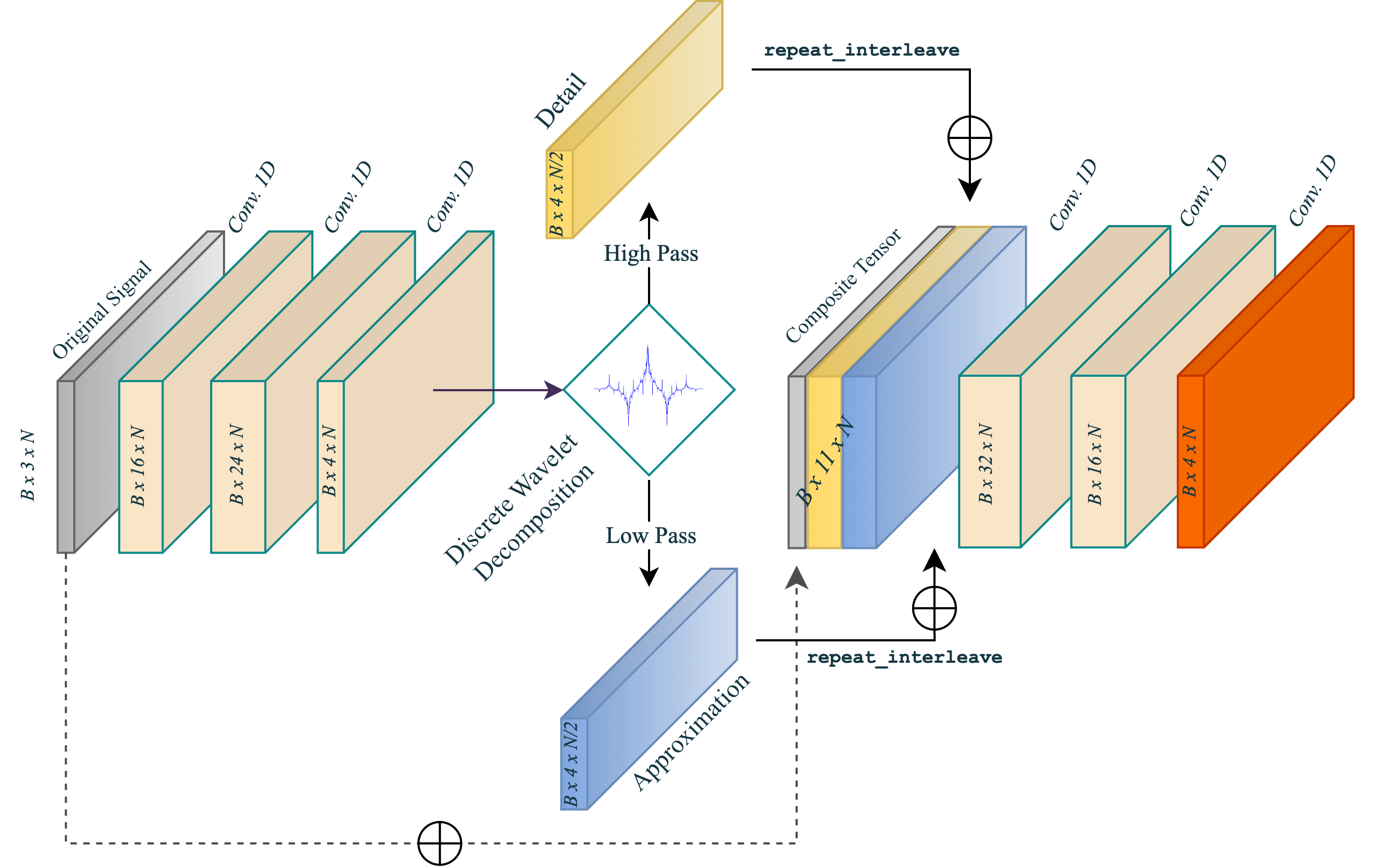}} \\
    \subfloat[Model Phase 2]{\includegraphics[width=\textwidth]{ 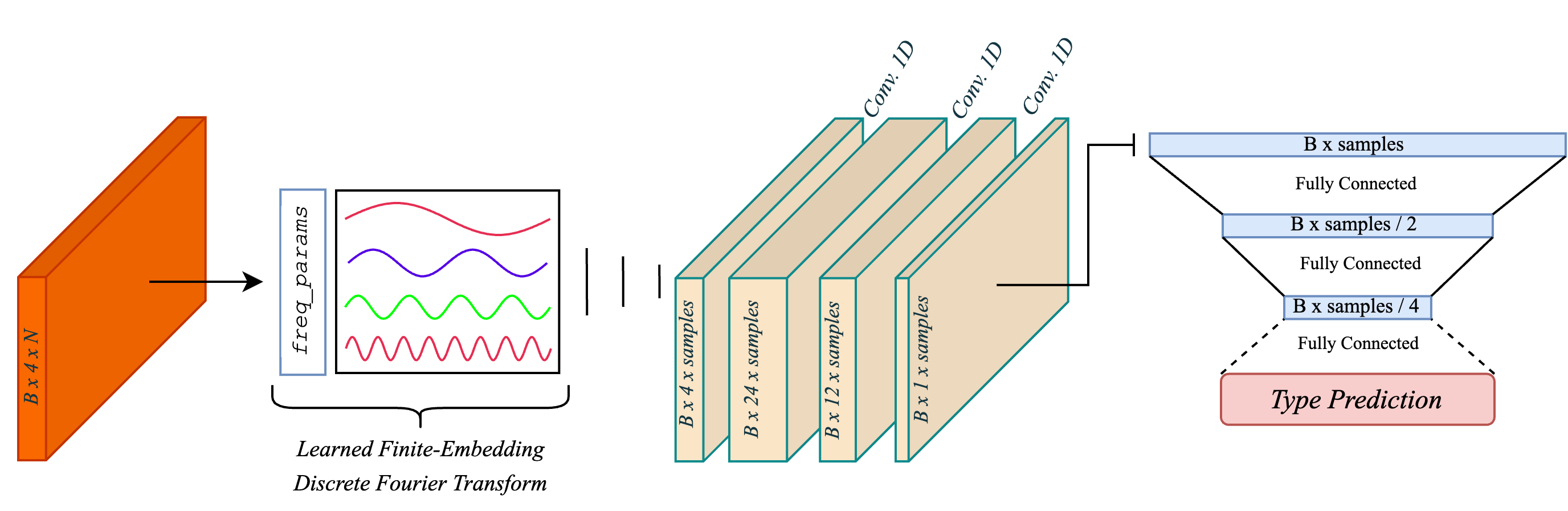}}
    \caption{The Complete VARnet Model}
\end{figure*}

\subsection{Overall Model} \label{subsec:complete-model}

The complete layout of the VARnet model can be seen in Figure \figref{8}. The input tensor is of shape $(B, 3, N)$, where $B$ is the number of batches inputted, $N$ is the length of the longest light curve, with each time having flux, uncertainty, and timestamp in dimension 1. Again, the ReLU \citep{ReLU} was chosen as the activation function all the way through the model. The model begins with $3$ convolutional layers, preparing for the wavelet transform layer. After the wavelet transform, the original signal is concatenated in the channels dimension along with the approximation and detail vectors. Three more layers of convolution are implemented, preparing the signal and compressing into $4$ channels for the Fourier extraction module. After performing the Fourier feature extraction, we are left with a vector of a definite size. Three more convolutional layers extract information regarding frequencies which are proximal, and finally compress down the $4$ channels. Finally, the vector undergoes three fully connected layers to the output vector. When running the model, the final vector undergoes the softmax operation so that the prediction can be interpreted probabilistically and sorted by confidence. 

\section{Results and Discussion} \label{sec:results}

\subsection{Training and Statistics} \label{subsec:stats}
\cc{
To train the model, several optimizers were tested, and a gridsearch was implemented to find optimal hyperparameters for those optimizers. We found the AdamW \citep{adamw} optimizer to be the most performant by a fair margin. We use weight decay as our preferred method of regularization for the network, using a value of $10^{-5.5}$. VARnet appears to train well using a learning rate near $0.0001$. We select $lr=0.000085$ for our training. After $53$ epochs, which translates to $1.06$M unique synthetic example light curves, the model converged to its highest F1 score and lowest cross entropy loss. The final confusion matrix can be viewed in Figure \figref{9}a. We opt to primarily measure the success of our model using the F1 score as a more robust metric than overall accuracy. In order to ignore the effects of our class imbalances in the true positive catalog, we take the macro averages of F1 score, precision and recall. As can be derived from this confusion matrix, the model achieves a precision of $0.918$, a recall of $0.910$, an accuracy of $92.2\%$ and an F1 score of $0.914$. These values are satisfactory for our studies. It should be noted that the confusion between the null class and all other classes is the most important to keep track of. Another confusion matrix is available in Figure \figref{9}b, which is the result of simply collapsing all the variable classes from the 4-class confusion matrix into one, in order to study the real-bogus distinction that VARnet is making. The result is a precision of $0.973$, a recall of $0.975$, an accuracy of $97.4\%$ and an F1 score of $0.974$. When observing the final confusion matrix, it is also apparent that there is the most confusion between the pulsator and transit classes. This is understandable, as some transits, particularly eclipsing binaries of the W Ursae Majoris type have smoother short-period fluctuations in brightness, very similar to short-period pulsators. If the distinction between the pulsator and transit class were to be made via other methods after a secondary classification step, we could combine these classes for this step and greatly improve performance. By combining both the synthesizers and true positives for the pulsators and transits and retraining the model, the final confusion matrix in Figure 9 showcases the result of this approach. It yields an improved precision, recall and F1 score of $0.980$ and an accuracy of $97.4\%$. 
}

\begin{figure}
    \centering
    \includegraphics[width=\linewidth]{ 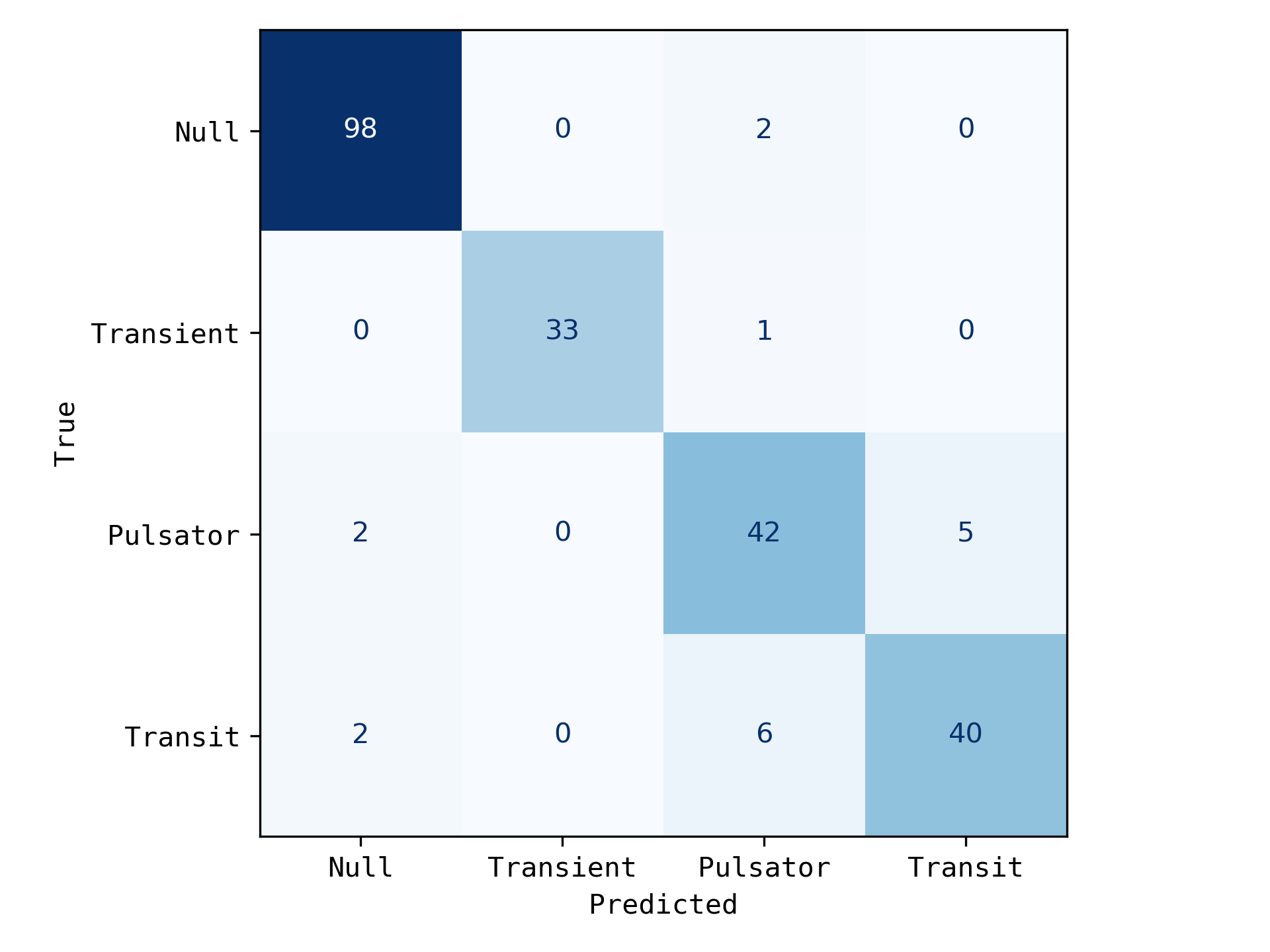}
    \parbox{\linewidth}{\centering (a) 4-class classification scheme confusion matrix}
    \vspace{0.5cm} 
    
    \includegraphics[width=\linewidth]{ 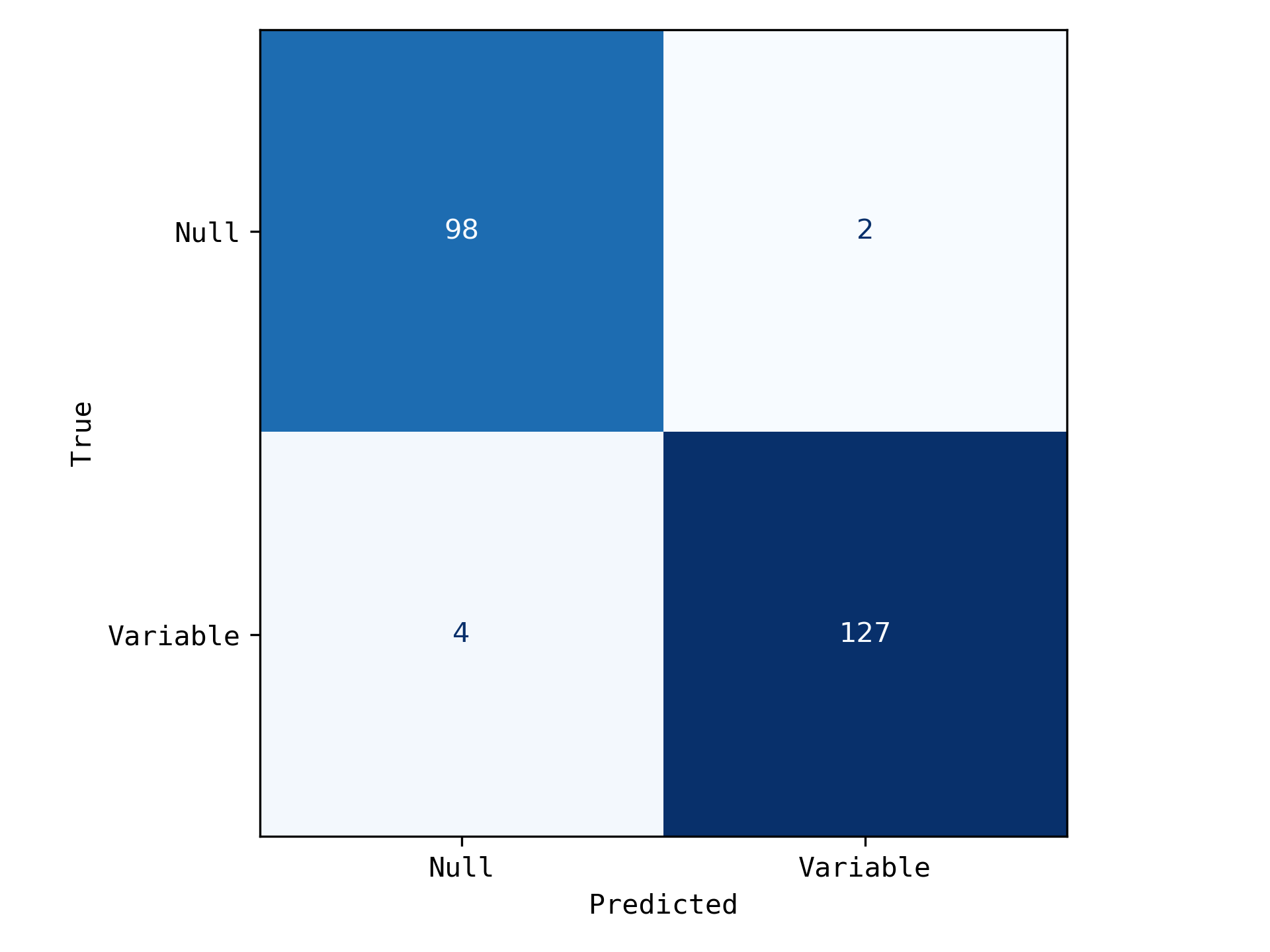}
    \parbox{\linewidth}{\centering (b) Real-bogus confusion matrix produced from (a)}
    \vspace{0.5cm} 
    
    \includegraphics[width=\linewidth]{ 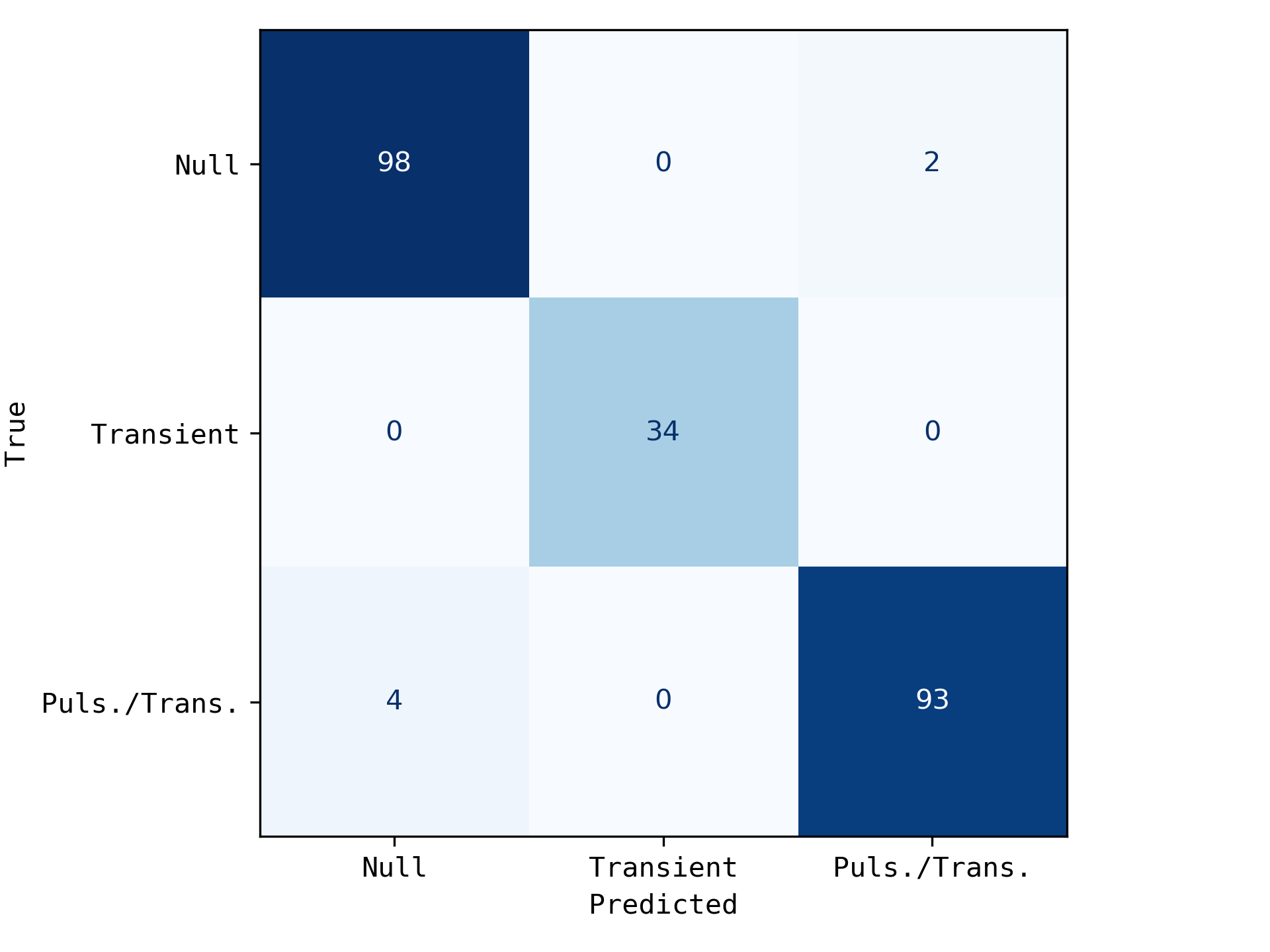}
    \parbox{\linewidth}{\centering (c) 3-class classification scheme confusion matrix, from a separately trained VARnet}
    
    \caption{\centering Confusion matrices produced by a trained VARnet Classifier}
    \label{fig:conf-matrix}
\end{figure}

\subsection{Hardware and Performance}
The computer used for this paper contains an NVIDIA Quadro RTX 6000 with $22$GB of VRAM, $200$GB of RAM and a 32-core Xeon CPU, courtesy of Caltech. \cc{In inference, the model can process a batch of $4096$ light curves in $216$ms, resulting in a per-source speed average of $52\mu s$. Notably, the amount of operations done is determined by the light curve in the batch with the greatest length because of padding. If the greatest light curve has fewer samples, more light curves will be able to fit in memory, thus leading to higher parallelization.} The DBSCAN implementation we use is part of the Scikit-Learn package. This configuration introduces some amount of overhead when running the model, but there is ample room for optimization and possibly a switch to a lower-level \cc{implementation} for speed improvements.

\subsection{Trial Run}
To sample the speed of an actual implementation of these methods, we inspected a $25$ square degree region in the orion nebula with the pipeline. These $25$ square degrees corresponded to $62,841,525$ rows of data. The clustering step took $287.47$ seconds to complete, and the classification with the model including data loading, GPU onloading and offloading, took $112.44$ seconds. However, the total time spent by the model performing operations amounted to $0.41$ seconds. The rest of the time was occupied by data management.

Currently, the total count of NEOWISE catalog apparitions without any sort of filtering is $188,876,840,852$. Given that all operations done by the network are linear $\mathcal{O}(n)$ time, the actual evaluation time for the dataset can be reasonably estimated by \[\frac{188876840852}{62841525} \times 0.41s \approx 1232s\] which is approximately $20.5$ minutes, more than satisfactory for a dataset of this size. This is assuming that every single row of the database will correspond to a legitimate source and be included in a cluster, which is far from the truth. Many regions have a high number of false catalog detections due to nebulae or ray impacts, which do not concentrate enough to be recognized as a clusters and thus are not ultimately processed by the model. Of course the time necessary for data loading and other sources of overhead present in the evaluation loop will dwarf this value. However the real value is subject to many variables and can be drastically reduced without major change to pipeline presented in this paper.

The primary clustering step is still the highest-cost part of the pipeline. It only must be performed once, but in order to process the entire NEOWISE database, a more sophisticated approach will have to be developed in order to segment the sky and perform the clustering in parts and possibly with greater parallelism. A setup with more CPUs and critically RAM would greatly improve the speed of the clustering. Others have approached this problem and achieved satisfactory results \citep{blockdbscan, paralleldbscan}, but such an implementation was out of scope for this paper. 

\subsection{First Found Sources} \label{subsec:firstfoundsources}
After running the pipeline on various different regions of the sky, it can be confirmed that the algorithm is precise in detecting the desired objects and is also capable of detecting more uncommon and irregular anomalies. Overall, the model proves effective. We will cover and analyze a few interesting detections from the top of the flag list \footnote{The highest value logit (\ref{subsec:ml-background}) after the softmax operation (Equation 7) is taken to be the confidence of the models prediction. The flag list is sorted by descending confidence.}

\subsubsection{Recovered sources} \label{subsubsec:recovered}
Beyond its performance on \cc{our catalog of true positives}, the algorithm proves capable of recovering many known objects with high confidence. For example, eclipsing binary \emph{V* V1403 Ori} is recovered at a confidence greater than $0.99$ (Figure \figref{9}). Another elegant detection is eclipsing binary system \emph{CRTS J054306.5-024247}, with a period of $3.89$ days (Figure \figref{10}). There are many other recoveries from this test with lower S/N features that are not given here; however, these prototypical objects simply demonstrate that we are able to find the desired objects with precision and confidence.

\begin{figure}[H]
    \label{fig:v1403-lc}
    \centering
    \includegraphics[scale=0.25]{ 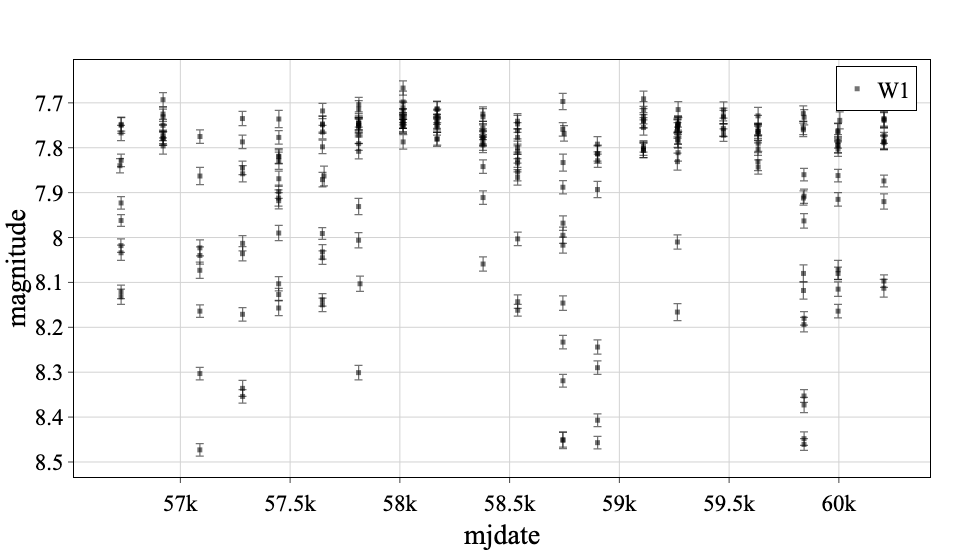}
    \includegraphics[scale=0.25]{ 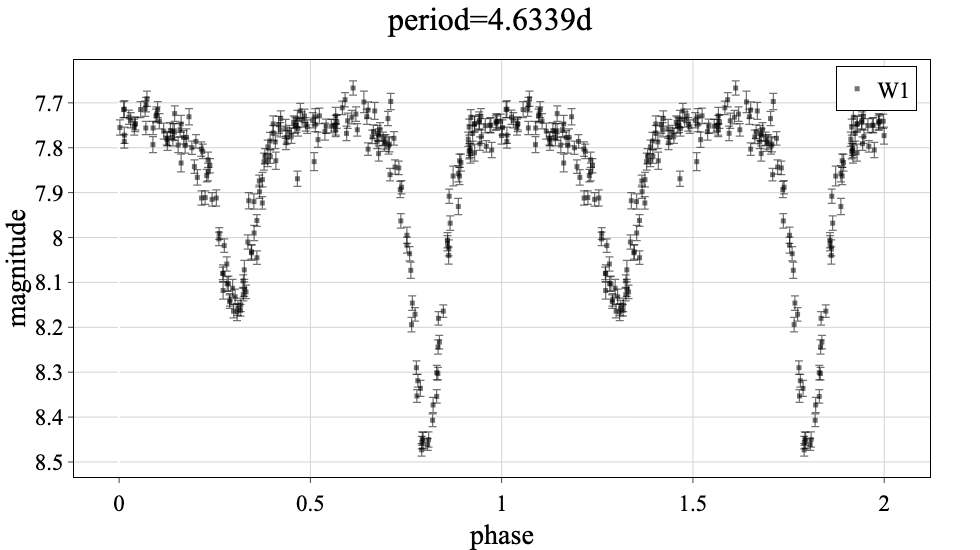}
    \caption{V1403 Ori as seen by WISE in the W1 band. Bottom curve is folded at period$\sim 4.63$ days.}
\end{figure}
\vspace{-1.5em}

\begin{figure}[H]
    \label{fig:crts-lc}
    \centering
    \subfloat{\includegraphics[scale=0.25]{ 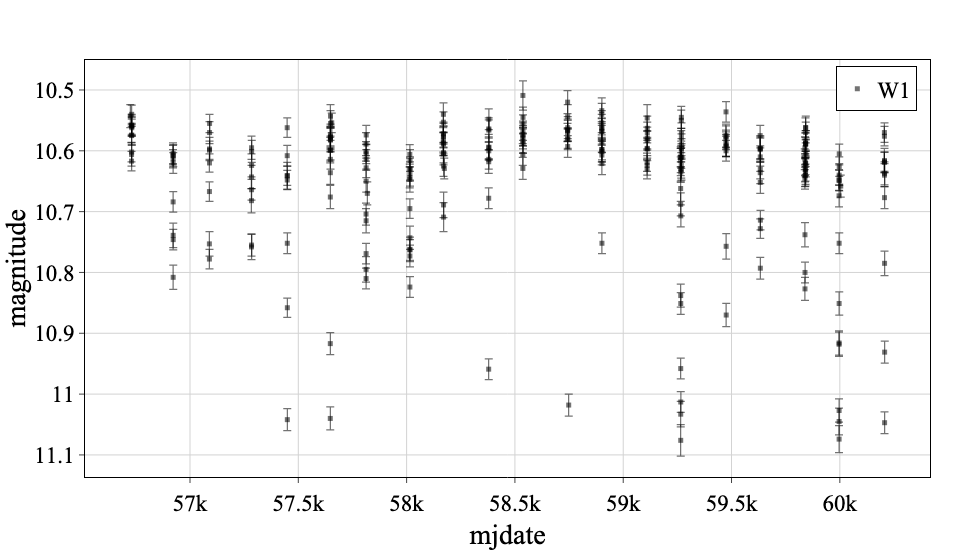}} \\
    \subfloat{\includegraphics[scale=0.25]{ 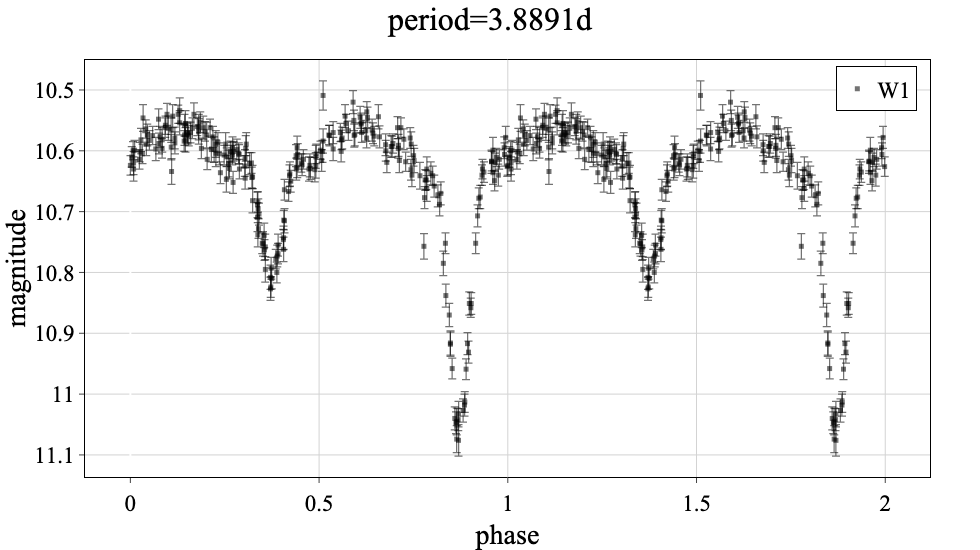}}
    \caption{CRTS J054306.5-024247 as seen by WISE in the W1 band. Bottom curve is folded at period$\sim 3.89$ days.}
\end{figure}
\vspace{-1.5em}

\subsubsection{Uncataloged Eclipsing Binary} \label{subsubsec:eb-cand}

An object centered at J2000 RA/Dec $1.53483$ deg, $-59.08751$ deg was flagged as a variable candidate by VARnet. A search reveals no literature or containment in any catalog. Via the Plavchan algorithm \citep{plavchanalg}, the period was determined to be approximately $5.877$ days.

\begin{figure}[H]
    \label{fig:eb-cand-lc}
    \centering
    \includegraphics[scale=0.25]{ 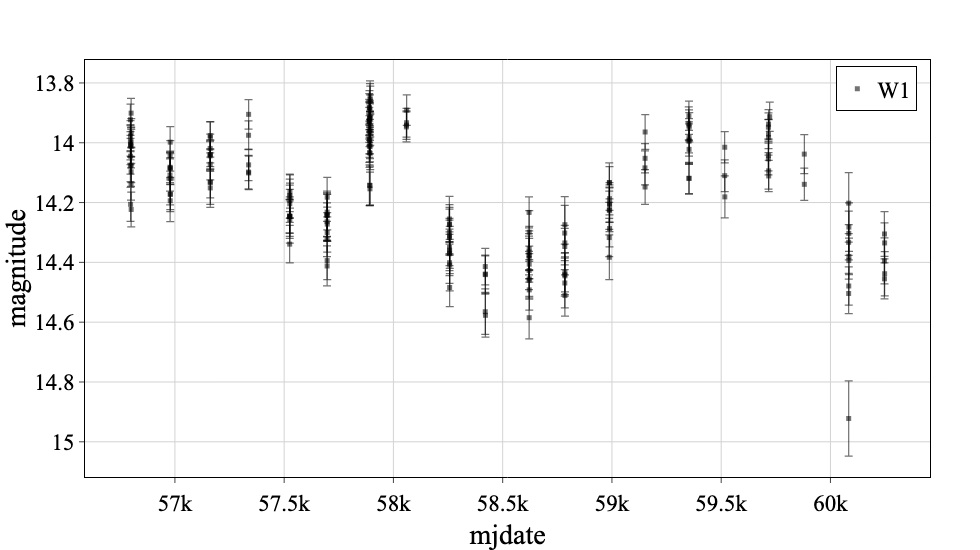}
    \includegraphics[scale=0.25]{ 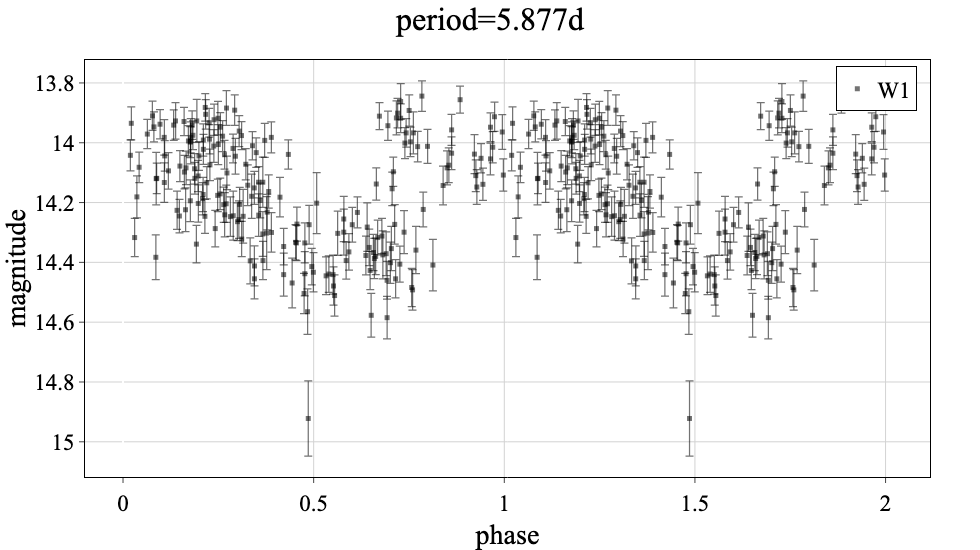}
    \caption{\centering Light curve of binary system at J2000 RA/Dec $1.53483,-59.08751$. Bottom curve is phase-folded at $5.877$ days}
\end{figure}

The phase-folded light curve (Figure \figref{11}) reveals a \cc{possible} transit signal.

\subsubsection{Detached Binary 2MASS J01542169-5944445} \label{subsubsec:chem-pec}

The pipeline gave a detection for variability centered at J2000 RA/Dec $28.59051$ deg, $-59.74571$ deg. We use the 2MASS designated name, but \emph{2MASS J01542169-5944445} was also detected and cataloged by the GALAH survey in its third data release \citep{galah_dr3}\cc{, not as a variable star but as a chemically peculiar star.} This raises the possibility of a companion body which has altered the chemistry and thus the observed spectrum of this source through mass transfer.

Again, using the Plavchan algorithm, the period of this system was determined to be $5.8061$ days (Figure \figref{12}). As seen in the phase-folded plot, there is a sharp transit signal. In imaging, and as can be inferred from the absence of a secondary transit, the companion body is dim and not visible with WISE. Notably we do not observe any variability in the flux of the system where there is no transit, suggesting that this is a detached binary system with little current mass transfer. \cc{One potential explanation is a companion white dwarf which has altered the chemistry of its companion via mass transfer, and now transits the visible star. Ancillary data may confirm or deny this candidacy.}

\begin{figure}[H]
    \label{fig:chem-pec}
    \centering
    \includegraphics[scale=0.25]{ 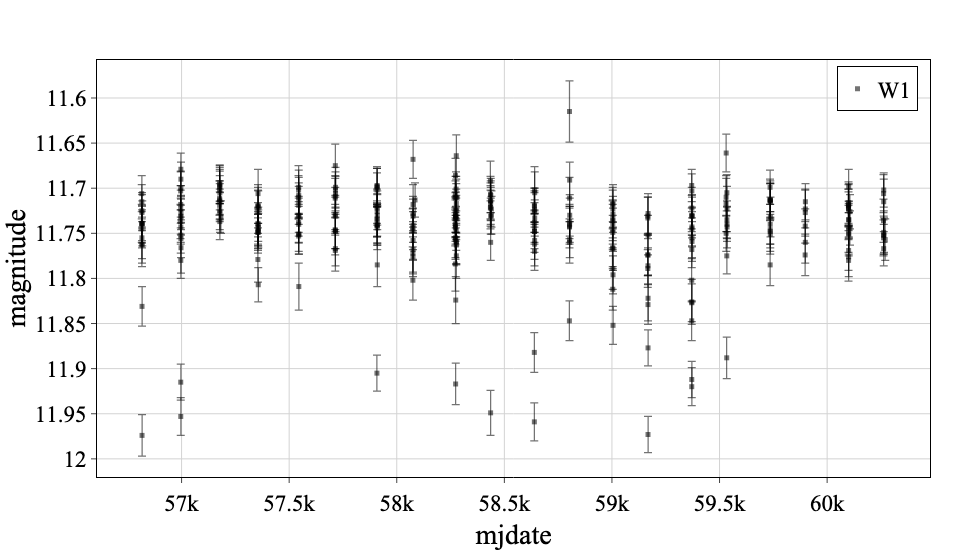}
    \includegraphics[scale=0.25]{ 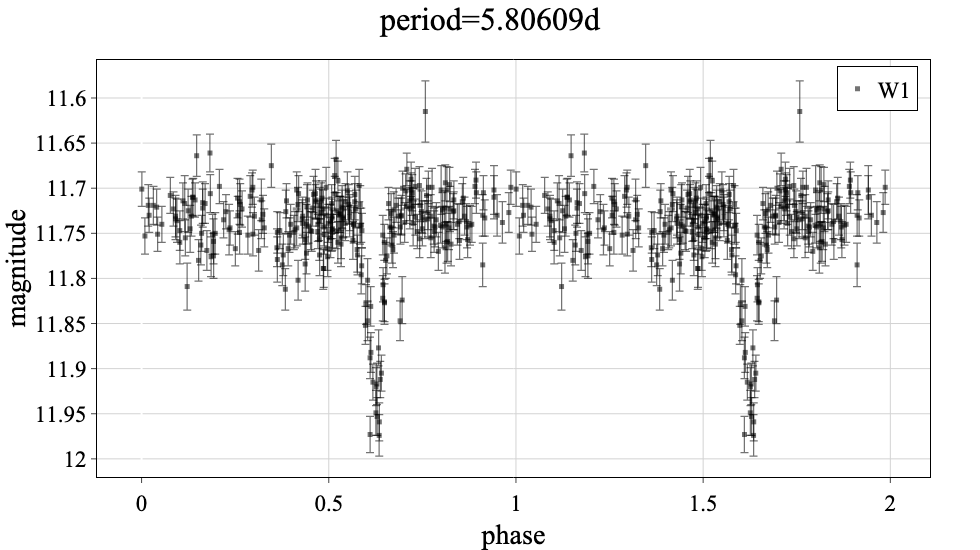}
    \caption{Light curve of \emph{2MASS J01542169-5944445}. Bottom curve is phase folded at $5.8061$ days.}
\end{figure}

\subsubsection{Variability of LEDA 174461} \label{subsubsec:quasarcand}

The pipeline gave a detection for variability at J2000 coordinates $29.30835$ deg, $-61.59386$ deg, matching to galaxy LEDA 174461. The variability was quite faint, but significant in comparison to the absolute flux, with a relatively atypical difference between the two available infrared bands.

\begin{figure}[H]
    \label{fig:quasarcand-lc}
    \centering
    \includegraphics[scale=0.25]{ 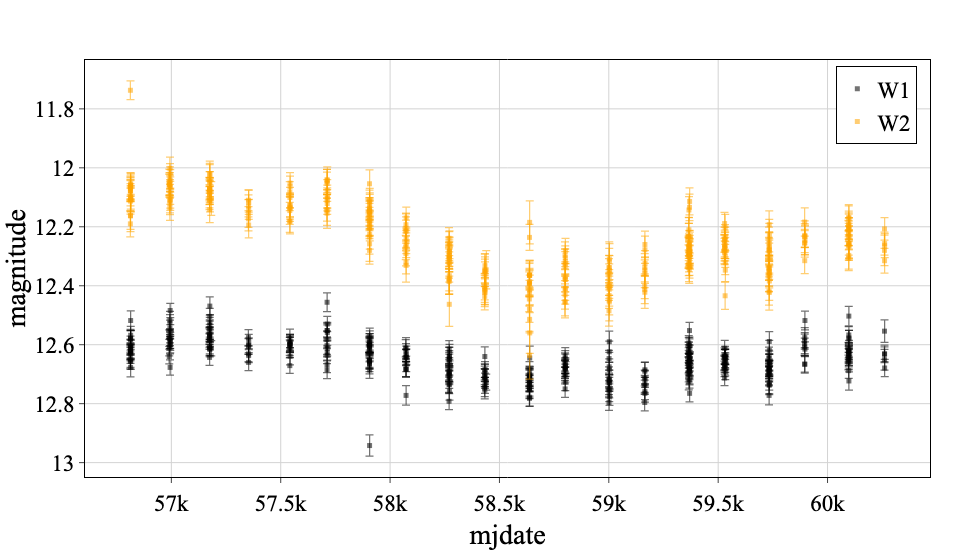}
\end{figure}

\begin{figure}[H]
    \label{fig:quasarcand-lc-folded}
    \centering
    \includegraphics[scale=0.25]{ 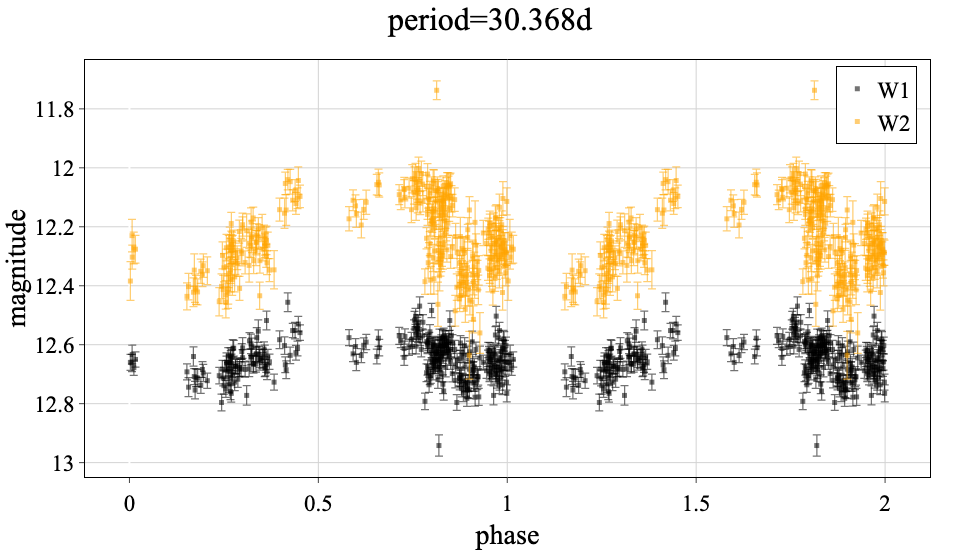}
    \caption{2-band light curve of galaxy LEDA 174461. Period was determined to be 30.368 days. Quality of phase folded curve is impacted, as the period is close to a clean factor of the observation cadence.}
\end{figure}

The Lomb-Scargle periodogram and manual inspection yielded a period of $30.368$ days, though the best fold provided is not overwhelmingly convincing. When viewed through WISE imagery, the object appears predominantly \cc{red} (dominant in the W2 infrared band). Additionally, there appears to be an aperiodic variability in the spectrum of the source, with the W2 band dimming in magnitude discrepant with the W1 band (Figure \figref{13}).


Using higher resolution surveys, two other objects can be resolved nearby this galaxy (Figure \figref{14}). However, when inspecting the variability through WISE tools it can be determined that the variability in flux is attributable to the galactic nucleus. \cc{From this information, we conclude that the variability of \emph{LEDA 340305} is attributable to an active galactic nucleus, although the variation in color of this source requires further analysis.}

\begin{figure}[H]
    \label{fig:quasarcand-imaging}
    \centering
    \mbox{
    \subfloat{\includegraphics[scale=0.21, trim={50px 0px 50px 0px}, clip]{ 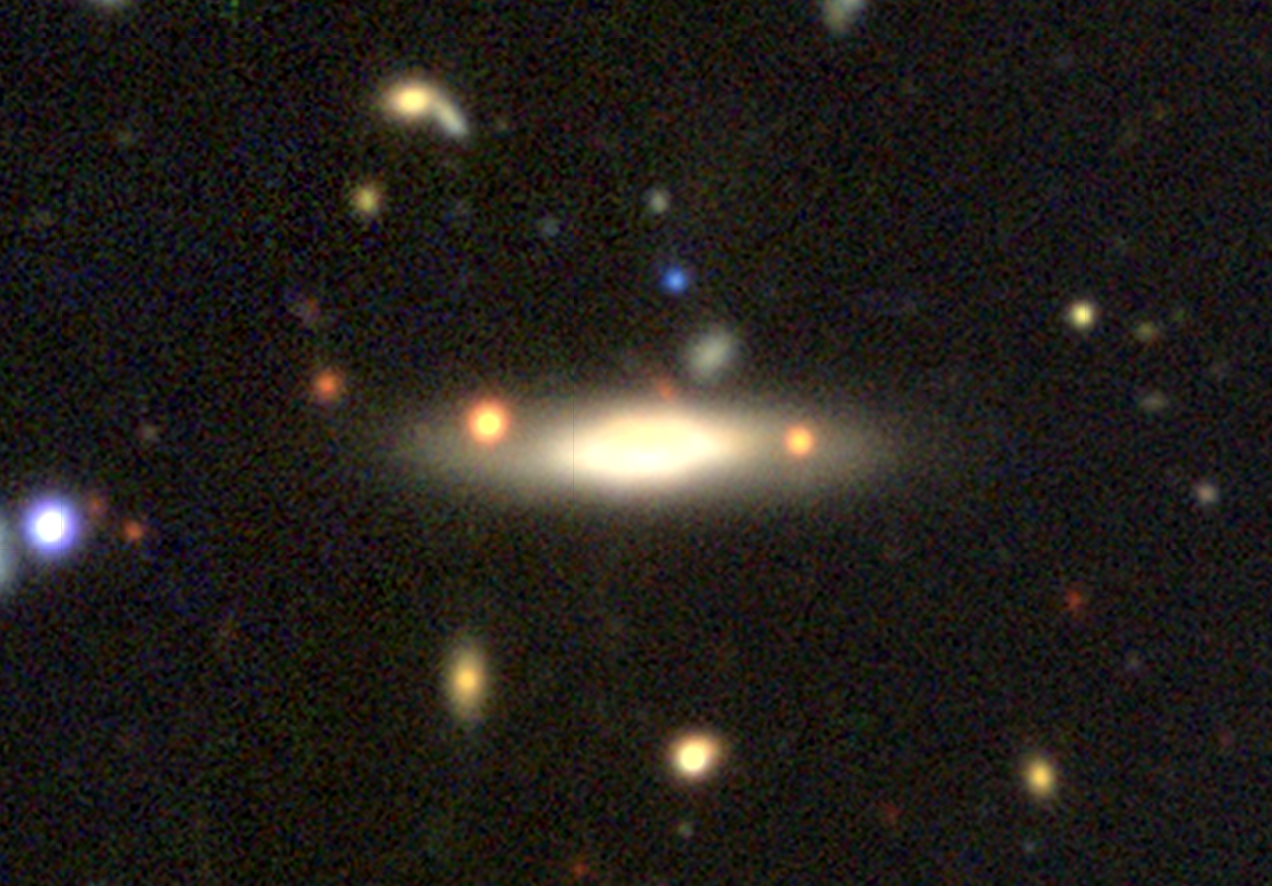}} \hspace{0.5em}
    \subfloat{\includegraphics[scale=0.21, trim={50px 0px 50px 0px}, clip]{ 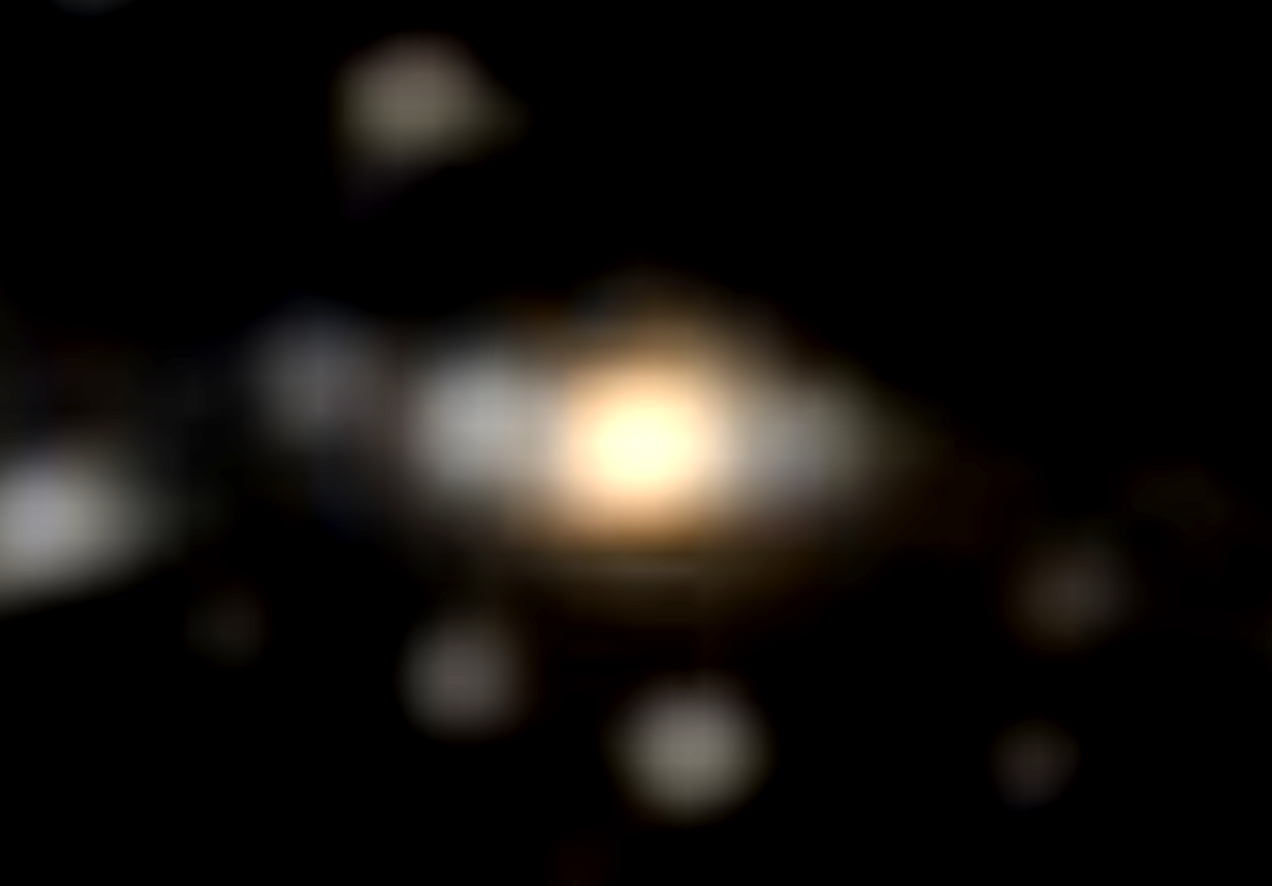}}
    }
    \caption{\centering Left - Legacy Survey Data Release 9 imaging of LEDA 174461. Right - unWISE W1/W2 NEOWISE Data Release 7 imaging.}
\end{figure}

\subsubsection{Supernova within LEDA 358365} \label{subsubsec:supernovacand}

The pipeline flagged a source for transient activity at J2000 RA/Dec $31.40235$ deg, $-61.05673$ deg, matching to cataloged galaxy \emph{LEDA 358365}. In June of 2023, the source appeared to quickly brighten. At the time of the next observation with WISE, it had returned to the mean (Figure 16).

\begin{figure}[H]
    \label{fig:supernovacand-lc}
    \centering
    \includegraphics[scale=0.25]{ 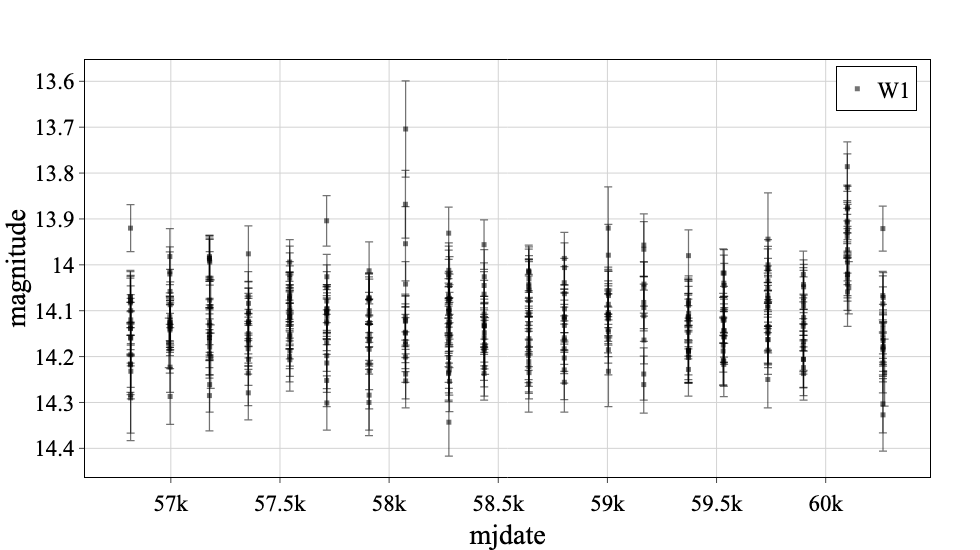}
    \caption{Light curve of \emph{LEDA 358365}. The transient event is visible through WISE in the second to last observation epoch, mid 2023.}
\end{figure}

After searching through transient catalogs, this event matches to detection \emph{AT 2023lkp}. Given the duration of the event and its origin within this galaxy, we conclude in favor of a supernova. The light curve is shown in figure \figref{15}.

\subsubsection{Active Galactic Nucleus of LEDA 340305} \label{subsubsec:agn}

The pipeline flagged a source at J2000 RA/Dec $358.45721$ deg, $-62.75215$ deg for transient activity, matching to cataloged galaxy \emph{LEDA 340305}. The source exhibited a significant increase in the observed brightness of the galaxy, peaking in May of 2016 (Figure \figref{16}).

\begin{figure}[H]
    \label{fig:agn-lc}
    \centering
    \includegraphics[scale=0.25]{ 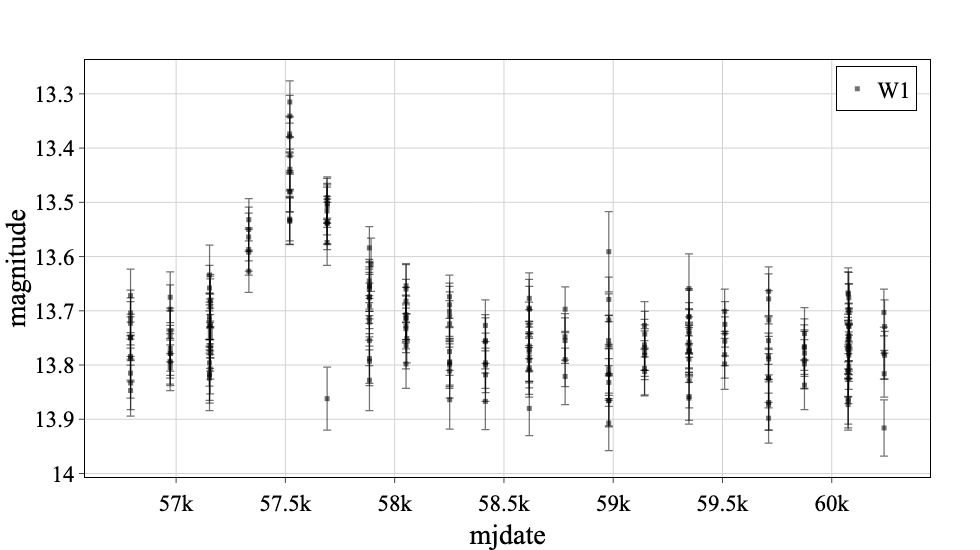}
    \caption{Light curve of \emph{LEDA 340305}. Transient activity in the first half of the WISE observation window.}
\end{figure}

Notably, the transient event is dissimilar to a supernova in that it has a relatively slow transition to its most excited state, although it does appear to dim exponentially thereafter. These observations are most consistent with an active galactic nucleus.

\begin{figure}[H]
    \label{fig:agn-imaging}
    \centering
    \includegraphics[scale=0.365]{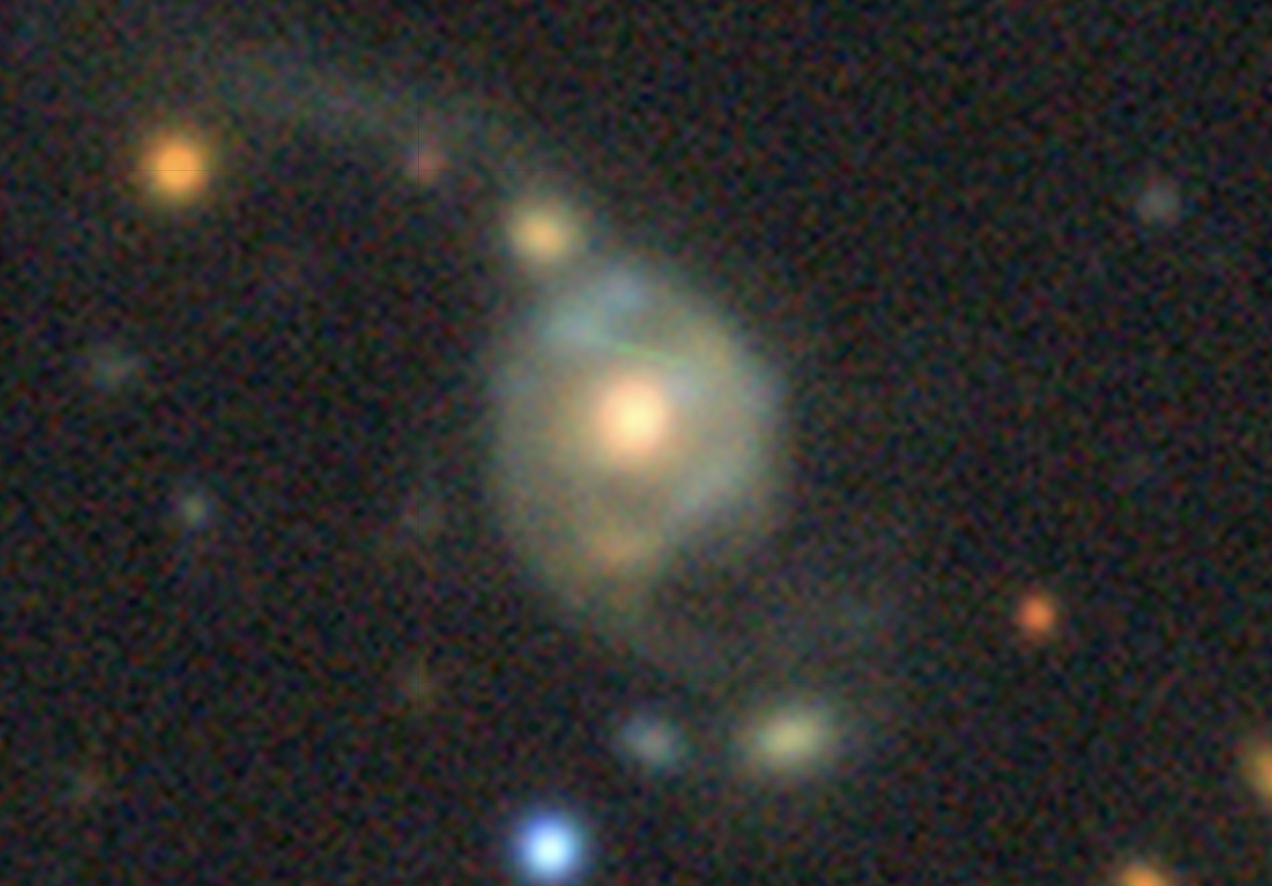}
    \caption{ \centering \emph{LEDA 340305} imaging from Legacy Survey Data Release 9. Note the extended spiral arms and irregular shape.}
\end{figure}

After inspecting higher resolution imaging, \emph{LEDA 340305} can be seen to have an irregular shape (Figure \figref{17}). In particular its extended spiral arms suggest that \emph{LEDA 340305} was involved in a collision or merger with another galaxy in the past. This activity conceivably has brought new matter within reach of the supermassive black hole, leading to this observed AGN.

\subsection{\cc{Summary and Conclusions}} \label{subsec:ending-notes}
\cc{This paper introduces VARnet as a signal processing model optimized for rapid astronomical timeseries analysis. VARnet incorporates tools from deep learning such as fully connected and convolutional neural networks in tandem with wavelet decomposition and the Finite-Embedding Fourier Transform to dynamically analyze timeseries information. We apply VARnet to the issue of detecting infrared source variability in the NEOWISE dataset, devising methodology to handle WISE data such as preprocessing and density-based clustering of apparitions. To train VARnet for this use case, we implemented light curve synthesizers for the objects we hope to identify. After training on $1.06$M unique generated light curves, VARnet achieves an F1 score of $0.91$ on a validation set of known variable objects, as well as a $0.97$ real-bogus precision score. VARnet operates in linear time at the same time as it leverages GPU parallelization, enabling it to achieve a per-source processing time of $52\mu s$ on our hardware. After running the model in several regions across the sky, we confirm that our methodology confidently recovers known variable sources and is capable of discovering previously undetected or unstudied infrared variability.}

\cc{This paper is intended to be proof of concept for methods to be used in an upcoming all-sky survey of variability using the entirety of the NEOWISE database. That survey will have a more dense taxonomy for variability, utilizing secondary and tertiary analyses to subdivide the initial classes provided by VARnet. From the studies done here, we have identified some issues to be addressed before carrying out a large-scale deployment of this pipeline. The primary concern is the clustering step. Given the sheer magnitude of data present in the catalog, without question the task will need to be spatially divided in some way, and parallelized over potentially several machines. It is likely that this will be the most time consuming and computationally intensive step. Additionally, in our current implementation, the pipeline is flagging an unusual amount of transient activity, which appears to be the fault of occasional observational artifacts which persist through the photometry pipeline. There is more information contained in the metadata of each apparition in the single-exposure database that will be used to further preprocess and improve the quality of data, reducing the effects of artifacting.}

\cc{As it pertains to VARnet, our results suggest that VARnet is a capable model architecture. With other goals in mind, VARnet could be leveraged on other surveys in search of other phenomena with the only change being the training and validation sets. In the case of WISE, VARnet proves promising for a full-sky pass.}

\newpage

\begin{acknowledgements}
    \section{ACKNOWLEDGEMENTS}
    
    I would like to acknowledge and thank deeply my mentor Davy (Dr. J. Davy Kirkpatrick) for introducing me to astronomy at IPAC and providing guidance throughout this project, aiding in data analysis and the collection of known objects for the test set.
    \vspace{1em}

    I would like to thank the Near-Earth Object Surveyor mission for covering the page charges for this paper through contract number 80MSFC20C0043.
    \vspace{1em}
    
    This publication makes use of data products from the Near-Earth Object Wide-field Infrared Survey Explorer (NEOWISE), which is a joint project of the Jet Propulsion Laboratory/California Institute of Technology and the University of Arizona. NEOWISE is funded by the National Aeronautics and Space Administration.
\end{acknowledgements}

\newpage
\appendix
\section{Wavelet Family}
\label{sec:appendixA}
\centering
Below are shown the biorthogonal and reverse biorthogonal variety of wavelets discussed in section \ref{subsec:wavelet-decomposition}.
\vspace{0.8em}

\includegraphics[scale=0.5]{ 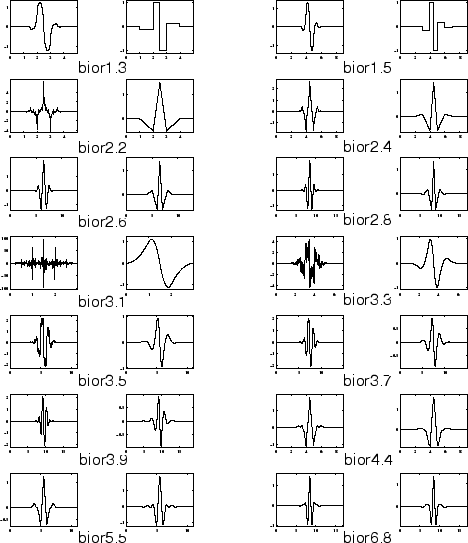} \\
\emph{The biorthogonal / reverse biorthogonal family of wavelets. The biorthogonal variety uses the left wavelet for decomposition and the right for reconstruction. The reverse biorthogonal variety uses the right wavelet for decomposition. Available at Mathworks website (\url{https://www.mathworks.com/help/wavelet/gs/introduction-to-the-wavelet-families.html})}

\newpage

\section{Data Generation Samples}
\label{sec:appendixB}
Below are shown 64 synthetic light curves as discussed in section \ref{subsec:datagen}. 

\begin{figure}[H]
    \label{fig:datagen-samples}
    \centering
    \subfloat{\includegraphics[scale=0.08]{  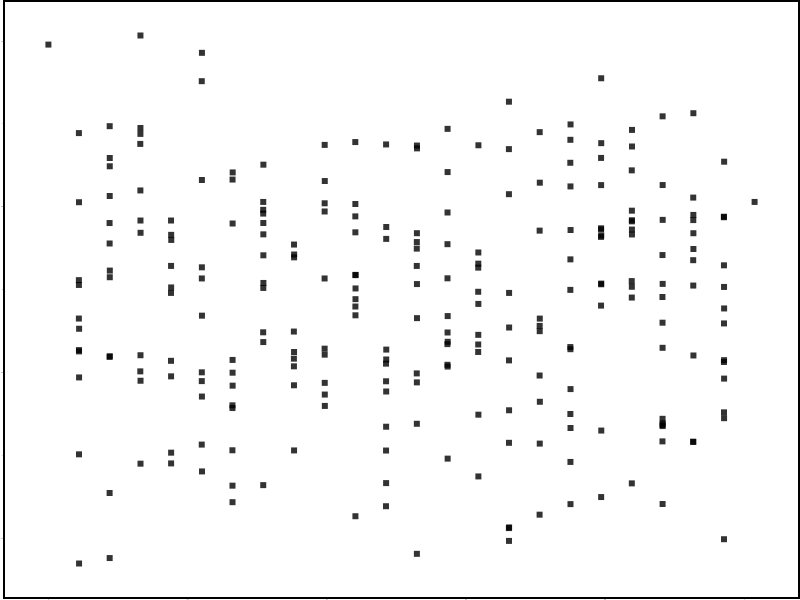}}
    \subfloat{\includegraphics[scale=0.08]{  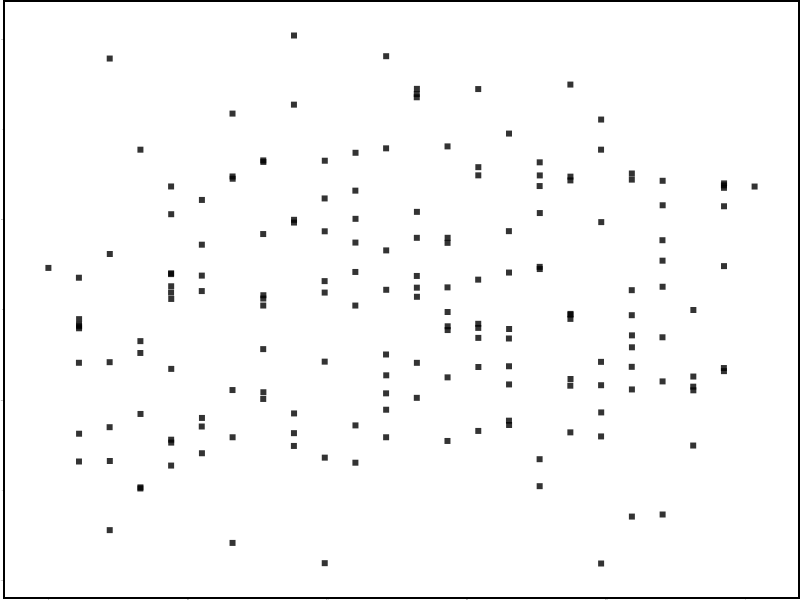}}
    \subfloat{\includegraphics[scale=0.08]{  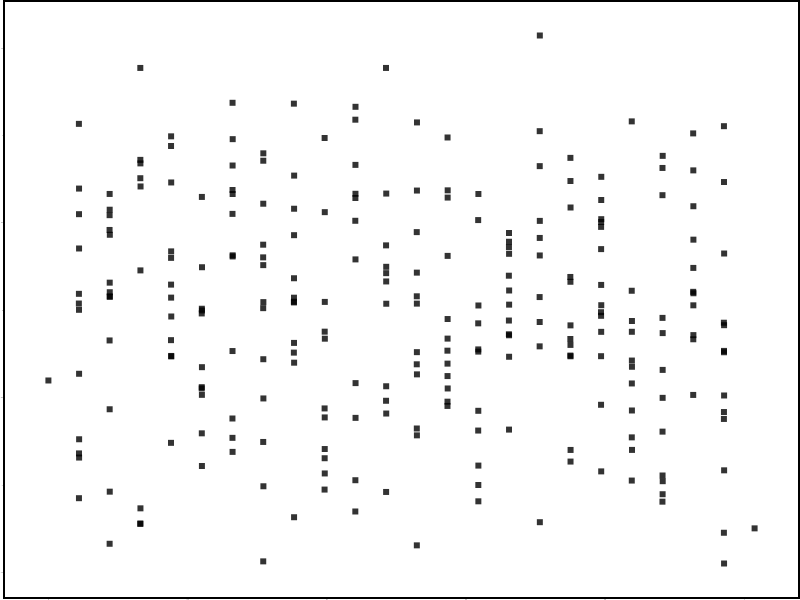}}
    \subfloat{\includegraphics[scale=0.08]{  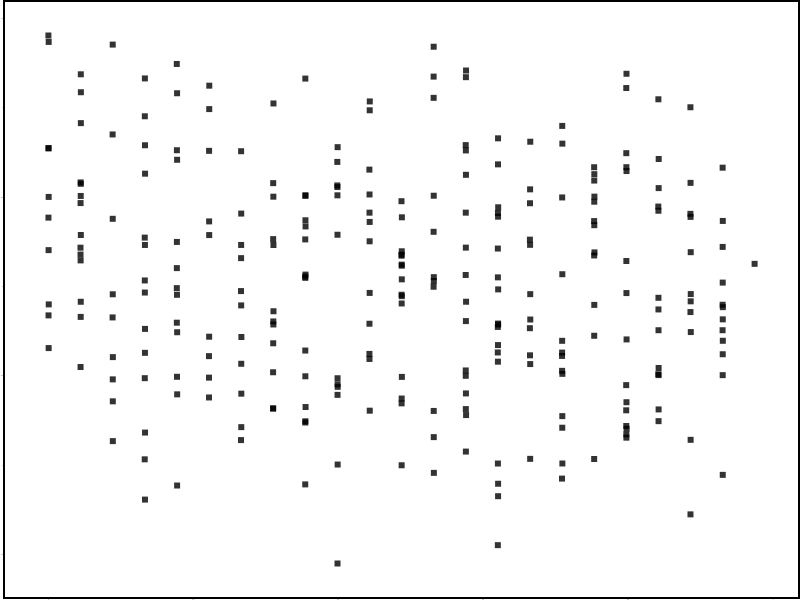}}
    \subfloat{\includegraphics[scale=0.08]{  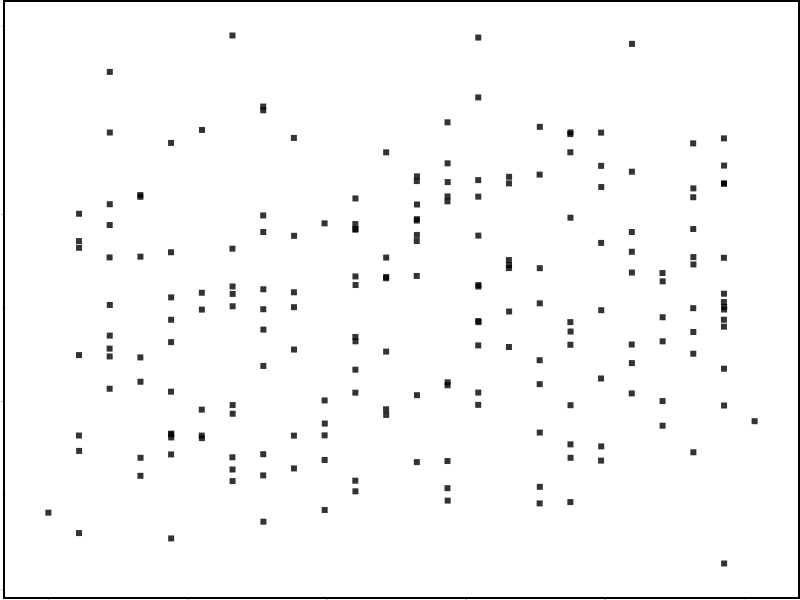}}
    \subfloat{\includegraphics[scale=0.08]{  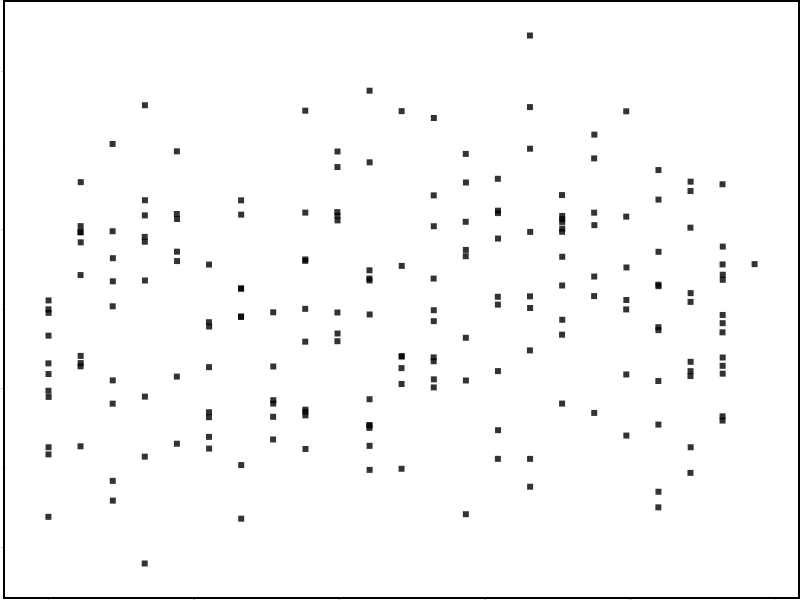}}
    \subfloat{\includegraphics[scale=0.08]{  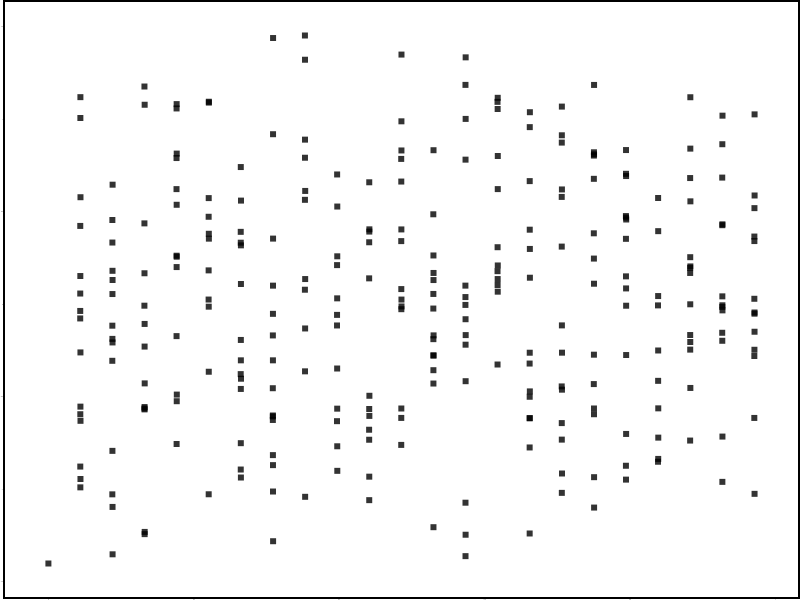}}
    \subfloat{\includegraphics[scale=0.08]{  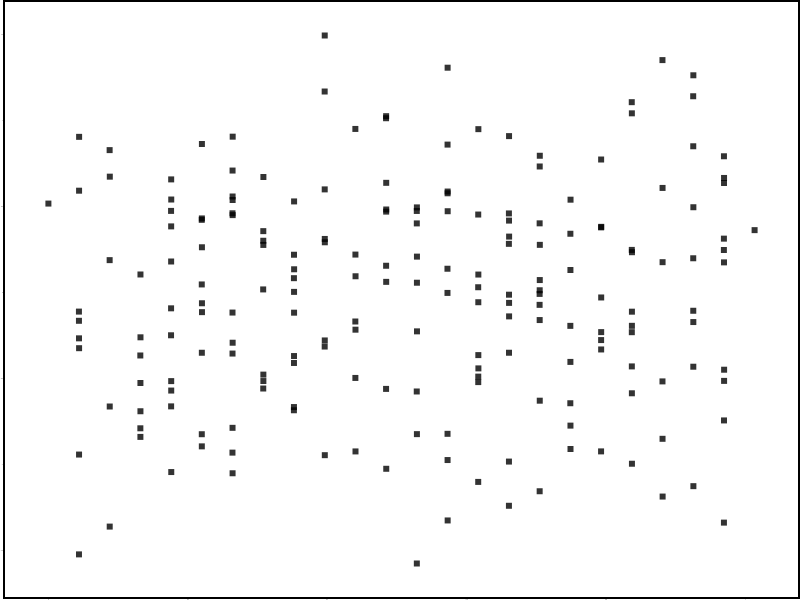}} \\
    \subfloat{\includegraphics[scale=0.08]{  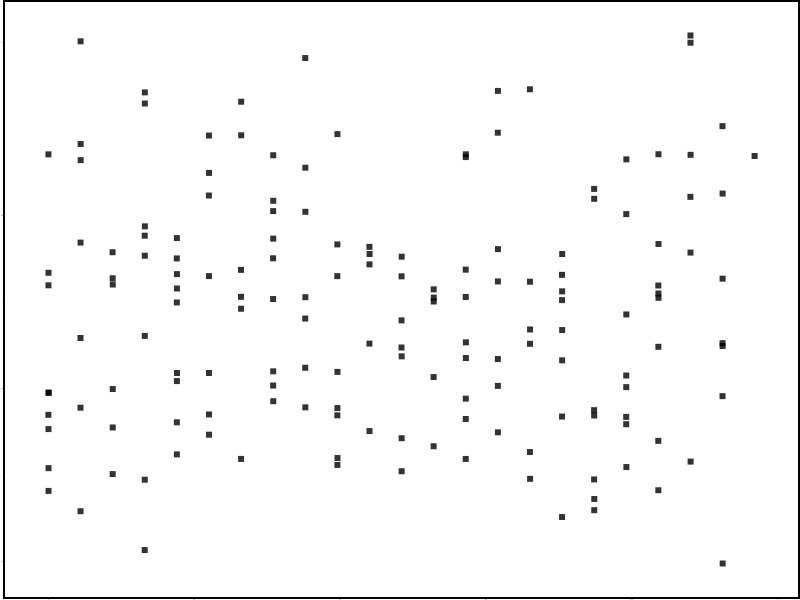}}
    \subfloat{\includegraphics[scale=0.08]{  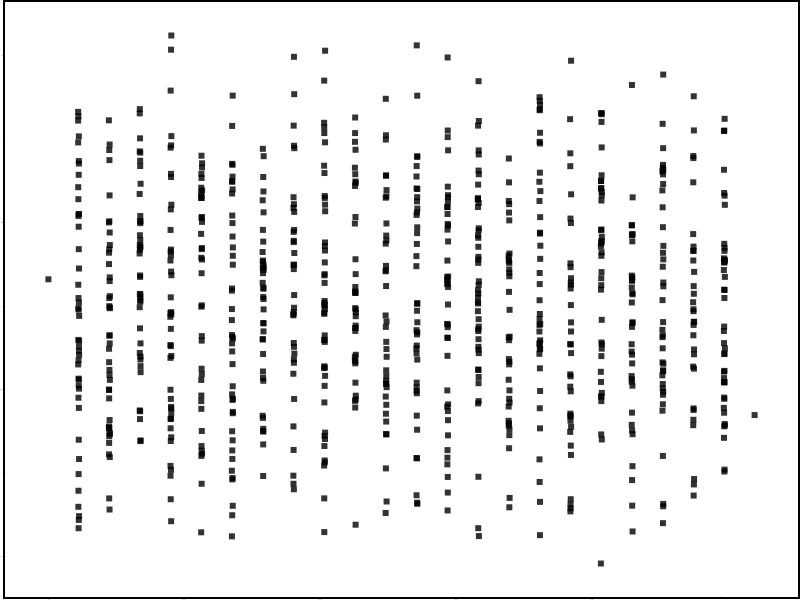}}
    \subfloat{\includegraphics[scale=0.08]{  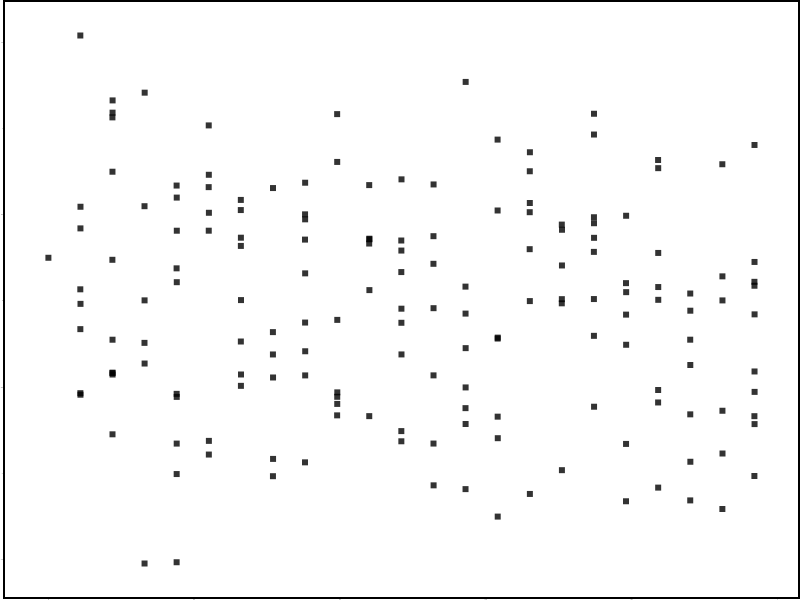}}
    \subfloat{\includegraphics[scale=0.08]{  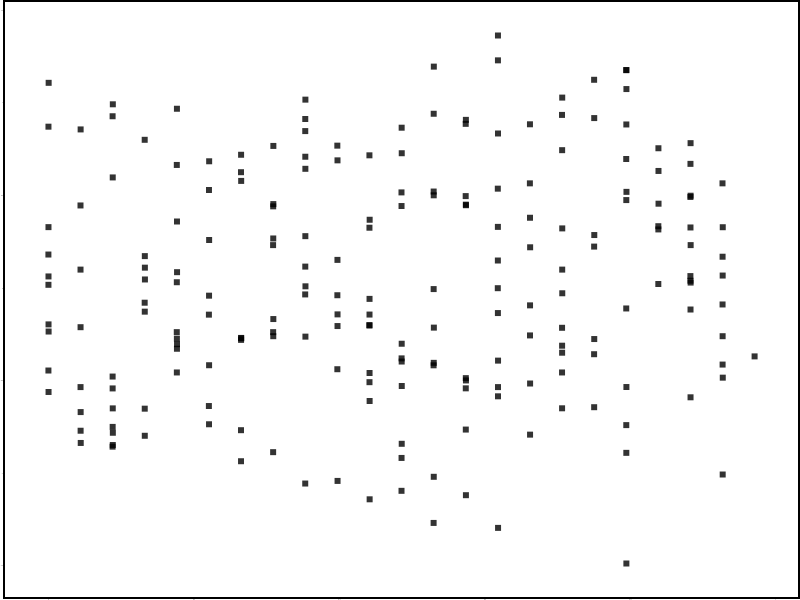}}
    \subfloat{\includegraphics[scale=0.08]{  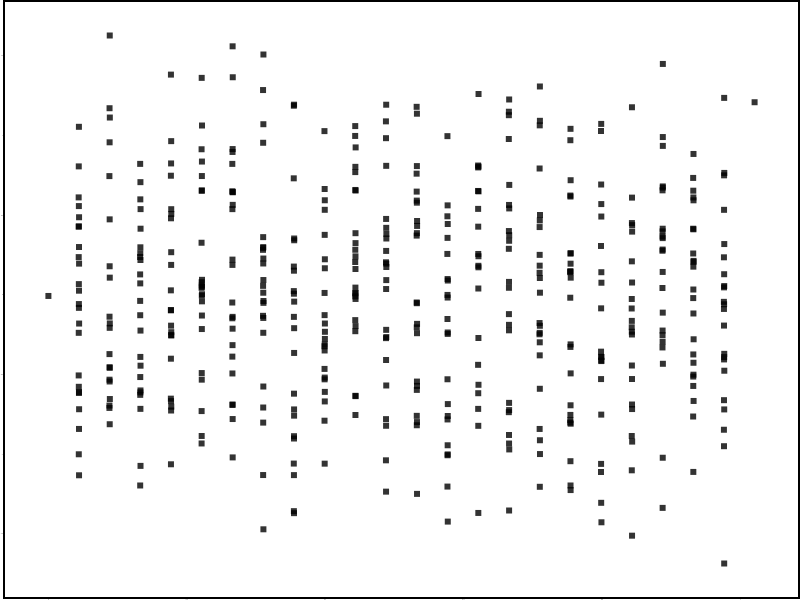}}
    \subfloat{\includegraphics[scale=0.08]{  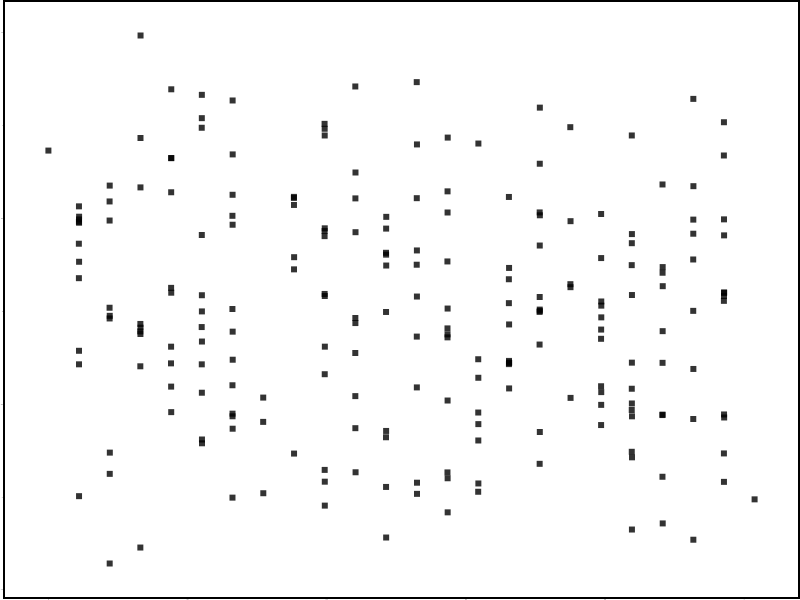}}
    \subfloat{\includegraphics[scale=0.08]{  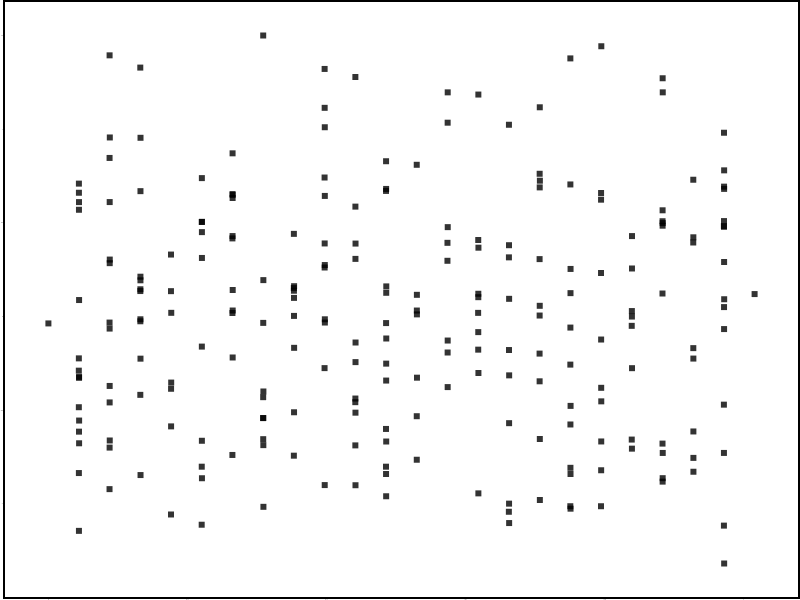}}
    \subfloat{\includegraphics[scale=0.08]{  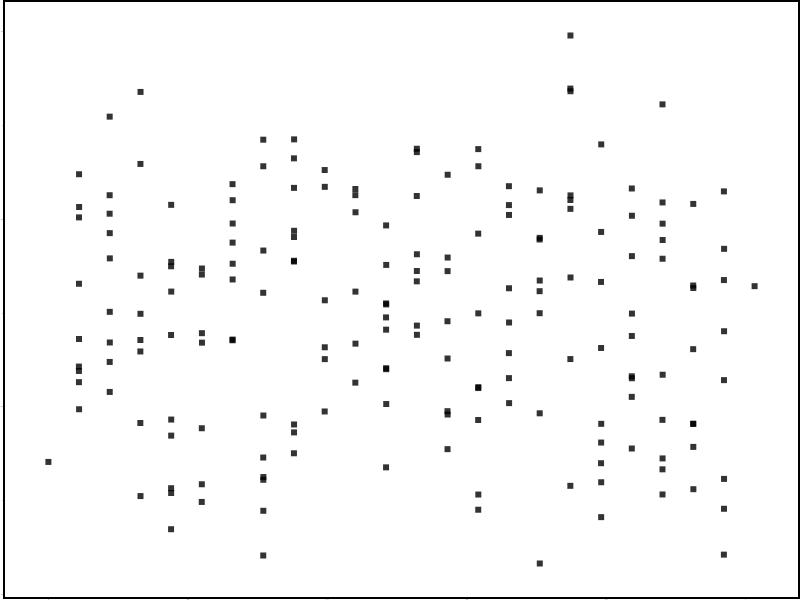}} \\
    \subfloat{\includegraphics[scale=0.08]{  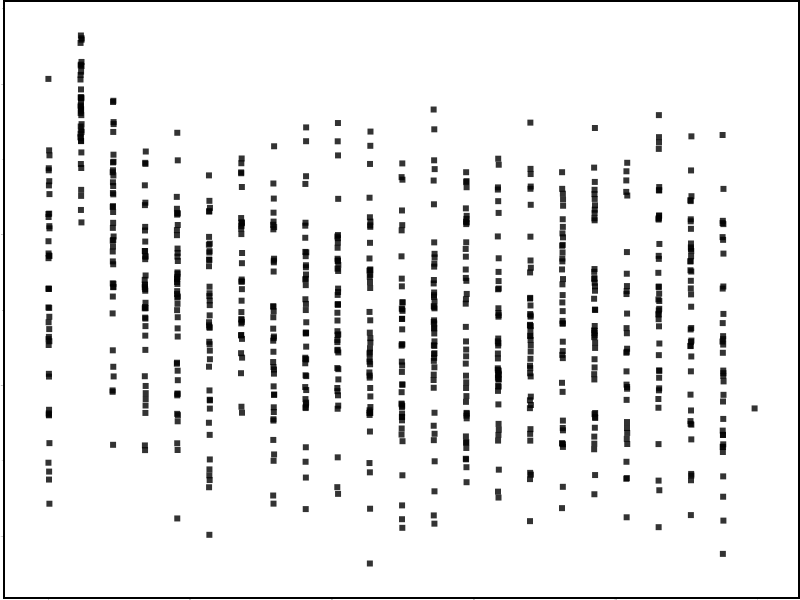}}
    \subfloat{\includegraphics[scale=0.08]{  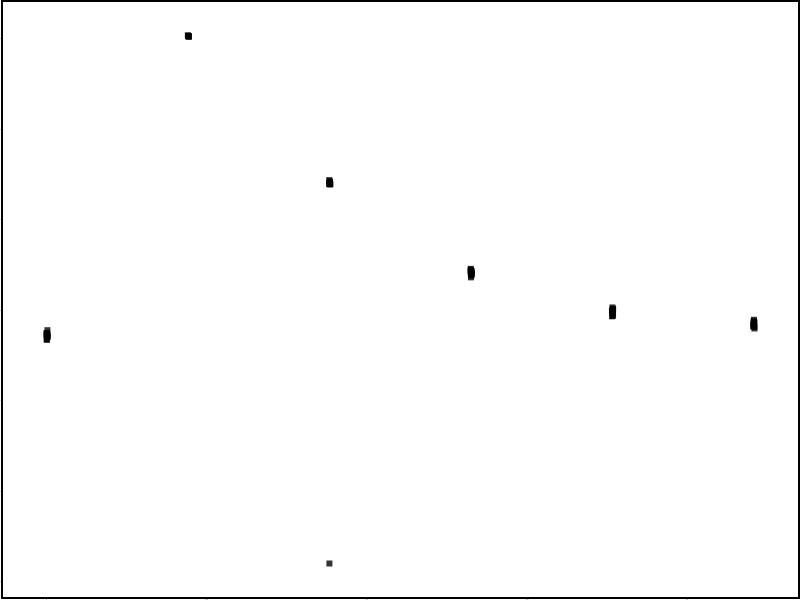}}
    \subfloat{\includegraphics[scale=0.08]{  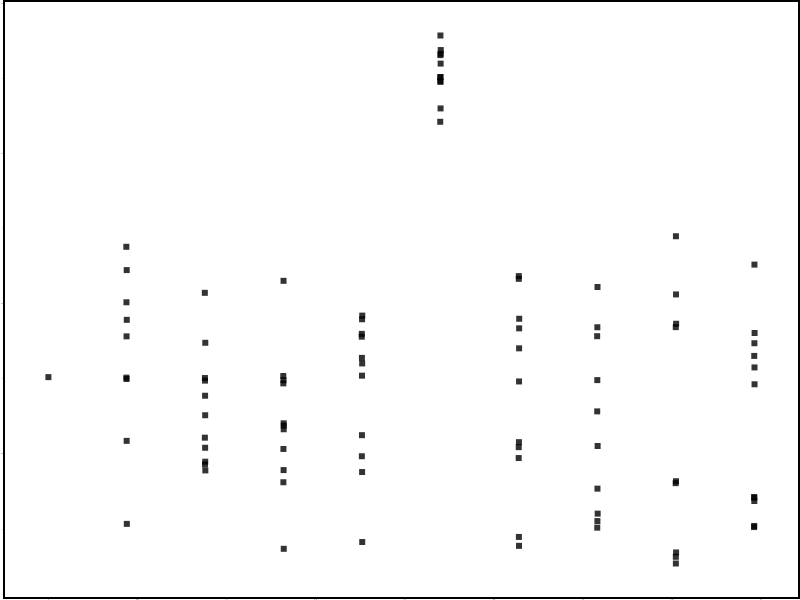}}
    \subfloat{\includegraphics[scale=0.08]{  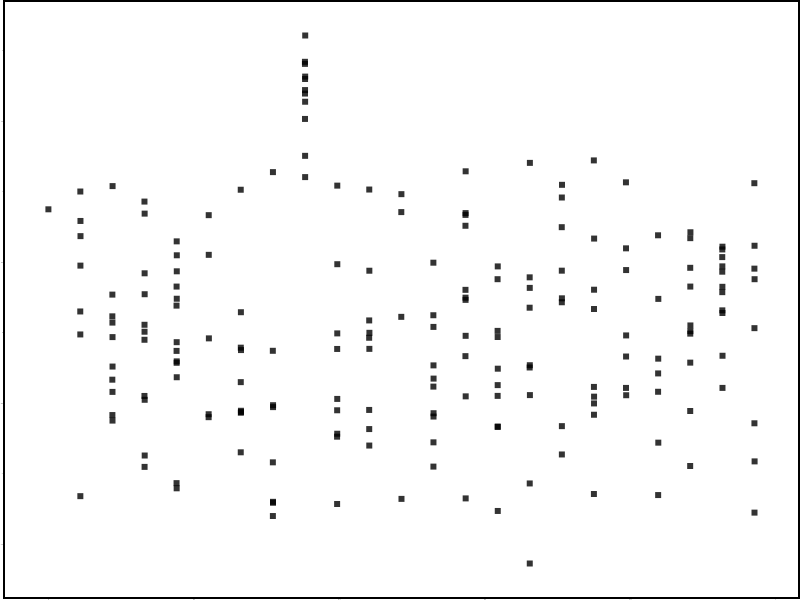}}
    \subfloat{\includegraphics[scale=0.08]{  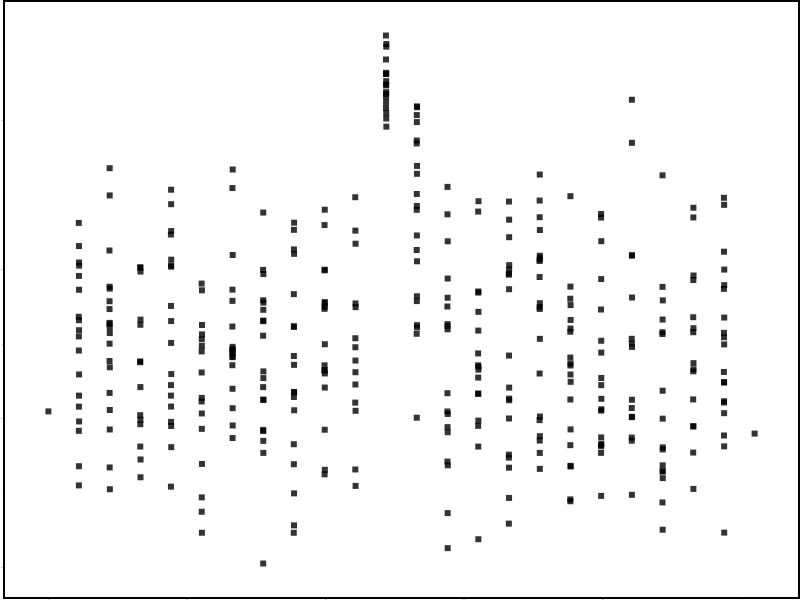}}
    \subfloat{\includegraphics[scale=0.08]{  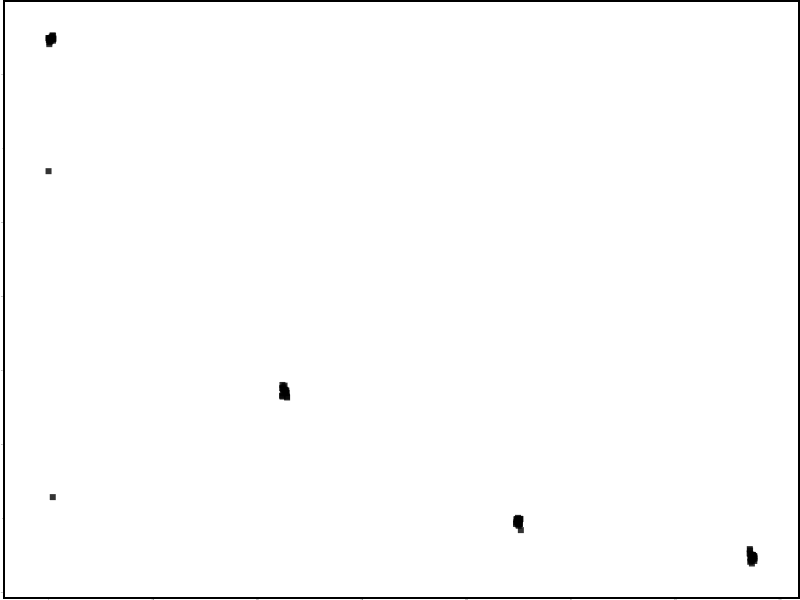}}
    \subfloat{\includegraphics[scale=0.08]{  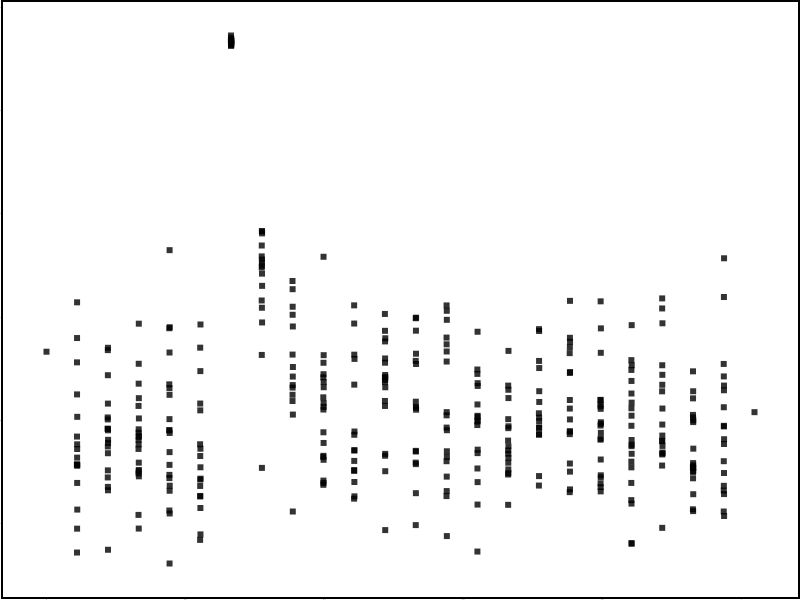}}
    \subfloat{\includegraphics[scale=0.08]{  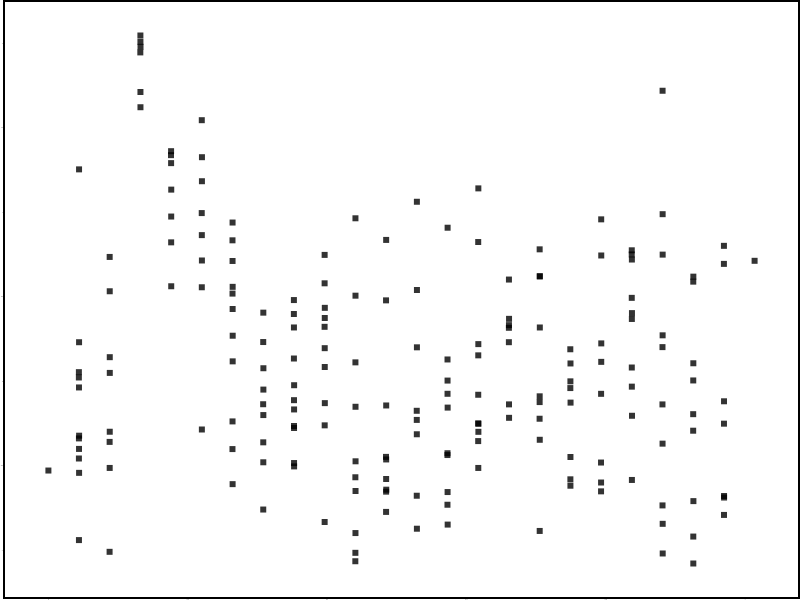}} \\
    \subfloat{\includegraphics[scale=0.08]{  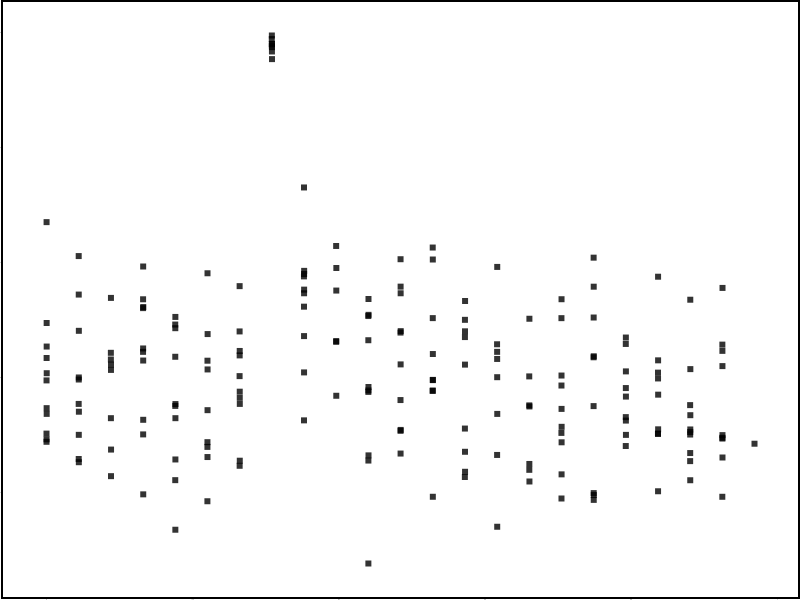}}
    \subfloat{\includegraphics[scale=0.08]{  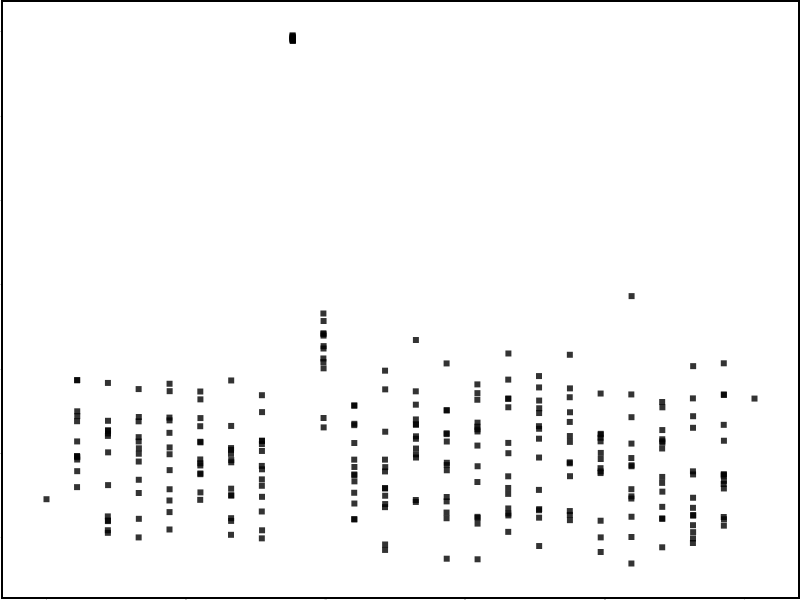}}
    \subfloat{\includegraphics[scale=0.08]{  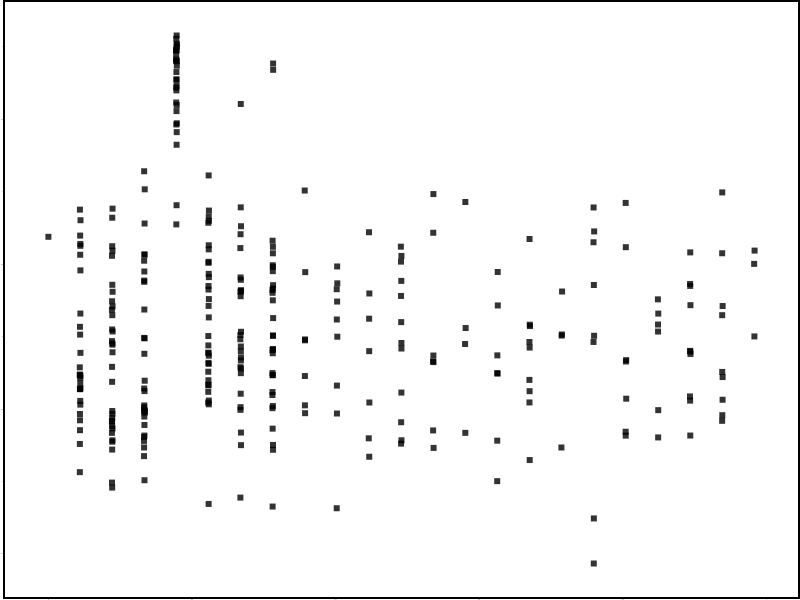}}
    \subfloat{\includegraphics[scale=0.08]{  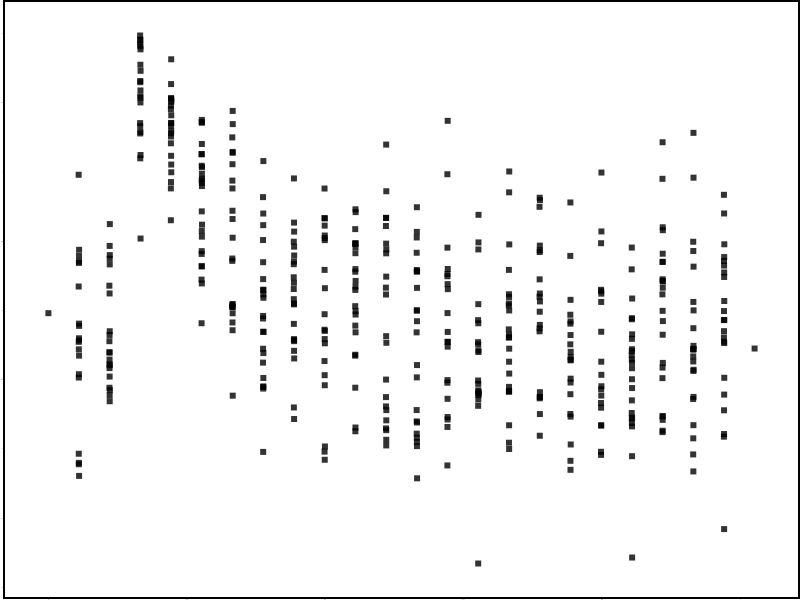}}
    \subfloat{\includegraphics[scale=0.08]{  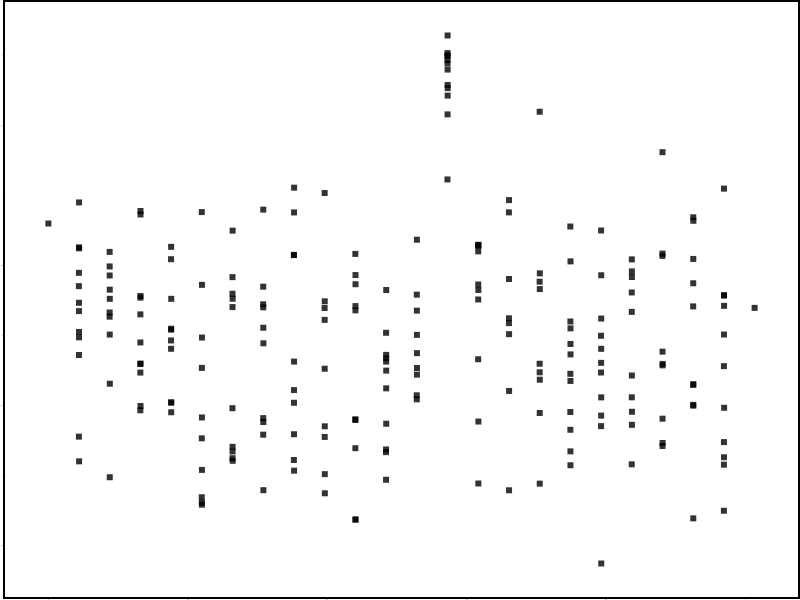}}
    \subfloat{\includegraphics[scale=0.08]{  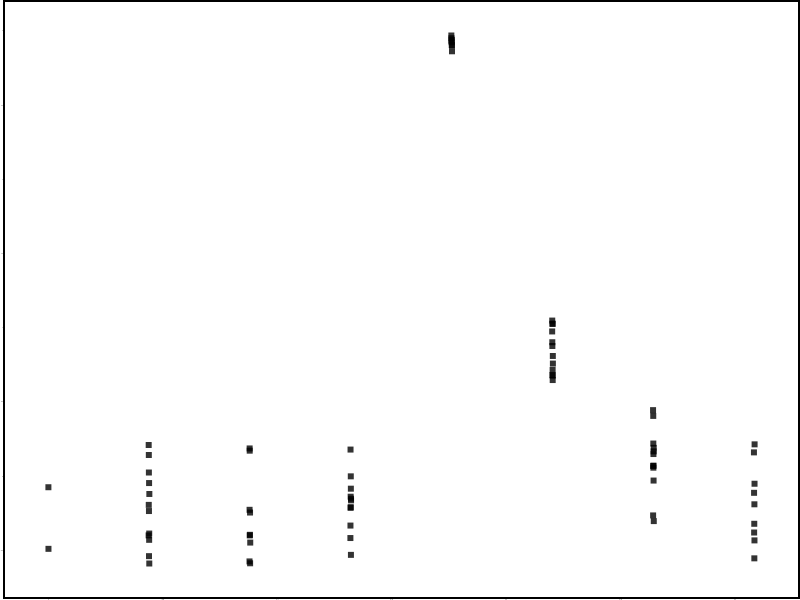}}
    \subfloat{\includegraphics[scale=0.08]{  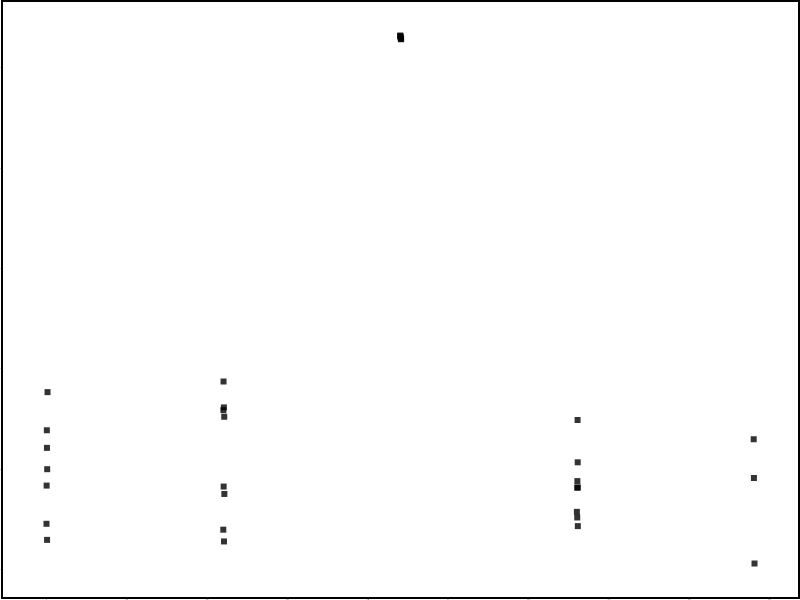}}
    \subfloat{\includegraphics[scale=0.08]{  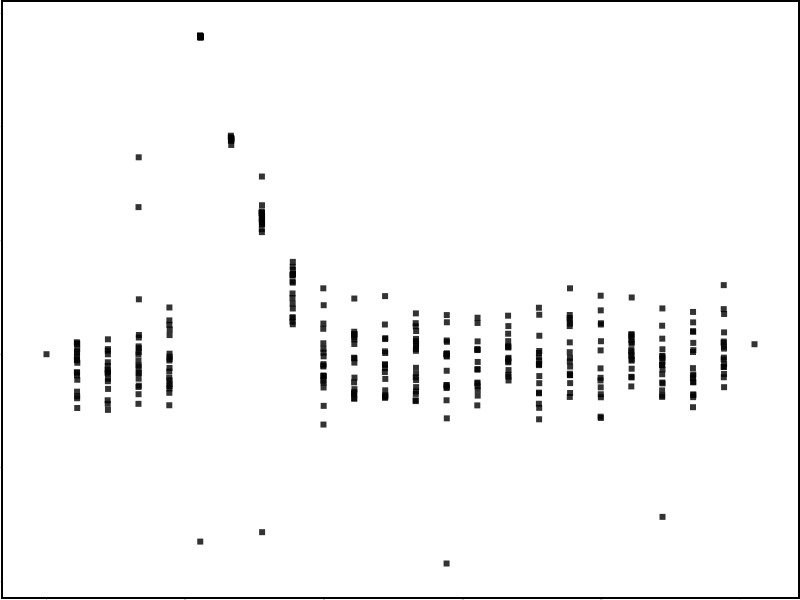}} \\
    \subfloat{\includegraphics[scale=0.08]{  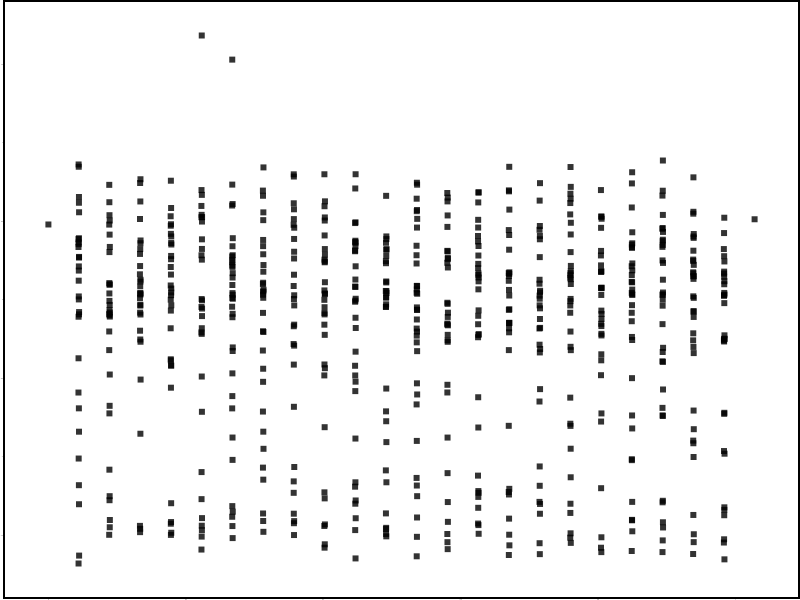}}
    \subfloat{\includegraphics[scale=0.08]{  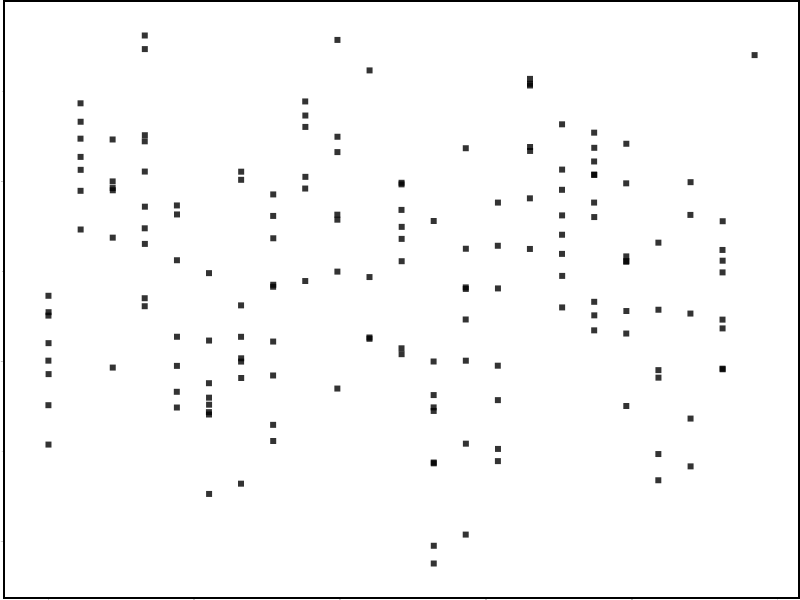}}
    \subfloat{\includegraphics[scale=0.08]{  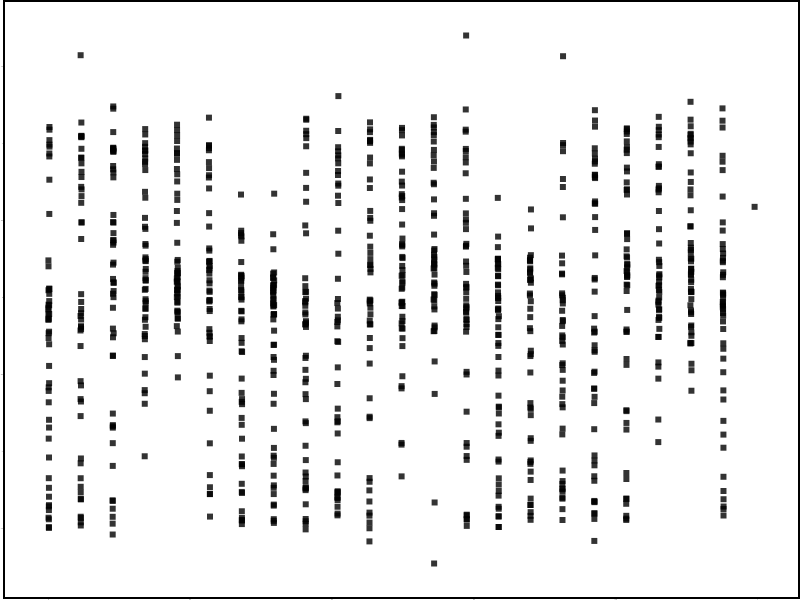}}
    \subfloat{\includegraphics[scale=0.08]{  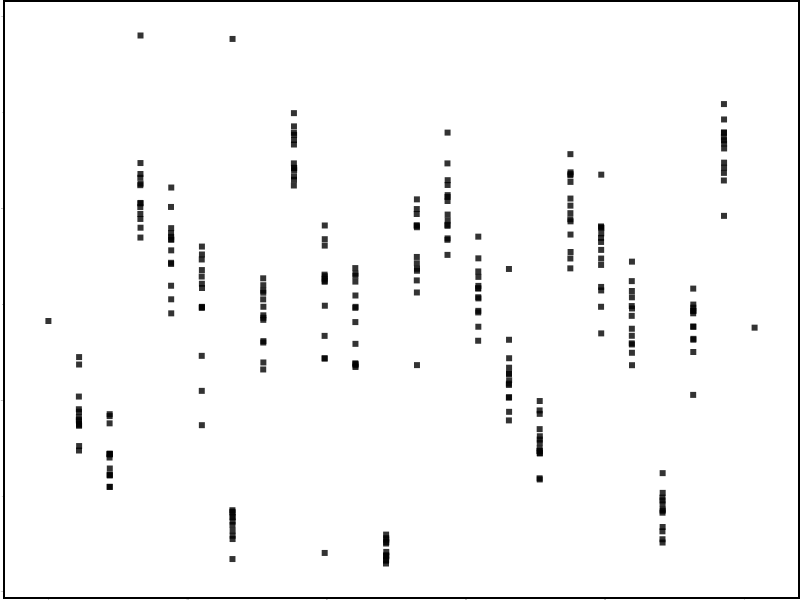}}
    \subfloat{\includegraphics[scale=0.08]{  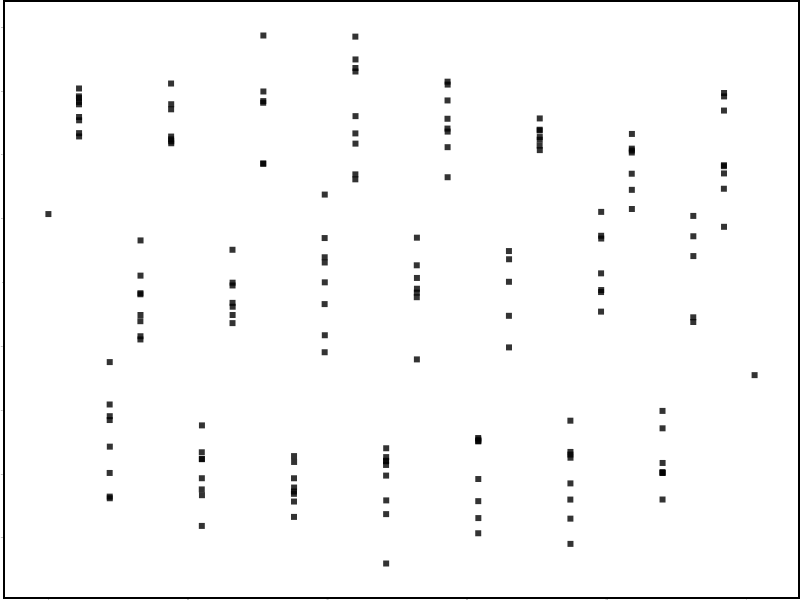}}
    \subfloat{\includegraphics[scale=0.08]{  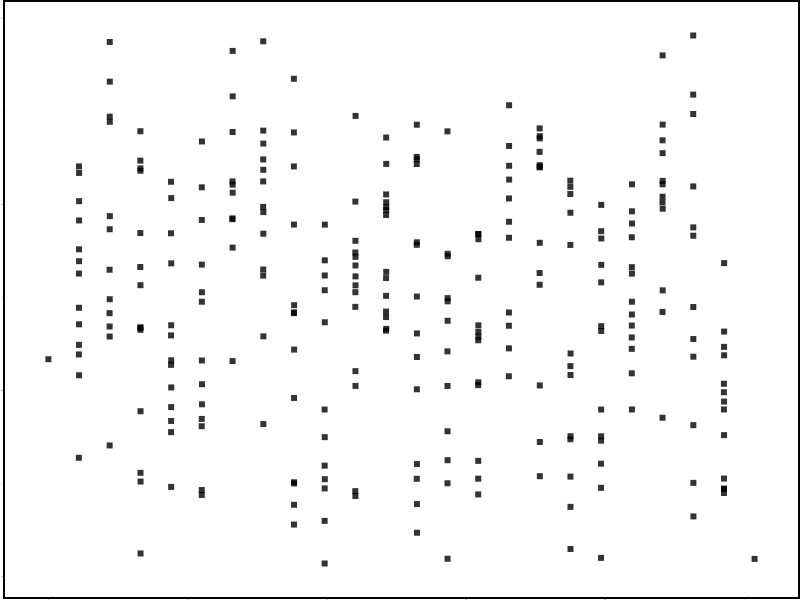}}
    \subfloat{\includegraphics[scale=0.08]{  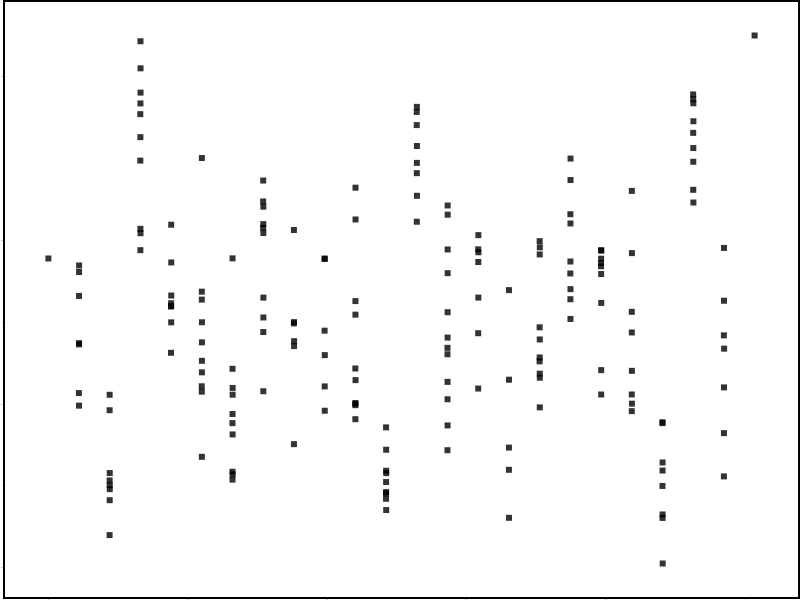}}
    \subfloat{\includegraphics[scale=0.08]{  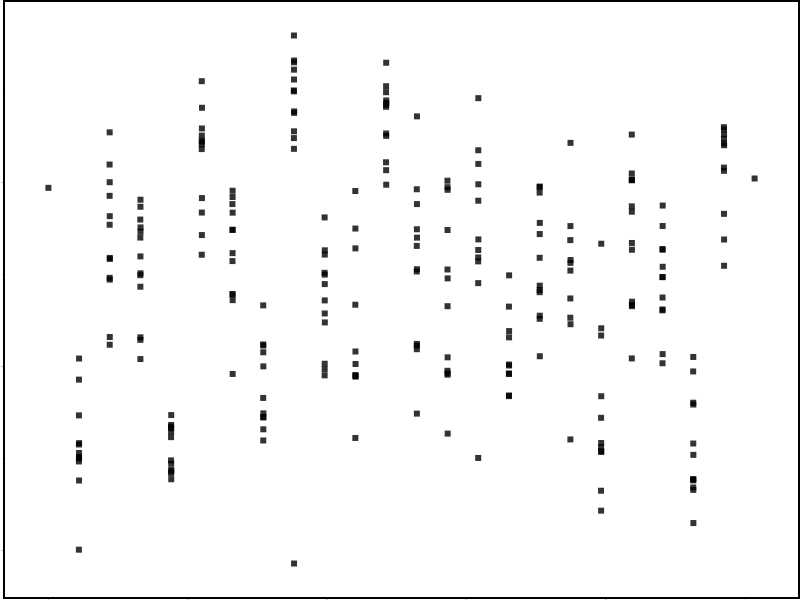}} \\
    \subfloat{\includegraphics[scale=0.08]{  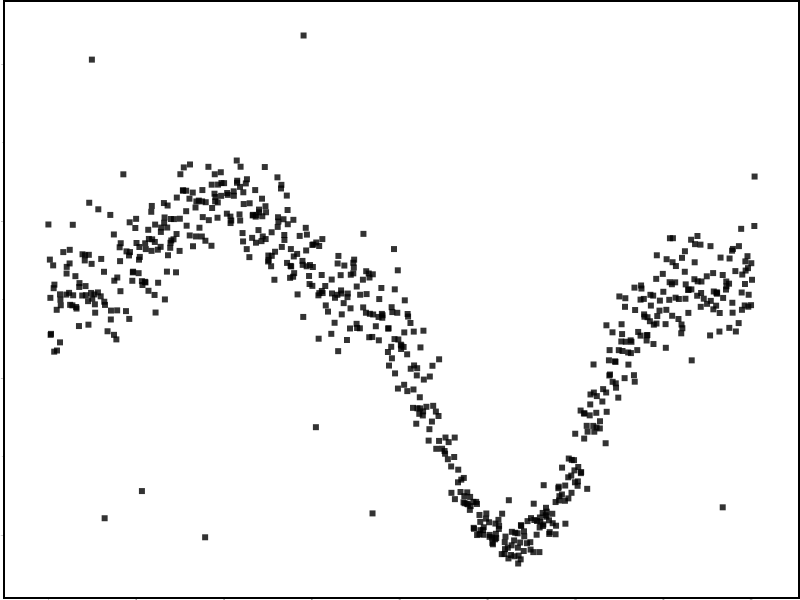}}
    \subfloat{\includegraphics[scale=0.08]{  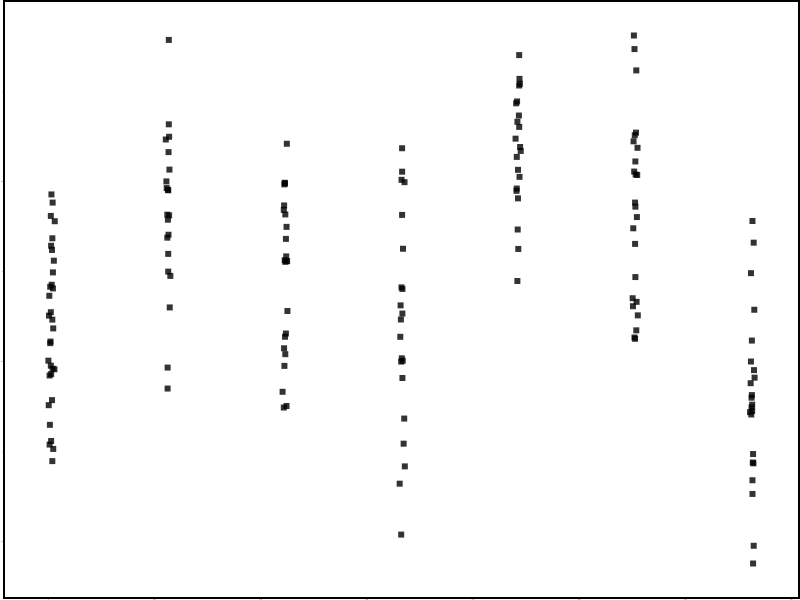}}
    \subfloat{\includegraphics[scale=0.08]{  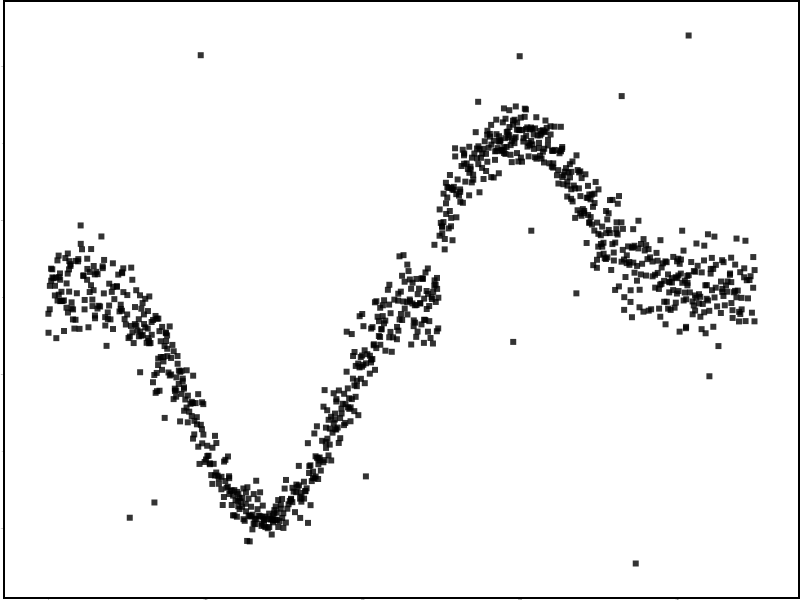}}
    \subfloat{\includegraphics[scale=0.08]{  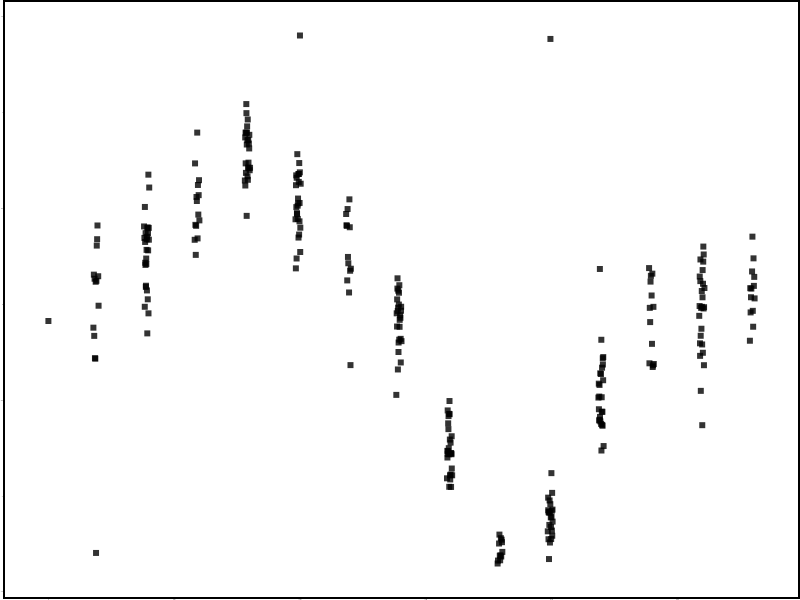}}
    \subfloat{\includegraphics[scale=0.08]{  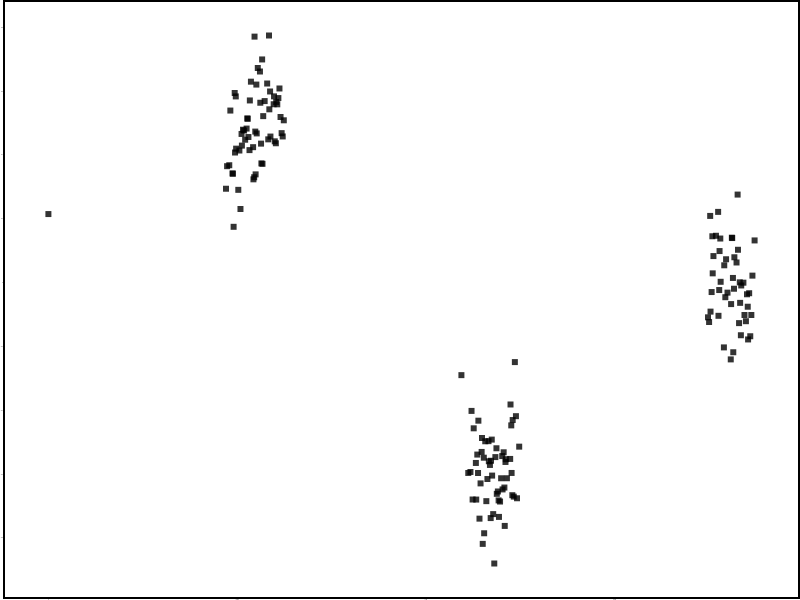}}
    \subfloat{\includegraphics[scale=0.08]{  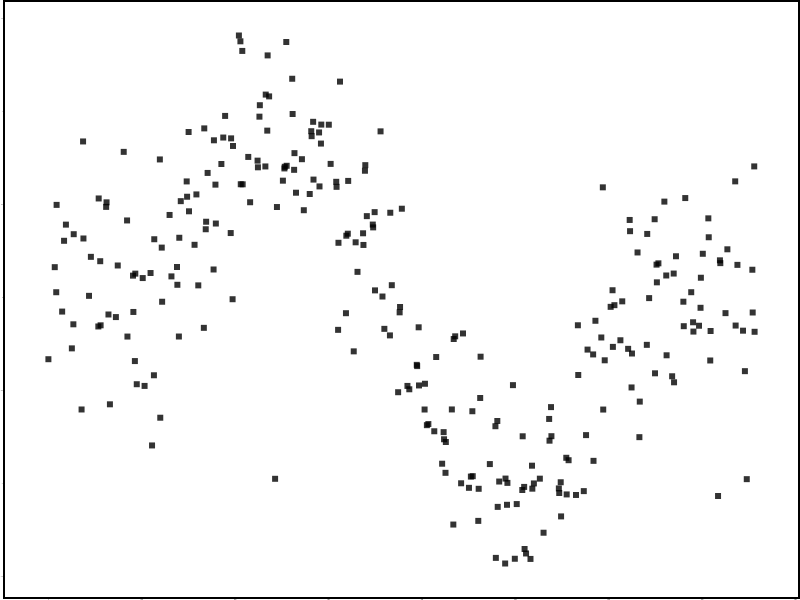}}
    \subfloat{\includegraphics[scale=0.08]{  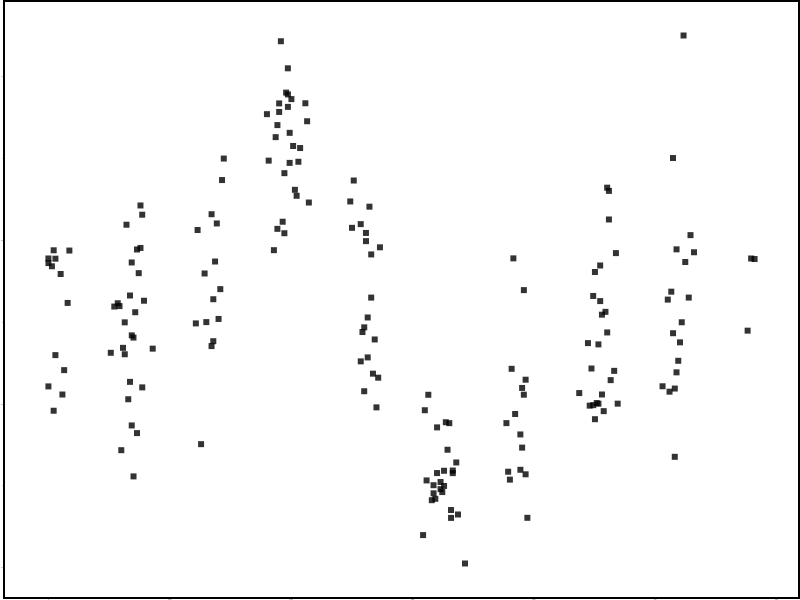}}
    \subfloat{\includegraphics[scale=0.08]{  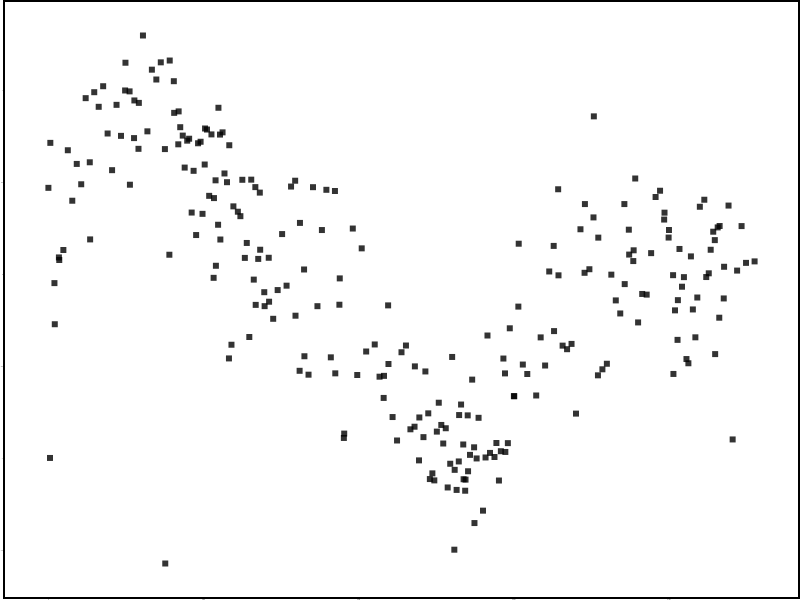}} \\
    \subfloat{\includegraphics[scale=0.08]{  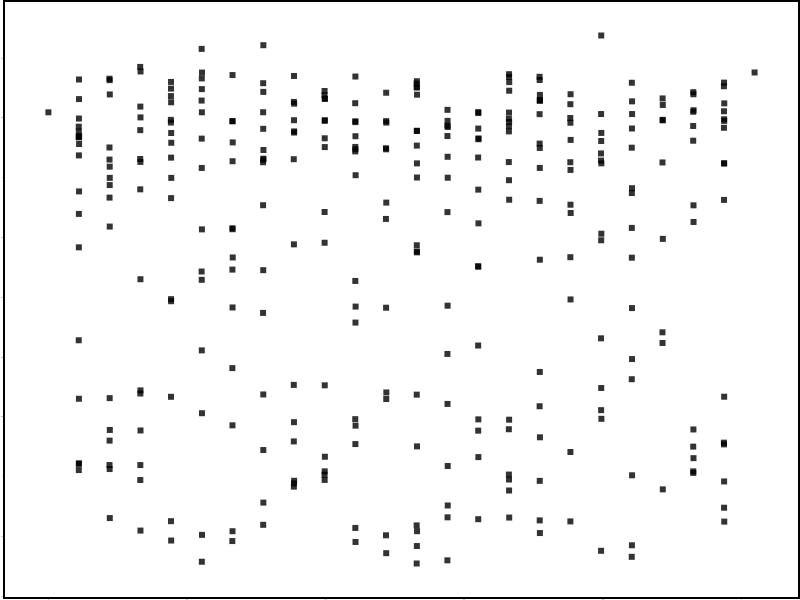}}
    \subfloat{\includegraphics[scale=0.08]{  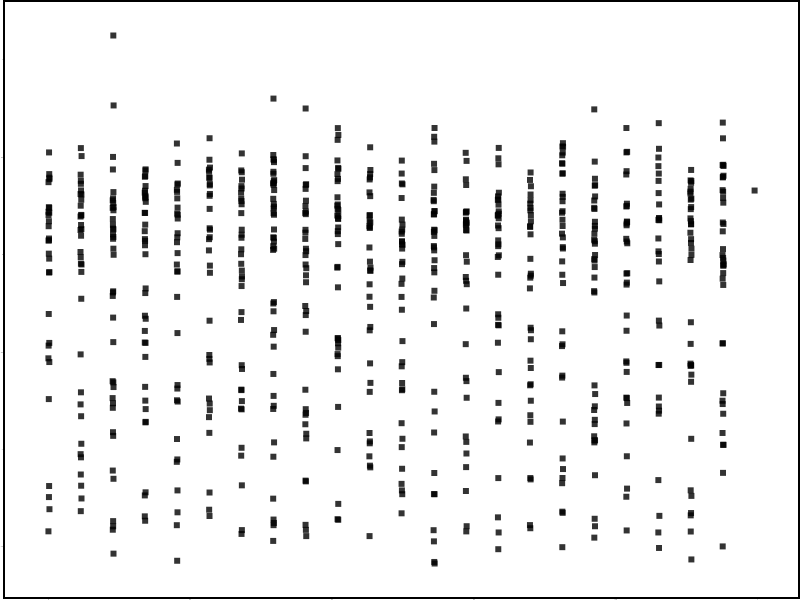}}
    \subfloat{\includegraphics[scale=0.08]{  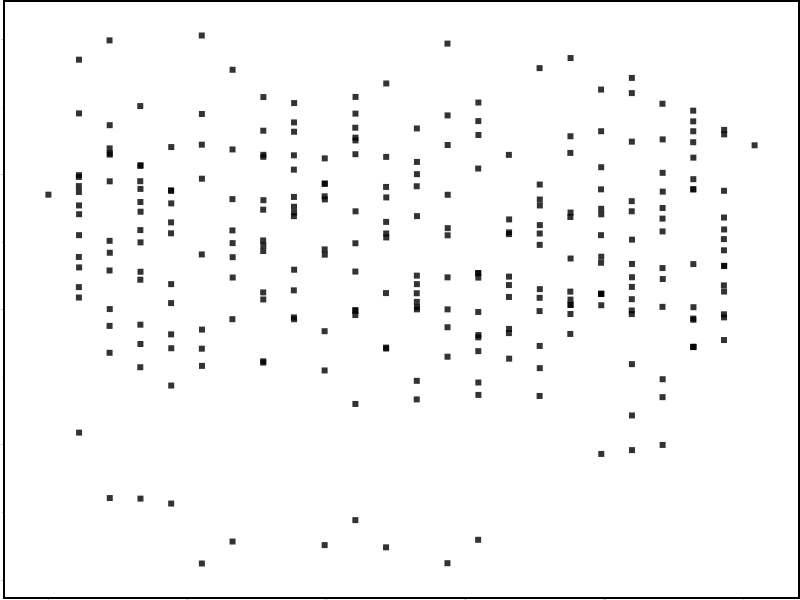}}
    \subfloat{\includegraphics[scale=0.08]{  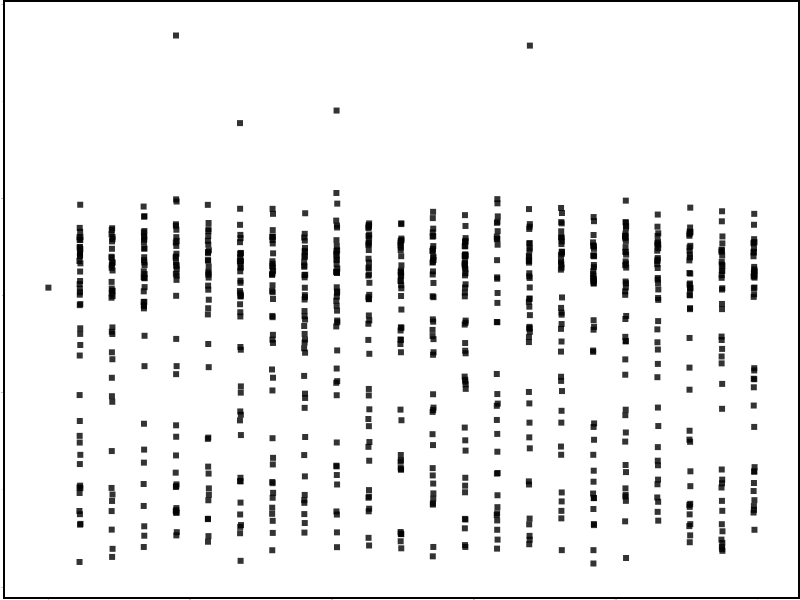}}
    \subfloat{\includegraphics[scale=0.08]{  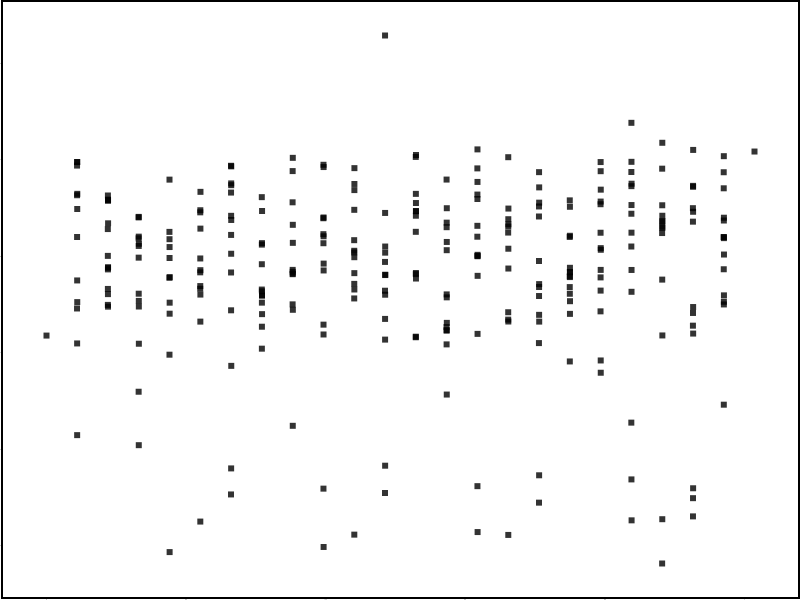}}
    \subfloat{\includegraphics[scale=0.08]{  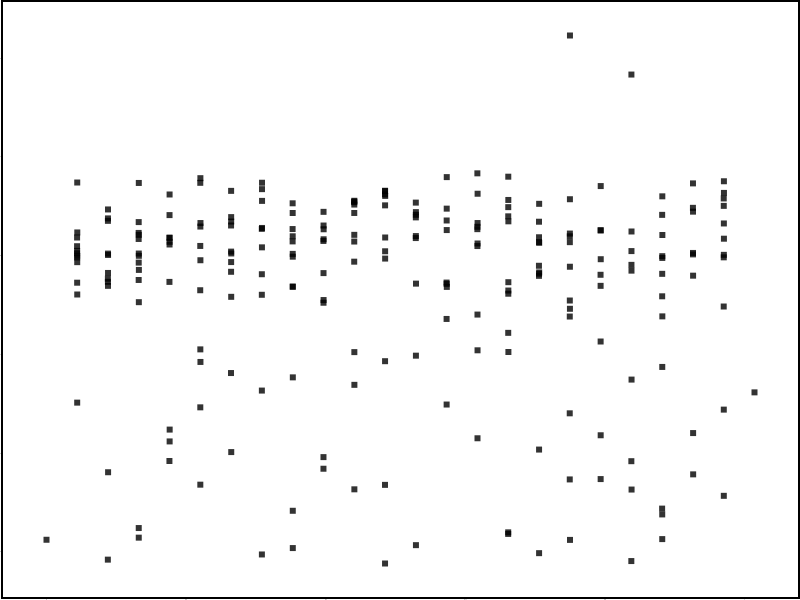}}
    \subfloat{\includegraphics[scale=0.08]{  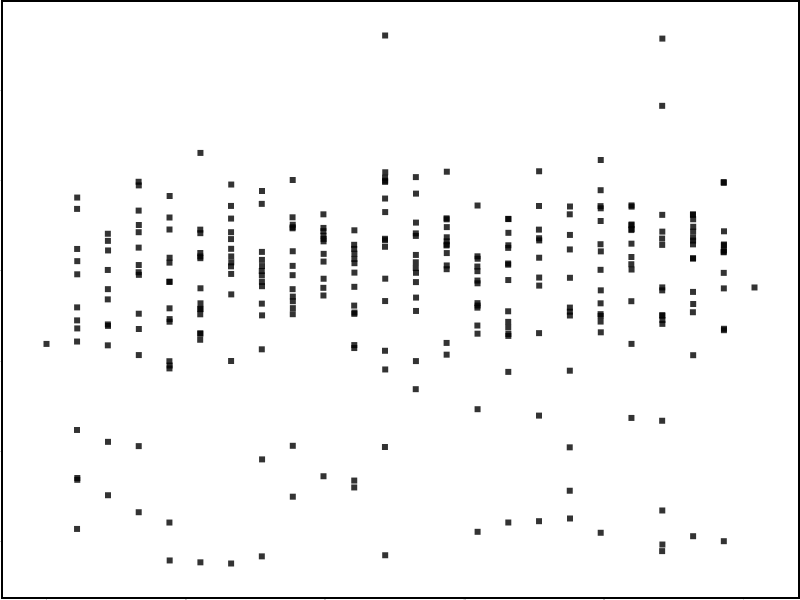}}
    \subfloat{\includegraphics[scale=0.08]{  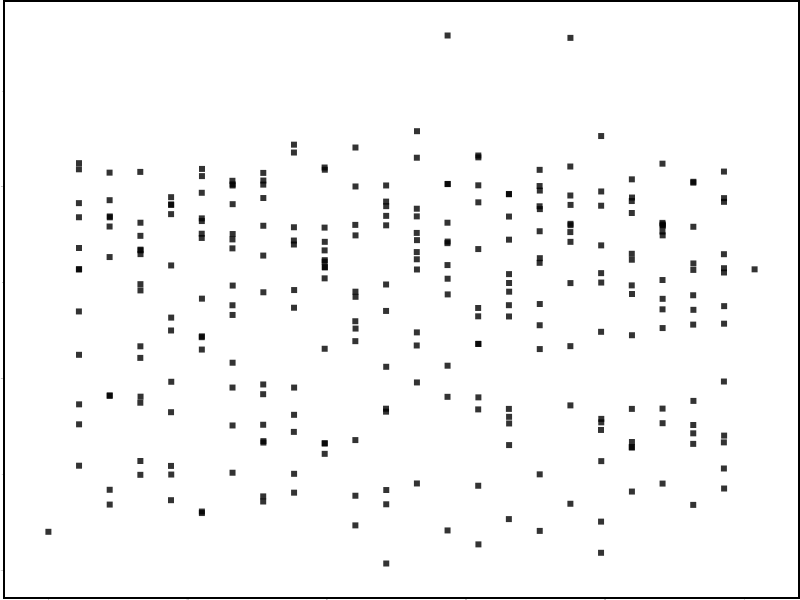}} \\
    \subfloat{\includegraphics[scale=0.08]{  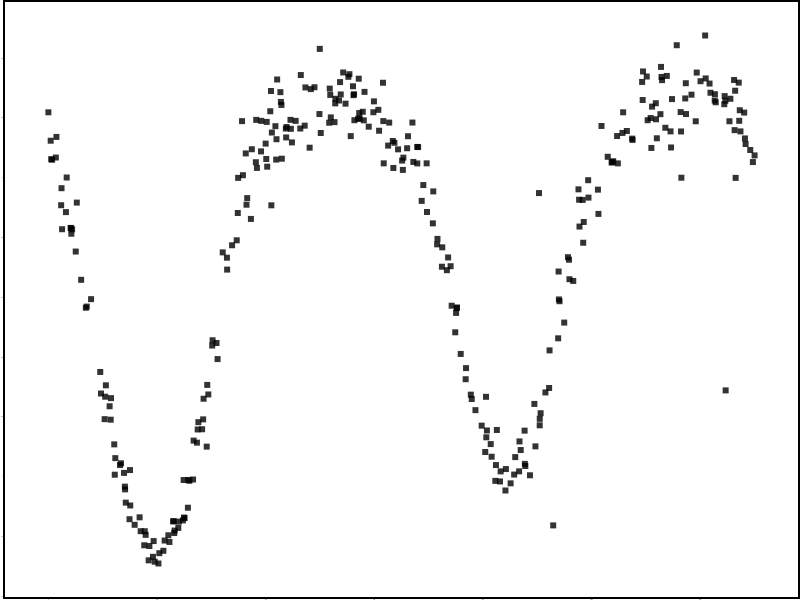}}
    \subfloat{\includegraphics[scale=0.08]{  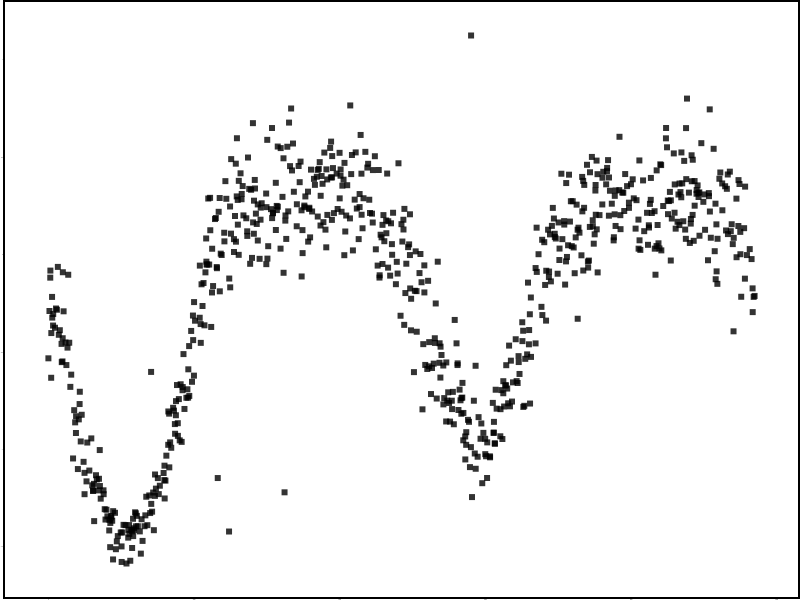}}
    \subfloat{\includegraphics[scale=0.08]{  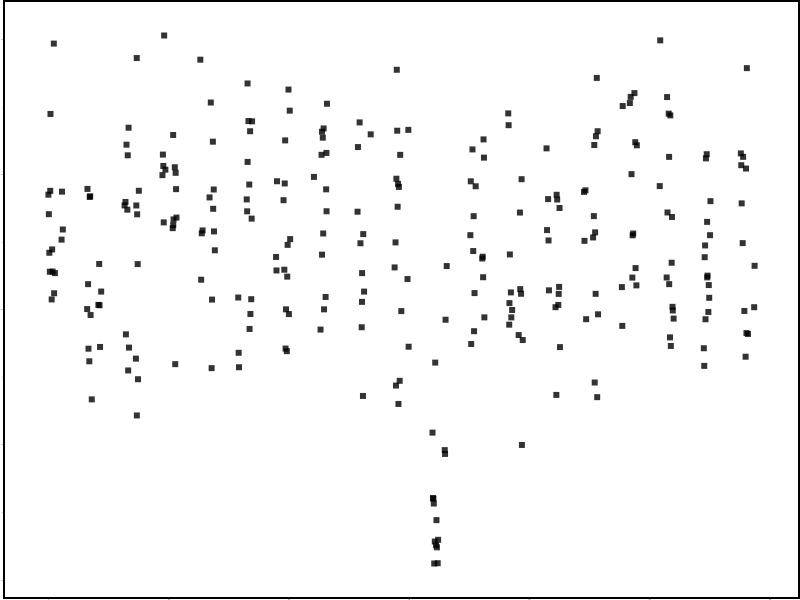}}
    \subfloat{\includegraphics[scale=0.08]{  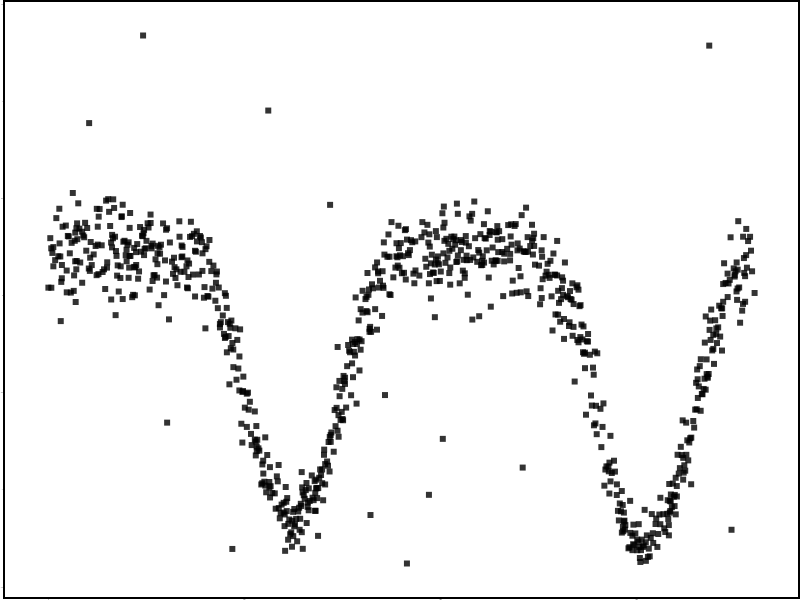}}
    \subfloat{\includegraphics[scale=0.08]{  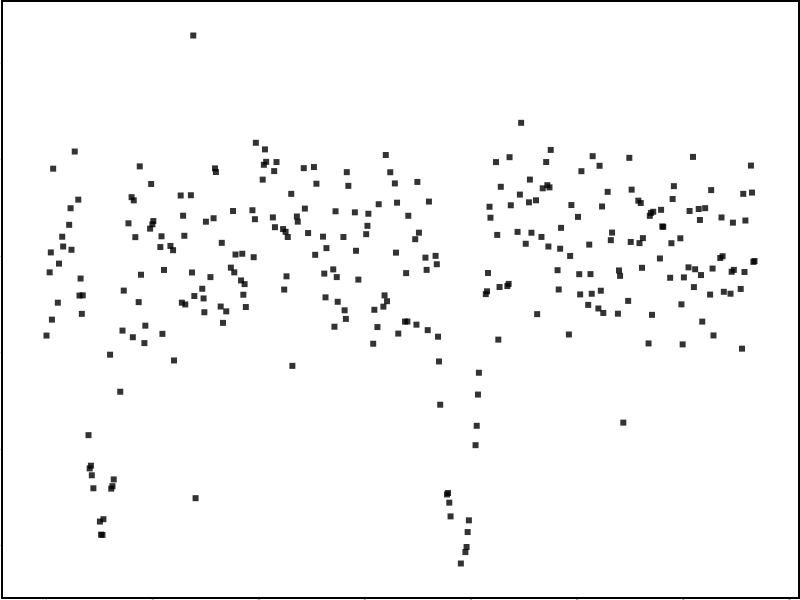}}
    \subfloat{\includegraphics[scale=0.08]{  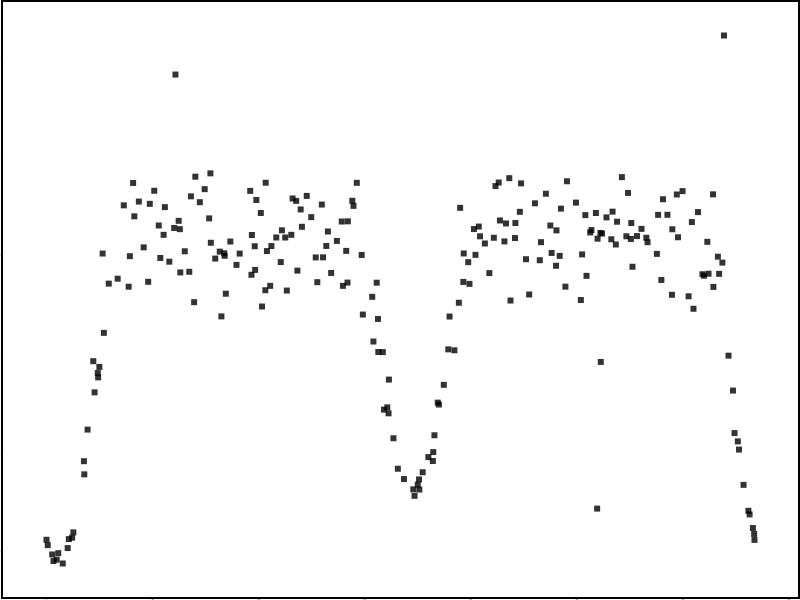}}
    \subfloat{\includegraphics[scale=0.08]{  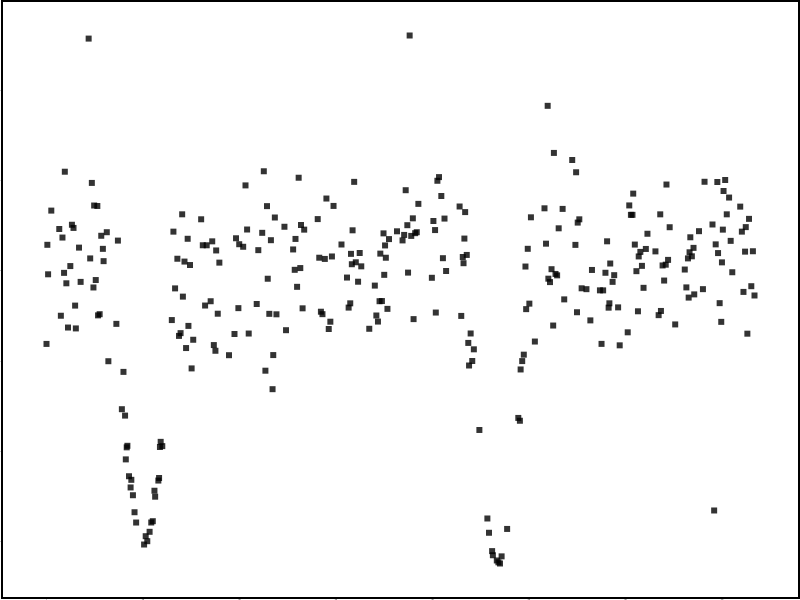}}
    \subfloat{\includegraphics[scale=0.08]{  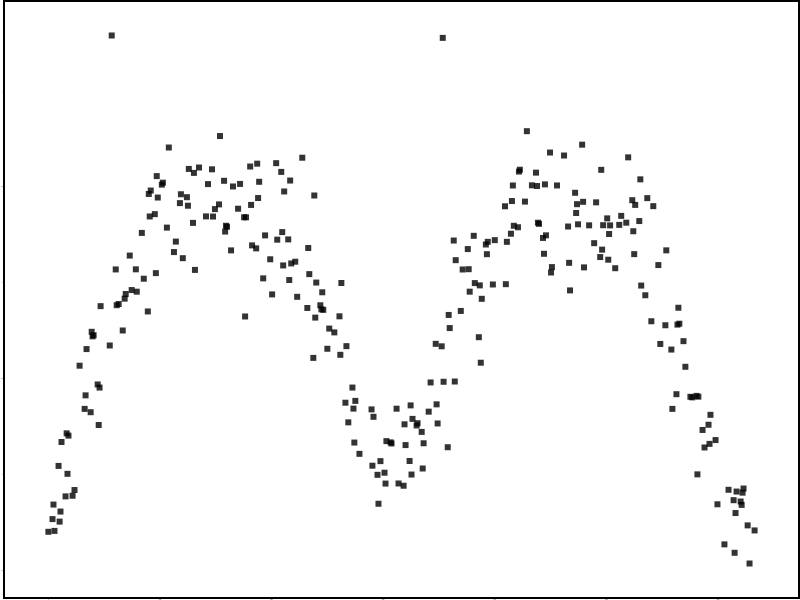}}

    \caption{\centering Sample of synthetically generated light curves. Row 1-2 are null stars, row 3-4 are novae. Light curves in row 5 are pulsators, with their phase folded waveforms directly below in row 6. Row 7 is transits, with their phase folded counterparts directly below.}
\end{figure}

\vskip 0.2in

\newpage
\bibliography{main}{}
\bibliographystyle{aasjournal}



\end{document}